\documentclass[twocolumn]{aastex631}
\usepackage[normalem]{ulem}
\usepackage{cancel}
\usepackage{soul}
\usepackage{graphicx}
\def\la{\hbox{\raise.35ex\rlap{$<$}\lower.6ex\hbox{$\sim$}\ }}
\def\ga{\hbox{\raise.35ex\rlap{$>$}\lower.6ex\hbox{$\sim$}\ }}

\def\water{H$_2$O }

\def\Eeff{\big< E_{{\rm eff}}\big>}
\def\beq{\begin{equation}}
\def\eeq{\end{equation}}
\def\beqa{\begin{eqnarray}}
\def\eeqa{\end{eqnarray}}

\def\sub#1{_{_{#1}}}

\newcommand{\sfrac}[2]{\small \mbox{$\frac{#1}{#2}$}}
\newcommand{\REVV}[1]{#1}
\newcommand{\REV}[1]{#1}

\include{epsf}
\include{epstopdf}
\DeclareGraphicsExtensions{.png,.jpg,.pdf,.eps}
\received{\today}
\shorttitle{A Near Surface Temperature Model of Arrokoth}
\shortauthors{Umurhan et al.}
\graphicspath{{./}{figures/}}

\begin{document}

\title{A Near Surface Temperature Model of Arrokoth}

\correspondingauthor{Orkan M. Umurhan}
\email{oumurhan@seti.org, orkan.m.umurhan@nasa.gov}

\author[0000-0001-5372-4254]{Orkan M. Umurhan}
\altaffiliation{New Horizons Science Team, Co-Investigator}
\affiliation{SETI Institute, 189 Bernardo Way, Mountain View, CA 94043, U.S.A.}
\affiliation{NASA Ames Research Center; Mail Stop 245-3, Moffett Field, CA 94035, USA}
\affiliation{Cornell Center for Astrophysics and Planetary Sciences, Cornell University, Ithaca, NY, USA}
\author[0000-0002-8296-6540]{William M. Grundy}
\altaffiliation{New Horizons Science Team, Co-Investigator}
\affiliation{Lowell Observatory, 1400 W. Mars Hill Road, Flagstaff, AZ 86001, USA}
\author[0000-0002-0308-8752]{Michael K. Bird}
\altaffiliation{New Horizons Science Team, Science Affiliate}
\affiliation{Argelander-Institut f\"ur Astronomie, Universit\"at Bonn, 53121 Bonn, Germany.}
\affiliation{Rheinisches Institut f\"ur Umweltforschung, Universit\"at zu K\"oln, 50931 Cologne, Germany.}
\author[0000-0003-4503-3335]{Ross Beyer}
\altaffiliation{New Horizons Science Team, Co-Investigator}
\affiliation{SETI Institute, 189 Bernardo Way, Mountain View, CA 94043, U.S.A.}
\affiliation{NASA Ames Research Center; Mail Stop 245-3, Moffett Field, CA 94035, USA}
\author[0000-0002-4803-5793]{James T. Keane}
\altaffiliation{New Horizons Science Team, Science Affiliate}
\affiliation{Jet Propulsion Laboratory, California Institute of Technology, Pasadena, CA 91125, USA}
\author[0000-0002-4832-4456]{Ivan R. Linscott}
\altaffiliation{New Horizons Science Team, Science Affiliate}
\affiliation{Stanford University, Stanford, CA 94305, USA}
\author[0000-0002-6840-7187]{Samuel Birch}
\affiliation{Department of Earth, Atmospheric and Planetary Science, Massachusetts Institute of Technology, Cambridge, MA 02139}
\author[0000-0002-6840-7187]{Carver Bierson}
\altaffiliation{New Horizons Science Team, Science Affiliate}
\affiliation{School of Earth and Space Exploration, Arizona State University, Tempe, AZ 85287, USA}
\author[0000-0002-7547-3967]{Leslie A. Young}
\altaffiliation{New Horizons Science Team, Co-Investigator}
\affiliation{Southwest Research Institute, 1050 Walnut Street, Suite 300, Boulder, CO 80302, USA}
\author[0000-0001-5018-7537]{S. Alan Stern}
\altaffiliation{New Horizons Science Team, Principal Investigator}
\affiliation{Southwest Research Institute, 1050 Walnut Street, Suite 300, Boulder, CO 80302, USA}
\author[0000-0002-9548-1526]{Carey M. Lisse}
\altaffiliation{New Horizons Science Team, Science Affiliate}
\affiliation{Space Exploration Sector, Johns Hopkins University Applied Physics Laboratory, 11100 Johns Hopkins Rd, Laurel, MD 20723, USA }
\author[/0000-0003-1869-4947]{Carly J.A. Howett}
\altaffiliation{New Horizons Science Team, Co-Investigator}
\affiliation{Southwest Research Institute, 1050 Walnut Street, Suite 300, Boulder, CO 80302, USA}
\author[0000-0001-8541-8550]{Silvia Protopapa}
\altaffiliation{New Horizons Science Team, Co-Investigator}
\affiliation{Southwest Research Institute, 1050 Walnut Street, Suite 300, Boulder, CO 80302, USA}
\author{John R. Spencer}
\altaffiliation{New Horizons Science Team, Co-Investigator}
\affiliation{Division of Space Science and Engineering, Southwest Research Institute, Boulder, CO 80302, USA}
\author[0000-0002-9995-7341]{Richard P. Binzel}
\altaffiliation{New Horizons Science Team, Co-Investigator}
\affiliation{Department of Earth, Atmospheric and Planetary Science, Massachusetts Institute of Technology, Cambridge, MA 02139}
\author[0000-0002-4131-6568]{William B. McKinnon}
\altaffiliation{New Horizons Science Team, Co-Investigator}
\affiliation{Department of Earth and Planetary Sciences and McDonnell Center for the Space Sciences, Washington University, St. Louis, MO 63130, USA}
\author[0000-0003-3234-7247]{Tod R. Lauer}
\altaffiliation{New Horizons Science Team, Co-Investigator}
\affiliation{National Optical- Infrared Astronomy Research Laboratory, National Science Foundation, Tucson, AZ 85726, USA}
\author[0000-0003-0951-7762]{Harold A. Weaver}
\altaffiliation{New Horizons Science Team, Co-Investigator}
\affiliation{Johns Hopkins University Applied Physics Laboratory, Laurel, MD 20723, USA}
\author[0000-0002-5846-716X]{Catherine B. Olkin}
\altaffiliation{New Horizons Science Team, Co-Investigator}
\affiliation{Division of Space Science and Engineering, Southwest Research Institute, Boulder, CO 80302, USA}
\author[0000-0003-3045-8445]{Kelsi N. Singer}
\altaffiliation{New Horizons Science Team, Co-Investigator}
\affiliation{Division of Space Science and Engineering, Southwest Research Institute, Boulder, CO 80302, USA}
\author[0000-0002-3323-9304]{Anne J. Verbiscer}
\altaffiliation{New Horizons Science Team, Co-Investigator}
\affiliation{Department of Astronomy, University of Virginia, Charlottesville, VA 22904, USA}
\author[0000-0002-6722-0994]{Alex H. Parker}
\altaffiliation{New Horizons Science Team, Co-Investigator}
\affiliation{Division of Space Science and Engineering, Southwest Research Institute, Boulder, CO 80302, USA}

 
\begin{abstract}
\REV{A near surface thermal model for Arrokoth is developed based on the recently released $10^5$ facet model of the body. This thermal solution takes into account Arrokoth's surface re-radiation back onto itself.  The solution method exploits Arrokoth's periodic orbital character to develop a thermal response using a time-asymptotic solution method, which involves a Fourier transform solution of the heat equation, an approach recently used by others. We display detailed thermal solutions assuming that Arrokoth's near surface material's  thermal inertia ${\cal I} = $ 2.5 W/m$^{-2}$K$^{-1}$s$^{1/2}$. We predict that at New Horizons' encounter with Arrokoth its encounter hemisphere surface temperatures were $\sim$ 57-59 K in its polar regions, 30-40 K on its equatorial zones, and 11-13 K for its winter hemisphere.  Arrokoth's orbitally averaged temperatures are around 30-35 K in its polar regions, and closer to 40 K near its equatorial zones. Thermal reradiation from the surrounding surface amounts to less than 5\% of the total energy budget, while the total energy ensconced into and exhumed out Arrokoth's interior via thermal conduction over one orbit is about 0.5\% of the total energy budget. As a generalized application of this thermal modeling together with other KBO origins considerations, we favor the interpretation that New Horizons' REX instrument's $29 \pm 5$K brightness temperature measurement is consistent with Arrokoth's near surface material's being made of sub-to-few mm sized tholin-coated amorphous \water ice grains with 1 W/m$^{-2}$K$^{-1}$s$^{1/2}$  $< {\cal I} < $10-20 W/m$^{-2}$K$^{-1}$s$^{1/2}$, and which are characterized by an X-band emissivity in the range 0.9 and 1.}

\end{abstract}


\section{Introduction}\label{sec:intro}

{The bilobate object} Arrokoth \citep[discovered by][as 2014 MU$_{69}$]{Buie_etal_2020} is a flattened ($\sim$ 10km$\times$20km $\times$30km), 15.9 hour rotating, high obliquity ($\sim 99^\circ$) Kuiper Belt Object (KBO) encountered by the New Horizons spacecraft on January 1, 2019 \citep[JD 2458485,][]{Stern_etal_2019}.  Owing to its relatively low eccentricity and inclination ($e=0.03, i=2.4^\circ$, respectively) and its location in the Kuiper Belt ($a=44.58$AU, and orbital frequency $\omega = 6.69\times10^{-10}$s$^{-1}$), it falls into the class of so-called cold classical KBOs (or CCKBO, for short) and, as such, is considered a Kuiper Belt planetesimal, being perhaps one of the oldest and relatively unprocessed relics of the solar system's formation era \citep{McKinnon_etal_2020}.  \REV{Its surface, with low mean hemispherical albedo \citep[$A \approx 0.063$,][]{Stern_etal_2019,Hofgartner_etal_2021},
 exhibits a relatively featureless H$_2$O-free infrared spectrum save for the possible presence of surface methanol ice (CH$_3$OH) observed
 in absorption \citep{Grundy_etal_2020}.} Arrokoth does not appear to exhibit signs of harboring any volatile species
 \citep{Lisse_etal_2021}. 
 \par
\begin{figure}
\begin{center}
\leavevmode
\includegraphics[width=7.25cm]{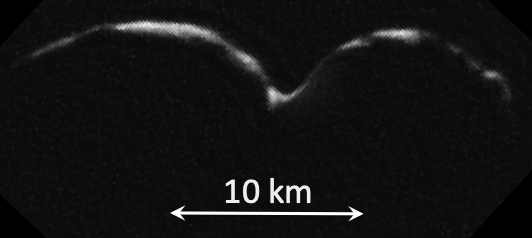}
\par
\end{center}
\caption{View of Arrokoth (CA07) taken by the LORRI camera on its departure trajectory approximately 10 minutes prior to the second REX scan \citep{Grundy_etal_2020}.  The lit crescent seen here would have appeared somewhat diminished during the actual REX scan ($\sim 10^\circ$ shift in viewing geometry).}
\label{CA07}
\end{figure}

 Furthermore, thermal X-band emission of Arrokoth (4.2-cm wavelength, 7.2 GHz) was observed by New Horizons' Radio Science Experiment (REX) in both its face on approach (low-phase angle) and look-back (high-phase angle) perspective.
 Fig.\ref{CA07} shows a visual view of Arrokoth a few minutes prior to the look-back phase REX scan.  Analysis of the winter night side thermal emission is consistent with a mean observed brightness temperature $T_{{\rm b,obs}} =  29 \pm 5$ K \citep[][also Bird et al. 2022, this volume]{Grundy_etal_2020}.  Although New Horizons was unable to directly measure Arrokoth's thermal inertia, based on other observed properties of CCKBOs 
 it is assumed to lie somewhere between 1-10 tiu 
 \citep[][also 1 tiu = 1 Wm$^{-2}$K$^{-1}$ s$^{1/2}$]{Lellouch_etal_2013,Muller_etal_2020}, indicative of highly insulating material -- at least for surface thicknesses corresponding to diurnal skin depths ($\sim$1-2 mm). For such a thermal inertia value applied to Arrokoth, together with an assumed low heat capacity material, predicts a seasonal (i.e., one orbit timescale $=297.6$yr) thermal skin depth of about 1 m \citep{Grundy_etal_2020}.
 Geomorphological
 analysis of Arrokoth's approach hemisphere seem to exhibit relatively bright surface units, especially in regions close to Arrokoth's neck \citep{Spencer_etal_2020}.  What relationship, if any, do these features have with the kinds of surface temperatures likely on Arrokoth's surface?  Furthermore, how does Arrokoth's relatively bright neck region correlate to both the seasonal cycle of received insolation and resulting surface temperatures there?  Lastly, can the REX brigthness temperature measurement be connected to and properly reconciled with a realistic temperature model for Arrokoth?
 \par
\begin{figure*}
\begin{center}
\leavevmode
\includegraphics[width=19.0cm]{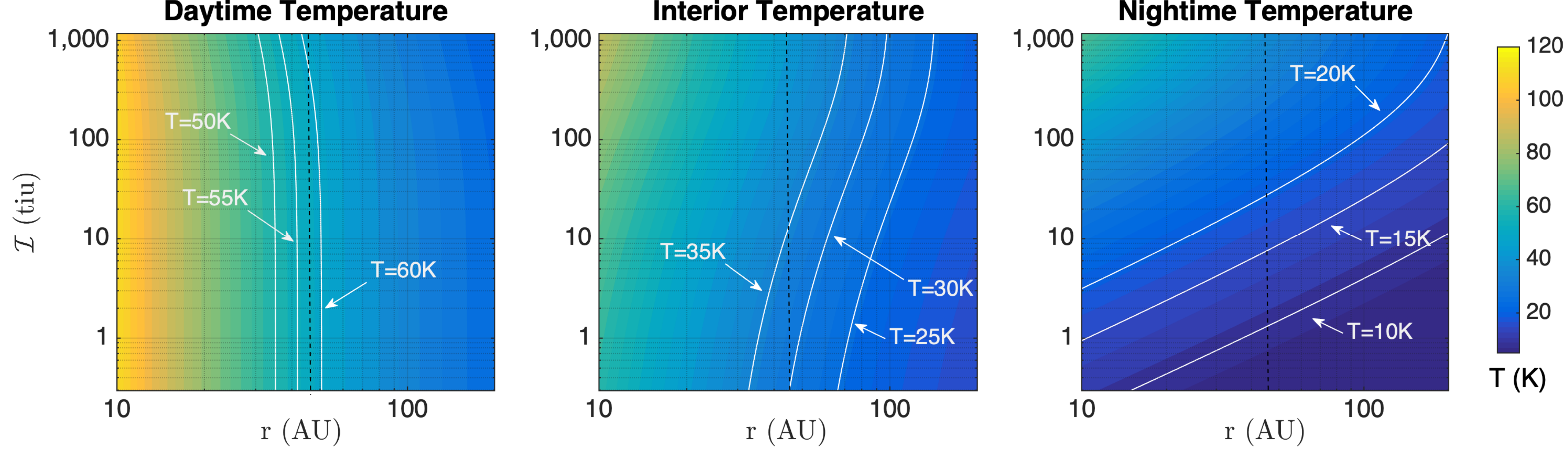}
\par
\end{center}
\caption{Predicted temperatures based on the simple theory of Section \ref{sec:simple_model}. Each panel
shows the predicted temperatures as a function of distance from Sun ($r$) and thermal inertia ($\cal I$).
All models assume $A=0.06$ and $\varepsilon = 0.9$.  The background flux consists entirely of the microwave background.
The panels depict (left panel) predicted summer season temperatures $T\sub{s,s}$, (middle panel) 
predicted interior temperatures $T\sub{{\rm int}}$, and (right panel) predicted winter season temperatures $T\sub{s,w}$.
The vertical hatched lines show the location of 44 AU and various iso-temperature lines of interest are shown
with white curves. }
\label{simple_theory}
\end{figure*}
 It is thus important to generate a thermophysical temperature map of Arrokoth to help address these questions.
  Stereo pair images
 taken by New Horizons' Long-Range Reconnaissance Imager (LORRI)
  \citet{Spencer_etal_2020} have been used to produce several cartographic data products including both
 a global shape model and a relatively detailed topographic map  
 of Arrokoth's closest approach day-side hemisphere.
 Together with this $\sim 10^5$ element shape model and Arrokoth's known orbital elements, Arrokoth's instantaneous distance from the Sun ($r$), as well as its subsolar latitude along every point along its orbit are also known \citep{Porter_etal_2018}. \par 
 \REV{Although New Horizons did not attempt a direct gravity measurement of Arrokoth, an estimate for the body's bulk mass density may be made by applying the gravitational slope analysis of \citet{Richardson_Bowling_2014} by using the shape/topographic model of \citet{Spencer_etal_2020}, which suggests that the body's mean density may be as little as $\rho = 150$ kg/m$^3$ and as high as $\rho = 650$kg/m$^3$\citep[][Keane et al. 2021]{McKinnon_etal_2020,Keane_etal_2020}.} For our purposes we will adopt
 a nominal value of $\rho = 250$ kg/m$^3$, also see detailed discussion in Keane et al. (2021).\footnote{There are no direct constraints on Arrokoth's density.  The quoted values are based on an empirical trend inferred from the observations of other small airless bodies of the solar system. For a body of given shape and rotation, the distribution of topographic slopes corresponds to maximum stability, i.e., the statistical slope distribution is furthest removed from the material's angle of repose \citep{Richardson_Bowling_2014,Richardson_etal_2019}.  \citet{McKinnon_etal_2020} finds this state of maximum stability by tuning the mean density to around  235 kg/m$^3$ with the quoted 1$\sigma$ error range.  We are cautioned that these are only indirect estimates based on geophysical inference and analogy to comets and asteroids. }  
 \REV{
 Fully compacted KBOs have typical densities of about $\sim 1500$kg/m$^3$ 
 \citep[e.g., see compilation found in][]{Bierson_Nimmo_2019},
 thus if Arrokoth is made of the same stuff, then its bulk porosity $p$ reasonably falls somewhere within the range of 0.6 to 0.85.
 Also, for reference we note the only direct measured density of a comet nucleus is that for 67P/Churyumov-Gerasimenko at 
 around 533 kg/m$^3$ \citep{Patzold_etal_2016}.
 However we do note that Arrokoth's topmost layers -- on the scale of a few seasonal thermal skin depths -- may have mean densities that deviate from the bulk average either due to compaction by impacts over geologic time or due to porosity increase due to volatile sublimation during the body's earliest times after formation.
 However, compaction appears to be less likely owing to the paucity of impact craters on Arrokoth
 \citep{Stern_etal_2019, Spencer_etal_2020,McKinnon_etal_2020}.
 }
 \par
 \REV{Another key input is the unconstrained value of Arrokoth's surface ice's heat capacity at constant pressure, $C_p$, a quantity not directly measured by New Horizons' flyby of the body.  For the thermal solutions developed in \citet{Grundy_etal_2020}, a value $C_p = 350$ J/K/kg was adopted based on estimates for cold H$_2$O Ih ice where  $C_p \approx 180$ J/K/kg \citep{Castillo-Rogez_etal_2012}, and 
 similar to adopted empirical forms like that found in \citet{Klinger_1980} and 
 \citet{Shulman_2004}.  The nearly doubled value can be rationalized on the argument that organic impurities and/or the presence of methanol in the matrix would substantially boost the ice complex's heat capacity.  This point is examined further in Section 6.1.3.}
 \par
 With these input ingredients, together with some assumptions regarding the surface thermal inertias, one may construct a current era temperature map of Arrokoth's surface, which is a vast improvement over the one presented in \citet{Umurhan_etal_2019} and \citet{Grundy_etal_2020}, which was based on a relatively coarse $\sim 2000$ element shape model.  
 \par
  {The main purpose of this study is detailing a general framework for generating temperature} 
  models of Arrokoth based on this 10$^5$ facet shape model.  Although we do not directly address the questions posed 
 earlier in this introduction -- especially connecting the predicted temperature profiles to the departure phase REX {\emph{brightness temperature}} measurement (see companion study by Bird et al. 2022) --
 the temperature solutions developed here will be employable toward answering all of them in future studies.\par
 The rest of this study is structured as following:
  Section 2 presents a simple thermal solution in order to ground our intuition as to what to expect.  Section 3 describes the thermal solution method, which includes a discussion of the shape model, the self-obscuration analysis of the shape model, a detailed description of how insolation is calculated over the course of one orbit, and finally a description of the Fourier transform based solution to the thermal diffusion equation.  Section 4 surveys our results in which we break up our discussion into encounter day thermal properties and orbitally averaged features.  
  \REV{ In Section 5 we develop the analysis framework that connecting the observed brightness temperature to the thermal solutions developed here.
  Section 6 discusses aspects of our results together with several speculations including a theoretical discussion that regarding the nature of Arrokoth's thermal inertia, as well as a few caveats.
 Section 7 concludes with a short summary.  The various appendices contain details of the methods used for our thermal solutions as well as details of the radiative transfer modeling employed here.}

\section{A Simplified Physics-Based Empirical Thermal Model} \label{sec:simple_model}
\REV{To guide our intuition we present a simple thermal model to ground expectations.  The model is one dimensional and is intended to crudely represent a thermal response of a facet over the course of one orbit with frequency $\omega$.  Its construction is 'physics-based' in the sense that
it is a pared-down casting of the more complete thermophysical model described in the next section.}
\par  The model is constructed in two parts and interpreted in terms of ``winter" and ``summer" time exposure given
Arrokoth's high obliquity, relatively flattened shape, and consequent extreme polar winters and summers.
Our view of this is guided by the observation that at high summer the daylit side of Arrokoth is close to a flat slab baking under the Sun, while at high winter the same slab is in total darkness while at the equinoxes, Arrokoth's surfaces are only tangentially illuminated.\par
  The first part describes the temperature
profile resulting on the surface and interior while the face is illuminated by the Sun with absorbed flux $f\sub{{\rm eff}} = (1-A){f}\sub\odot$, in which  ${f}\sub\odot$ is the peak summer solstice illumination flux, and where $A$ is the material albedo. 
If $r$ is the body's orbital radius (in AU) with $f\sub{{\rm SC}} = 1366$W/m$^2$ (defined as the solar constant), then the peak solar irradiance 
is ${f\sub\odot} = f\sub{{\rm SC}}/r^2$.
In this statement of the problem $\omega = \omega\sub\oplus r^{-3/2}$
where $\omega\sub\oplus = 1.99\times 10^{-7}$ s$^{-1}$.
The summer season's subsurface temperature model, $T\sub s(z)$, is 
\beq
T\sub s = T\sub{{\rm int}} + \left(T\sub{s,s}-T\sub{{\rm int}}\right)e^{kz},
\eeq
in which $T\sub{s,s}$ and $T\sub{{\rm int}}$ are the summer season surface temperature and deep interior temperature (respectively).  The thermal skin depth scale ($\ell$) is given by $2\pi/k$ where 
\beq
k \equiv \sqrt{\rho C_p \omega/K},
\eeq
with $K$ being the effective material conductivity, $\rho$ the ice mass density, $C_p$ the ice specific heat at constant pressure.  We will often refer to the effective thermal inertia of the medium, defined as 
\beq
{\cal I} \equiv 
\sqrt{K \rho C_p}.\label{thermal_intertia_definition}
\eeq  
The above postulated energy balance at the surface during the day entails
\beq
\underbrace{(1-A){f}\sub\odot + f\sub{{\rm bg}}}_{{\rm received \ energy \ flux}} = 
\underbrace{Kk\left(T\sub{s,s}-T\sub{{\rm int}}\right)}_{{\rm conducted \ thermal \ flux}} + 
\underbrace{\epsilon_{{\rm ir}} \sigma 
T\sub{s,s}^4}_{{\rm radiative \ losses}},
\label{simple_model_eqn1}
\eeq
in which $\epsilon_{{\rm ir}}$ is the material's thermal infrared emissivity, $\sigma$ is the Stefan-Boltzmann constant and 
$f\sub{{\rm bg}}$ represents a constant background illumination source.\par
\REV{The second part} of the model describes the temperature response during the winter season, in which
the surface is illuminated only by the background radiation field $f\sub{{\rm bg}}$, which we here 
adopt to be just the cosmic microwave background, i.e., $f\sub{{\rm bg}} \approx f\sub{{\rm cmb}} \approx
3.15\times 10^{-6}$W/m$^2$, corresponding
to $T_{{\rm cmb}} = 2.725$K.  
The corresponding winter season temperature solution is $T\sub w(z)$
where
\beq
T\sub w = T\sub{{\rm int}} + \left(T\sub{s,w}-T\sub{{\rm int}}\right)e^{kz},
\eeq
where $T\sub{s,w}$ is the night time surface temperature.  The corresponding surface energy balance, 
analogous to Eq. (\ref{simple_model_eqn1}), is given by
\beq
 f\sub{{\rm bg}} = Kk\left(T\sub{s,w}-T\sub{{\rm int}}\right) + \epsilon_{{\rm ir}} \sigma 
T\sub{s,w}^4.
\label{simple_model_eqn2}
\eeq
These coupled sets of equations are supplemented by an expression linking the winter and summer season solutions.  For this
we impose the condition that all of the energy conducted into the surface during summer returns to the surface in
winter.  Therefore we have
\beq
  Kk\left(T\sub{s,w}-T\sub{{\rm int}}\right) = -Kk\left(T\sub{s,w}-T\sub{{\rm int}}\right)
\label{simple_model_eqn3}
\eeq

Thus, Eqs. (\ref{simple_model_eqn1},\ref{simple_model_eqn2},\ref{simple_model_eqn3})
constitute a complete set of (algebraic) equations for the three unknowns 
$T\sub{s,w},T\sub{s,s}$ and $T\sub{{\rm int}}$.
From Eq. (\ref{simple_model_eqn3}) it immediately follows that
the interior temperature is simply the average of the day and night side temperatures, i.e., $T\sub{{\rm int}}
=\big(T\sub{s,s} + T\sub{s,w}\big)/2$.
The solution curves are governed by two non-dimensional numbers.  The first of these is $\eta$, given by
\beq
\eta \equiv \left(\frac{f\sub{{\rm bg}}}{(1-A){f}\sub\odot + f\sub{{\rm bg}}}\right)^{1/4},
\eeq
where $\eta$ is the ratio of the surface flux temperatures between the night and day side. From this one may
define averaged summertime and wintertime flux temperatures, $\overline T_s$ and $\overline T_w$ (respectively), in which
\beq
\overline T_s \equiv \left(\frac{(1-A){f}\sub\odot + f\sub{{\rm bg}}}{\epsilon_{{\rm ir}} \sigma}\right)^{1/4},
\qquad
\overline T_w \equiv \left(\frac{f\sub{{\rm bg}}}{\epsilon_{{\rm ir}} \sigma}\right)^{1/4}.
\label{scaled_characteristic_temperatures}
\eeq
\REV{The second non-dimensional parameter $\Gamma$ is the thermal parameter introduced by Spencer, which we hereafter refer to as the Spencer Number, and is defined here as}
\beq
\Gamma \equiv \frac{(\rho C_p K \omega)^{1/2} \overline T_s}{(1-A){f}\sub\odot + f\sub{{\rm bg}}}
\longrightarrow
\frac{(\rho C_p K \omega)^{1/2} (\epsilon_{{\rm ir}}\sigma)^{-1/4}}{\left[(1-A){f}\sub\odot + f\sub{{\rm bg}}\right]^{3/4}},
\label{Spencer_Number}
\eeq
estimating the relative contribution of downward directed thermal flux and radiative losses to space.\footnote{This non-dimensional parameter is often denoted in the literature by the symbol ``$\Theta$", but we use $\Gamma$ here because $\Theta$ is used elsewhere in this manuscript to denote Fourier temperature components.}
The simplified non-dimensional equations that must be simultaneously solved are
\beqa
1-\sfrac{1}{2}\Gamma(\theta\sub s - \eta \theta\sub w) - \theta\sub s^4 &=& 0, \\
1+\sfrac{1}{2}\Gamma \eta^{-4}(\theta\sub s - \eta \theta\sub s) - \theta\sub w^4 &=& 0,
\label{non-d_winterside-temperature}
\eeqa
where $\theta\sub s \equiv T\sub {s,s}/\overline T\sub s$ and $\theta\sub w \equiv T\sub {s,w}/\overline T\sub w$.
We observe that holding all input quantities constant but only varying the orbital radius, so long
as $(1-A){f}\sub\odot \gg f\sub{{\rm bg}}$ it follows that $\Gamma \sim r^{3/4}$, which says
that the importance of surface thermal conduction increases as an object like Arrokoth moves further away from the Sun.

\par
In Figure \ref{simple_theory}, we plot the predicted day and night side temperatures, as well as the predicted deep interior temperature,
for conditions representative for those of Arrokoth, wherein we adopt
$\epsilon_{{\rm ir}} = 0.9$, $A=0.06$, $\rho = 250$kg/m$^3$, and $C_p \approx 350$ J/K/kg.  For the sake of 
academic completeness, we show what the predicted values of
these varied temperature quantities for various radial positions (10 AU $<r<$150 AU) and thermal inertias
$0.1 \ {\rm tiu} < {\cal I} < 100$ tiu, while keeping in mind
that the likely values of Arrokoth's ${\cal I}$ is in the range of 1-10 tiu, together with a nominal value of $r=44$ AU.
\REV{This relatively uncomplicated physics-based empirical model} predicts a range of averaged daytime surface temperatures in the vicinity of 55-60 K, deep interior temperatures between 34 K and 38 K, with typical averaged nighttime temperatures in the range of 10-20 K.  
The winter side temperature reflects the subsurface heat flux returning to the surface.  In this respect it represents the balance between the emerging thermal conductive flux and blackbody radiative losses of Eq. (\ref{simple_model_eqn2}).
We will revisit these predictions in light of the full calculation we develop in the next two sections.\par

\begin{figure}
\begin{center}
\leavevmode
\includegraphics[width=8.5cm]{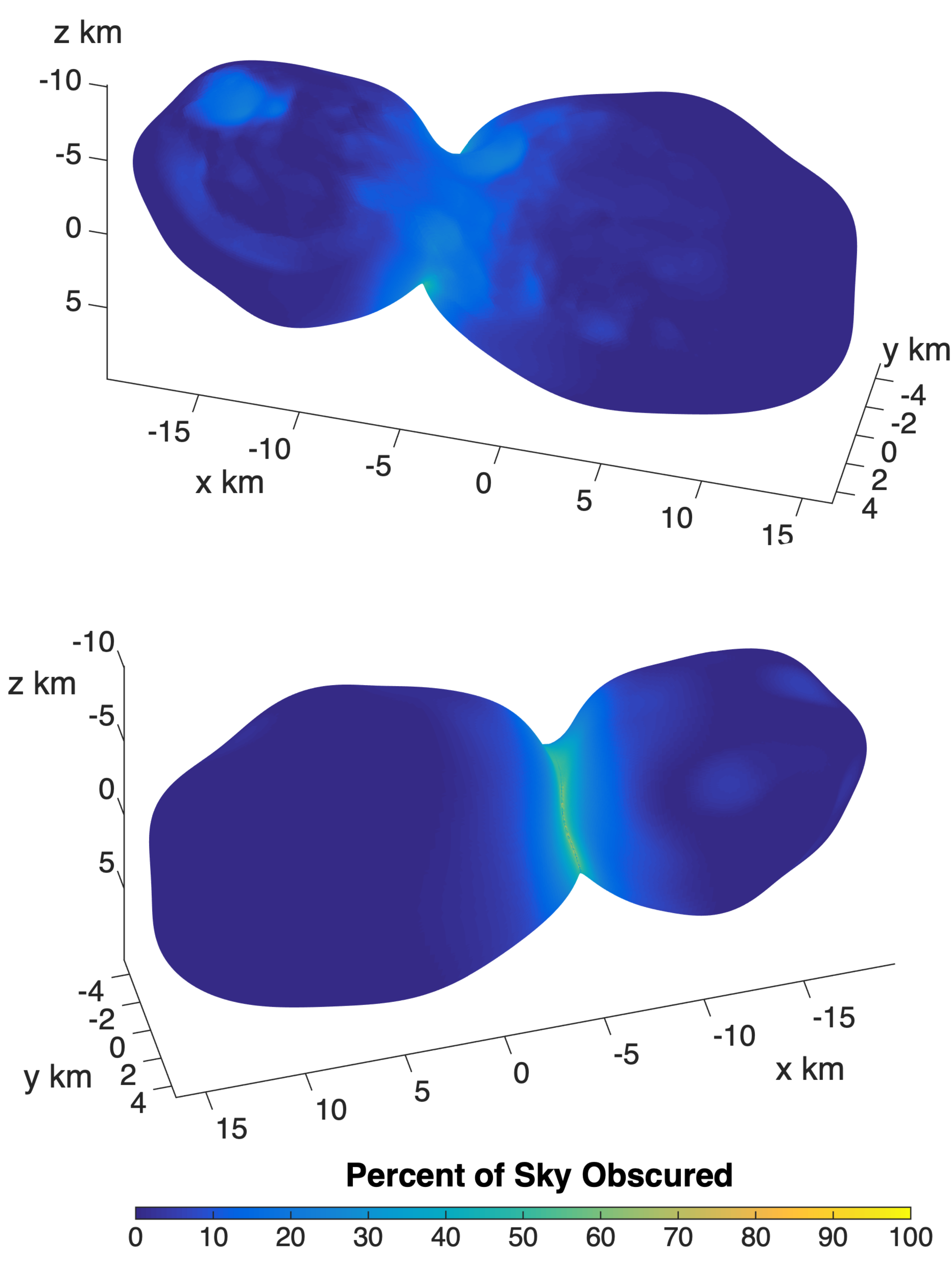}
\par
\end{center}
\caption{Percent of sky self-obscured of the 10$^5$ merged shape model. The views shown are approximately from the approach trajectory (CA03, top panel) and departure trajectory (CA07, bottom panel).  This references discussions found in both Secs. \ref{sec:shape_model_description}-\ref{sec:shape_analysis}, as well as Appendix A.}
\label{sky_coverage}
\end{figure}

\begin{table*}
\caption{Various quantities and parameters, and their adopted values where appropriate.}
\centering
\begin{tabular}{l c l}
\hline
Quantity & Value & Reference \\
\hline
Mass density of surface materials, $\rho$ & 150-650 kg/m$^3$ & \citet{McKinnon_etal_2020,Keane_etal_2020}  \\
Specific heat at constant pressure, $C_p$ & 350 J/K/kg & \citet{Castillo-Rogez_etal_2012} + \citet{Grundy_etal_2020} \\
 Thermal inertia, ${\cal I}$ & 1-10 W/m$^2$/K/s$^{-1/2}$  & \citet[][]{Lellouch_etal_2013}  \\
  Diurnal timescale thermal skin depth, $\ell_{{\rm day}}$ & $\sim 1$mm   & \citet[][]{Grundy_etal_2020}  \\
 Arrokoth's semimajor axis, $r$ & 44.58 AU & \citet{Buie_etal_2020} \\
 Orbital frequency of Arrokoth, $\omega$ & $6.69\times 10^{-10} {\rm s}^{-1}$ (298 yr orbit) & \citet{Porter_etal_2018} \\
  Diurnal frequency of Arrokoth, $\omega_{{\rm d}}$ & $1.10\times 10^{-4} {\rm s}^{-1}$ (15.9 hr day) & \citet{Buie_etal_2020}\\
  Obliquity & 99.1$^\circ$ & \citet{Spencer_etal_2020} \\
Mean hemispherical albedo, $A$  & $\sim   0.063 \pm 0.015$& \citet{Hofgartner_etal_2021} \\
thermal infrared emissivity, $\epsilon_{{\rm ir}}$ & $\sim 0.9$ &  \citet{Grundy_etal_2020} \\
\hline
\end{tabular}
\par  
\end{table*}

\section{Full Body Thermal Model: Formulation}
The simple equilibrium model presented in the previous section is useful for understanding the basic thermophysics of Arrokoth, but it assumes a slab-like surface where each facet is only in simple radiative equilibrium with the sky, and does not see other parts of Arrokoth's warm surface. Here we enhance the modeling by including the detailed shape and illumination information gained by the New Horizons flyby of Arrokoth. 
\par
Given a body's shape model with a set of N facets labeled by ``$i$" each of whose centers are given by the vector ${\bf r}_i$ in the coordinate frame of the body's center of mass (see section \ref{sec:shape_analysis} below), we solve for its temperature profile as a function of depth normal to the surface.  This is justified on the assumption that the seasonal thermal skin depth ($\sim 1-2$m) is
small compared to the horizontal scale of each facet ($\sim 50-100$m) and on the assumption there
are no other heat sources or sinks deep in Arrokoth's interior after 4.56 Gyr of evolution \citep[c.f.,][]{Lisse_etal_2021}.  
We therefore define the variable $\Theta_i(z,t)$ to be the temperature in the z-direction underneath each facet i, where $z=0$ is the surface.  We furthermore identify $T_i(t) = \Theta_i(z=0,t)$ as the surface temperature of facet i. For this 
initial study we assume the thermal conductivity to be independent of temperature, although the methods described here are generalizable to variable $K$.  The temperature response therefore satisfies the linear heat equation
\beq
\rho C_p \partial_t \Theta\sub i = \partial_z \big(K\partial_z\Theta_i\big),
\label{full_eqn_1}
\eeq
subject to boundary conditions, one of which is nonlinear.  We assume that the deep interior of Arrokoth no longer has
any active heat sources (an assumption likely not true during right after it was formed) and, as such, we assume the thermal flux goes to zero with large enough depth, i.e.,
\beq
\lim_{z\rightarrow -\infty}{ K \partial_z \Theta\sub i} = 0.
\label{full_eqn_2}
\eeq
The nonlinearities in this problem are expressed in the upper boundary condition, which is statement of the balance of received, emitted and interior transmitted energies, like embodied in the similar expressions
utilized in the previous section, namely Eqs. (\ref{simple_model_eqn1},\ref{simple_model_eqn2}).  We do not
consider the sublimation of volatile species given no observation of any gas emission.  We furthermore assume the albedo is uniform across the surface.
Thus at each facet surface we write
\beq
\epsilon_{{\rm ir}} \sigma T_i^4 = (1-A)f\sub{\odot,i} + f_{{\rm bg}} + K\partial_zT_i +
\int_{\partial S} \varepsilon \sigma T_j^4 S_{ij} d\hat{\bf s}_j, \ \ 
\label{full_eqn_3}
\eeq
where $K\partial_zT_i \equiv K\partial_z\Theta_i\big|_{z=0}$.  We note that the received solar insolation on
facet i, $f\sub{\odot,i}$,
is time dependent (see sec. \ref{insolation_calculation}).  The last term in the above expression represents the reradiated infrared radiation integrated
over
all facets $j$, with infinitesimal surface element $d{\bf s}_i$, that are visible to facet $i$. We assume
that the infrared albedo is 0.
The actual amount of reradiated radiation received at facet $i$ is contained in the matrix $S_{ij}$, which includes information about the relative inclination angle
between each facet pair, inverse square law effects and other mitigating factors (see further below).  
\REV{We have assumed that a given facet only radiates infrared
light, contained in the frequency integrated expression $\epsilon_{{\rm ir}} \sigma T_j^4 S_{ij}$, with no contribution due to reflected light. This neglect is justified (see Sec. 5.3.1) on account of the relatively low value of $A$.  A proper treatment of this contribution should be included in future analyses of Arrokoth. }
\par
Eqs. (\ref{full_eqn_1}-\ref{full_eqn_3}) constitute the nonlinear partial differential equations to be solved for every $T\sub i$ and generating a solution over one orbital period.  In the following subsections we provide a high-level view of the solution method we here employ.  The methodology used here is described in part in the Supplemental Materials of \citet{Grundy_etal_2020}
and were used to derive the solutions presented also in \citet{Umurhan_etal_2019}.

\begin{figure}
\begin{center}
\leavevmode
\includegraphics[width=8.0cm]{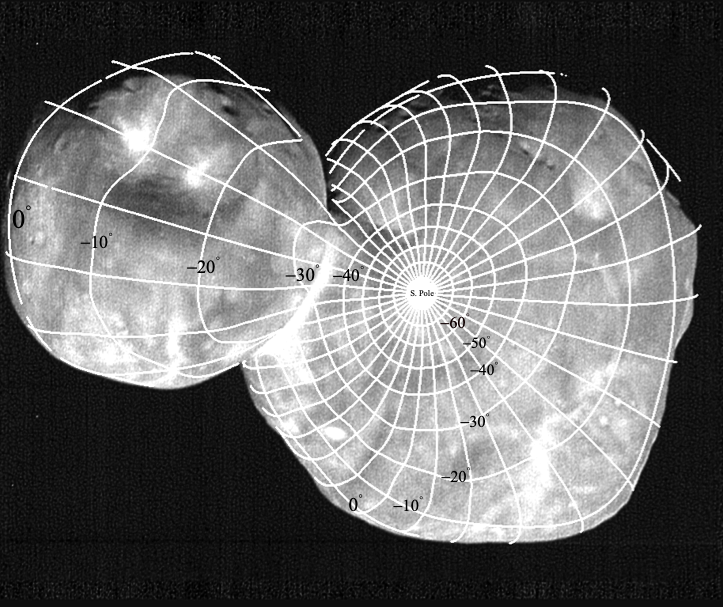}
\par
\end{center}
\caption{A Lucy-Richardson stacked deconvolved image from the LORRI CA06 sequence (images LOR$\_$0408626328 through LOR$\_$0408626336).  A single longitude/latitide graticule centered on the spin-pole is overlaid.  The negative-spin pole faced the Sun during the New Horizons flyby.  Thus, this view corresponds to the ``southern" hemisphere.}
\label{Arrokoth_With_Grid}
\end{figure}

\subsection{The Shape Model}\label{sec:shape_model_description}

We use the merged shape model described in \citet{Spencer_etal_2020}.
That model consists of a closed three-dimensional best-fit shape
derived from monoscopic images \citep{Porter_etal_2019}, upon which a
topographic surface of the spin-pole negative face derived from
stereogrammetry is fitted and merged \citep{Beyer_etal_2019}.  The model
contains 107,506 facets, which have an average area of 0.013 km$^2$ (with a standard deviation of 0.006 km$^2$), which corresponds to 
a typical facet diameter of about 100m -- this being the highest resolution achieved during New Horizons closest
approach.  A flavor of the shape model
is displayed in Fig. \ref{sky_coverage}.  
The shape model is saved as a meshfile (\texttt{.obj} file format) and will be a part of a forthcoming New Horizons PDS Small Bodies Node (SBN) release\footnote{\texttt{https:\slash\slash pds-smallbodies.astro.umd.edu\slash data\_sb\slash missions\slash newhorizons\slash index.shtml}}
scheduled for late 2021 (in preparation).  
Fig. 4 shows a Lucy-Richardson stacked deconvolved LORRI image of Arrokoth overlain with a single latitude-longitude graticule.
Given Arrokoth's $> 90^\circ$ obliquity, the approach view is from the ``southern" hemisphere, i.e., those of the negative latitudes.  \REV{Because of Arrokoth's concave shape, a latitude-longitude coordinate representation is impractical for our purposes as it can lead to degeneracies (see also Keane et al, 2021).  We typically refer to such conventions
for qualitative discussion purposes only.
For all of our calculations this work uses a Cartesian system centered on the body's center of Fig. 4 \citep[see][]{Beyer_etal_2019_LPSC} .}

\begin{figure}
\begin{center}
\leavevmode
\includegraphics[width=8.5cm]{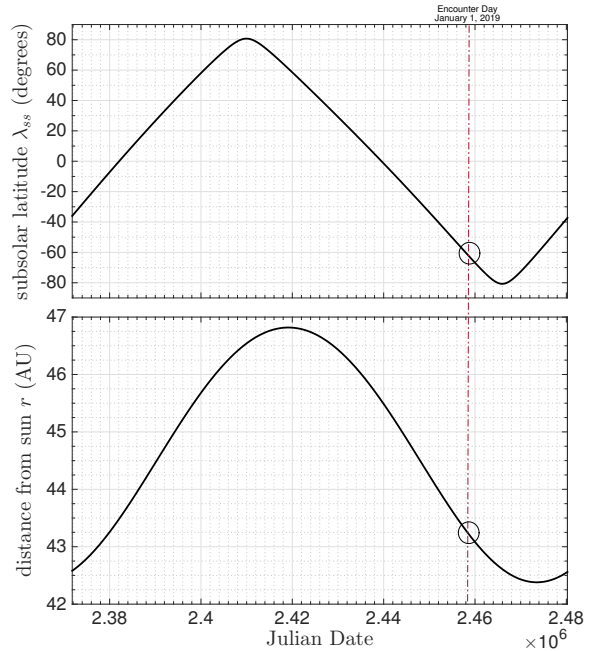}
\par
\end{center}
\caption{Subsolar latitude on Arrokoth ($\lambda\sub{ss}$, top panel) and it instantaneous distance from the Sun ($r$, bottom panel)
over the course of one Arrokoth orbit and consistent with the body's 99.1$^\circ$ obliquity.  The encounter day properties are shown with a vertical hatched line.
These are solutions based on the study found in \citet{Porter_etal_2018}.  }
\label{Orbital_Elements}
\end{figure}

\begin{figure*}
\begin{center}
\leavevmode
\includegraphics[width=17.5cm]{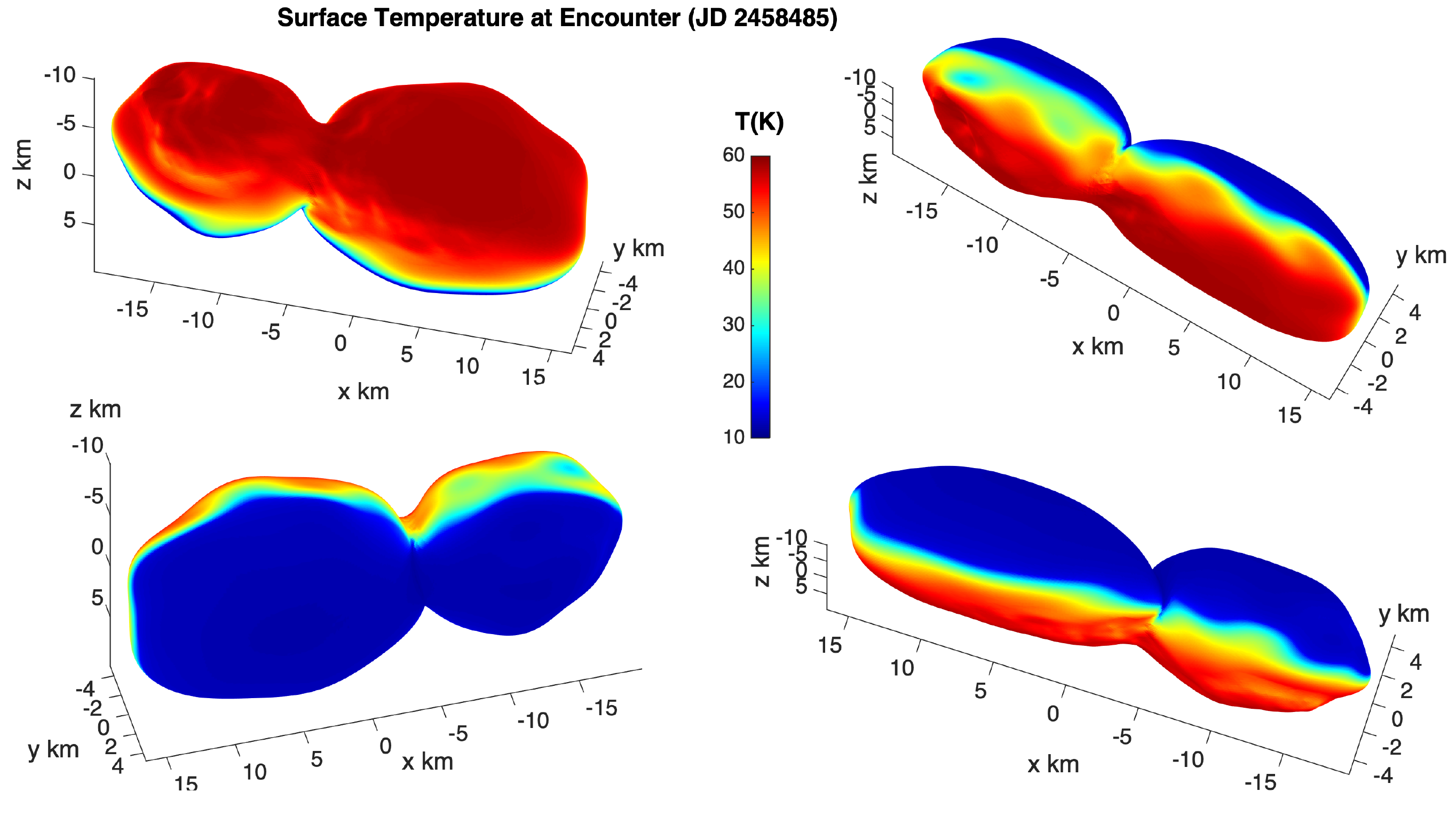}
\par
\end{center}
\caption{Predicted surface temperatures on encounter day (January 1, 2019) on the assumption ${\cal I} = 2.54$ tiu
and $\varepsilon = 0.9$. 
The upper left shows a view of the body on approach while the lower left depicts the view of Arrokoth during the CA08 observation.  The upper and lower right panels are the same model but rotated to emphasize the view of the equatorial regions.}
\label{Surface_Temperatures}
\end{figure*}

\begin{figure*}
\begin{center}
\leavevmode
\includegraphics[width=19.5cm]{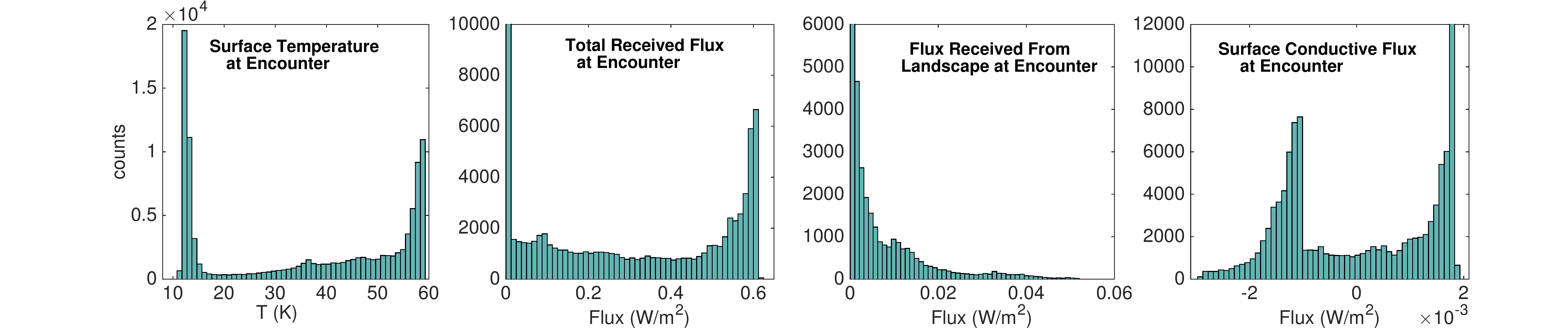}
\par
\end{center}
\caption{Histogram plots for various predicted quantities on encounter day (January 1, 2019).  Model assumes
${\cal I} = 2.54$ tiu and $\varepsilon = 0.9$.  Vertical axis corresponds to number of facets. Of the shown panels: (far left) 
Distribution of surface temperatures, (middle left)  Total received flux, (middle right) flux received as reradiation from surrounding landscape and, (far right) surface conductive flux in which a negative value corresponds to surface directed flow of energy from the interior.  In the last of these, note the trailing wing corresponding to the intense flux ring seen evident in Fig. \ref{Subsurface_Flux}.}
\label{encounter_day_histograms}
\end{figure*}

\begin{figure*}
\begin{center}
\leavevmode
\includegraphics[width=17.5cm]{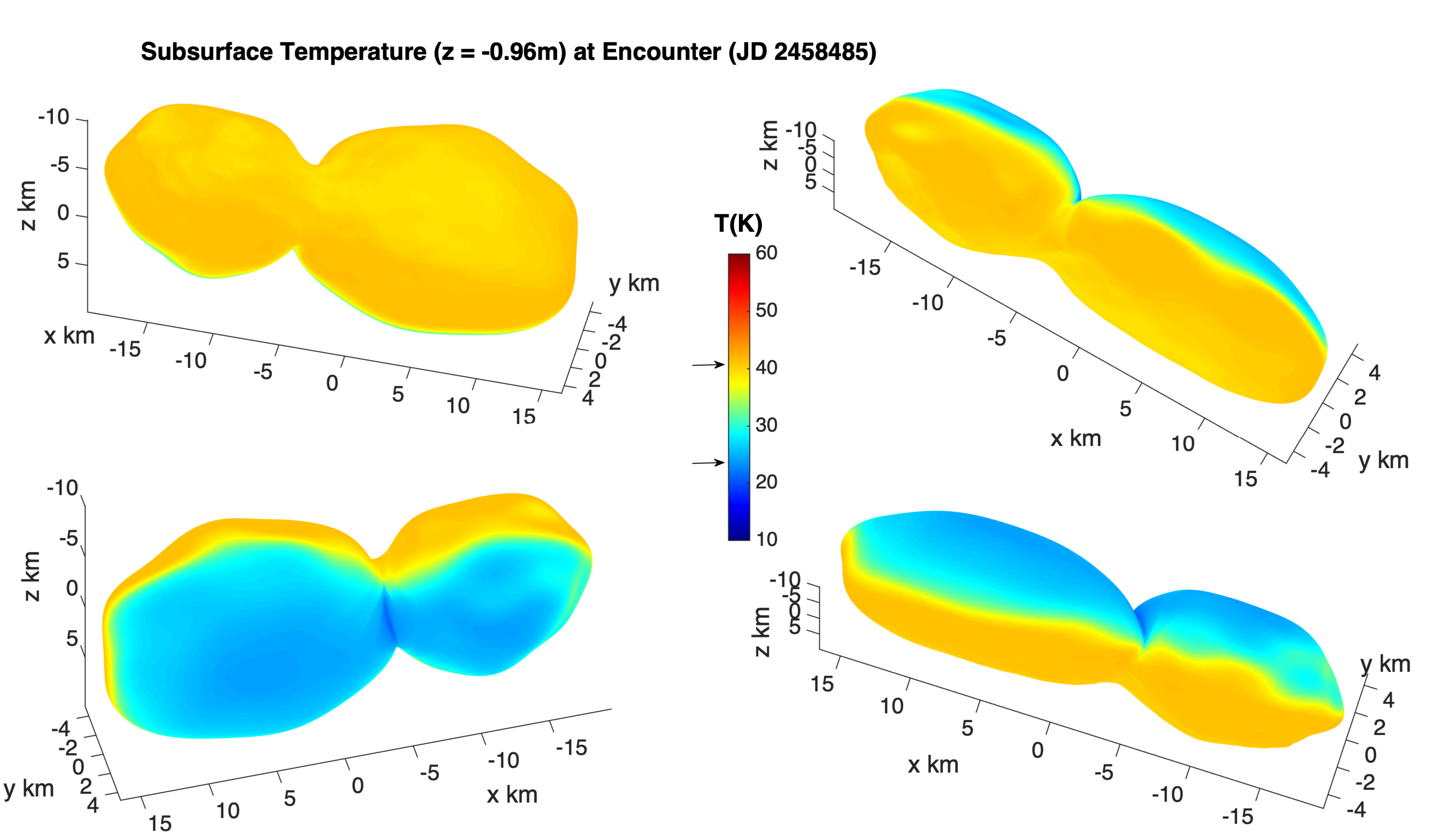}
\par
\end{center}
\caption{Four views of temperature at a depth of $\sim 1$ m at encounter day.  The arrows on the colorbar show
the range of temperature values at this depth.  The four views are the same
orientation as those shown in Fig. \ref{Surface_Temperatures}. }
\label{Subsurface_Temperatures}
\end{figure*}

\begin{figure*}
\begin{center}
\leavevmode
\includegraphics[width=17.5cm]{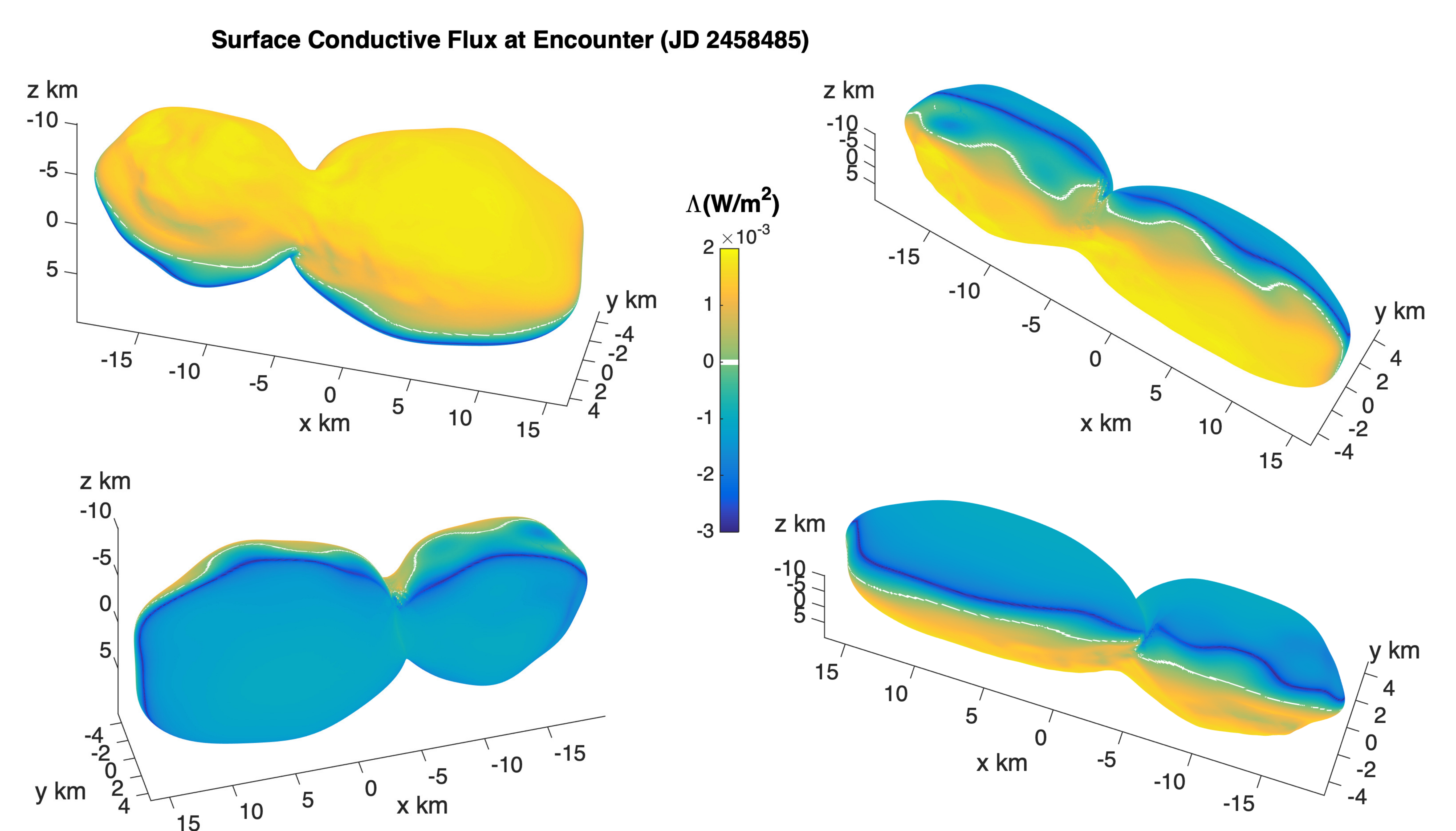}
\par
\end{center}
\caption{Four views of conductive flux ($\Lambda \equiv K\partial_z T\sub i$) at encounter day.  The arrows on the colorbar show
the range of flux values (also see Fig. \ref{encounter_day_histograms}).  The white line shows
the transition from inward and outward conductive flux.  The four views are the same
orientation as those shown in Fig. \ref{Surface_Temperatures}.  The intense surface directed ring follows the
temperature terminator (cf., Fig. \ref{Surface_Temperatures}).}
\label{Subsurface_Flux}
\end{figure*}

\begin{figure*}
\begin{center}
\leavevmode
\includegraphics[width=17.5cm]{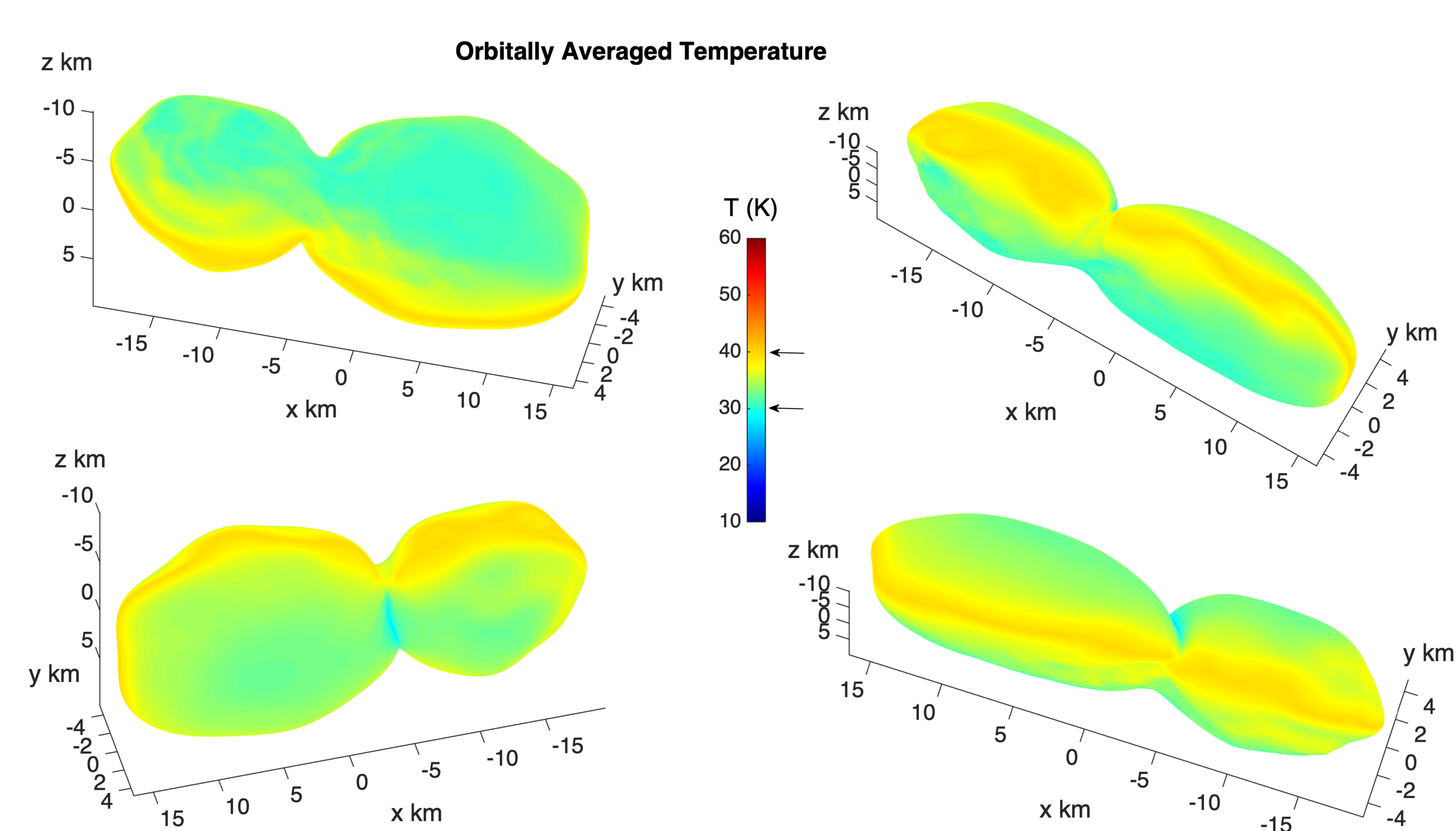}
\par
\end{center}
\caption{Four views of orbitally averaged surface temperature.   The four views are the same
orientation as those shown in Fig. \ref{Surface_Temperatures}.  The range of temperatures are shown with arrows, see also Fig. \ref{orbital_timescale_histograms}.}
\label{Avg_Temperature}
\end{figure*}

\begin{figure*}
\begin{center}
\leavevmode
\includegraphics[width=17.5cm]{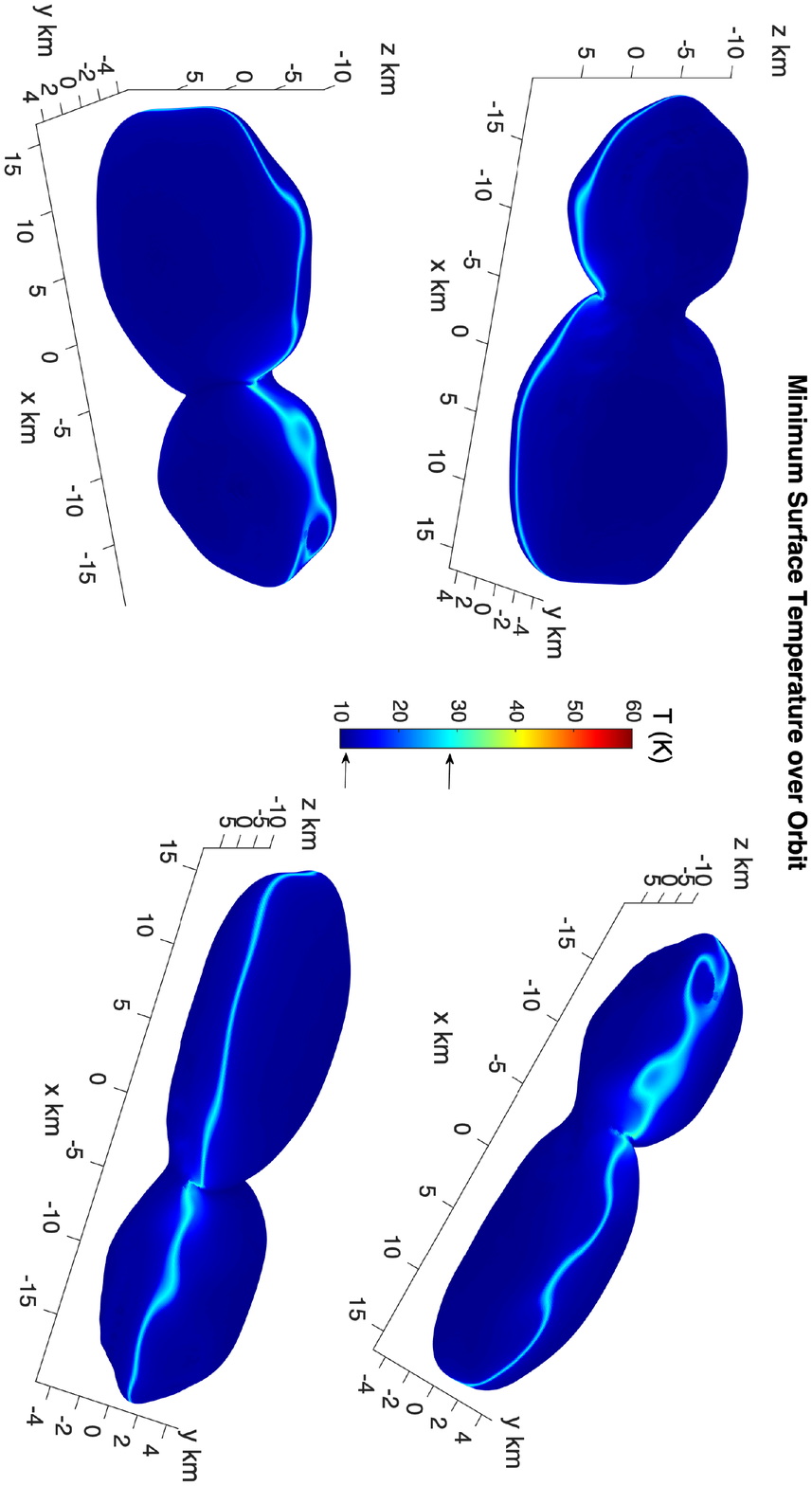}
\par
\end{center}
\caption{Four views of absolute minimum surface temperature of each facet over the course of one orbit.   The four views are the same
orientation as those shown in Fig. \ref{Surface_Temperatures}.  The range of temperatures are shown with arrows, see also Fig. \ref{orbital_timescale_histograms}.  Note the higher temperatures running within the equatorial zone and occasionally outlining the putative rims of cratered morphology lying in the tropical zone.}
\label{Min_Temperature}
\end{figure*}

\subsection{Shape Analysis: $S\sub{ij}$ and ${\rm K}\sub{ij}$}\label{sec:shape_analysis}
\REV{A shape analysis here is concerned with assessing a ``who-sees-who" network, identifying which facets $j$ of a shape are visible to a given facet $i$.  This information is then used to determine how much re-radiation is subsequently received.  We have constructed our own shape analysis algorithm, which is detailed in Appendix A, that largely mirrors the approach and philosophy utilized in several similarly motivated examinations of other solar system bodies, e.g., like that used for the Moon \citep{Glaser_Glaser_2019}, 67P \citep{Tosi_etal_2019}, and asteroids \citep{Rozitis_Green_2011}.  The shape and transfer model converts the continuous integral of Eq. (\ref{full_eqn_3}) into a discrete matrix operation, i.e.,
\beq
\int_{\partial S} \epsilon_{{\rm ir}} \sigma T_j^4 S\sub{ij} d\hat{\bf s}_j
\rightarrow \sum_{\forall j \ {\rm visible}} \epsilon \sigma T_j^4 {\rm K}\sub{ij},
\eeq
where ${\rm K}\sub{ij}$ contains knowledge of the ``who-sees-who" network as well as the amount of radiative transfer communicated between element pairs.}

\subsection{Calculating Diurnal Averaged Insolation $f\sub{\odot,i}$}\label{insolation_calculation}
Because the rotation rate is so short compared to the orbital time, we calculate
the diurnally averaged insolation received by each facet $i$.  Figure \ref{Orbital_Elements} shows both Arrokoth's subsolar latitude, $\lambda\sub{ss}(t)$, and its instantaneous distance from the Sun, $r(t)$ (in AU), over the course of one nearly 300 year orbit \citep{Porter_etal_2018}.
The instantaneous flux of solar radiation crossing Arrokoth's location is therefore $f\sub r \equiv f\sub{{\rm SC}}/r^2$.
With respect to the shape model described in the previous section, the direction of the Sun is given by the unit vector
$\hat{\bf n}\sub{{\rm sun}}$ whose individual components are
\beqa
n_{x,{{\rm sun}}} &=& \cos\varphi \sin \big(\pi/2 - \lambda\big), \nonumber \\
n_{y,{{\rm sun}}} &=& \sin\varphi \sin \big(\pi/2 - \lambda\big), \nonumber \\
n_{z,{{\rm sun}}} &=& \cos \big(\pi/2 - \lambda\big),
\eeqa
wherein $\varphi$ is the longitude corresponding to \REV{peak illumination} (``high noon") on Arrokoth.
Then with respect to facet $i$ with unit normal $\hat{\bf n}_i$, and where high noon occurs over longitude $\varphi$ with subsolar latitude $\lambda\rightarrow \lambda\sub{ss}$, the local flux of solar radiation received at facet is
\beq
f\sub{\oplus,i}(\varphi,\lambda) = \left\{\begin{array}{cc}
0, & {\rm if \ ray \ blocked \ or \ } \hat {\bf n}_i \cdot \hat{\bf n}\sub{{\rm sun}} < 0, \\
f\sub r \hat {\bf n}_i \cdot \hat{\bf n}\sub{{\rm sun}}, & {\rm otherwise}.
\end{array}
\right.
\eeq
where the condition $\hat {\bf n}_i \cdot \hat{\bf n}\sub{{\rm sun}} < 0$ means the Sun is below the horizon.
Note that like the shape analysis described in the previous section, we must execute ray-tracing routine
to determine if other facets block the Sun's rays from reaching the given facet.  
While this can be extremely expensive, in practice we reduce the actual number of ray-tracing
determinations by taking into account knowledge of the maximum landscape altitude angle $\beta\sub i$ determined
in the shape analysis: if the sun vector $\hat{\bf n}\sub{{\rm sun}}$ points with an angle with respect to the local facet horizontal
that exceeds the angle $\beta\sub i$, then no ray-tracing calculation is needed.  In other words
no ray-tracing is needed if
\beq
\hat {\bf n}_i \cdot \hat{\bf n}\sub{{\rm sun}} > \cos\big(\pi/2 - \beta\sub i\big) \ge 0.
\eeq
This criterion considerably speeds up calculations.
Formally then we define $f\sub{\odot,i}$ to be the daily average of $f\sub{\oplus,i}$, which involves
an integral over one Arrokoth 15.9 hour day.  We sample this integral using $N\sub k$ equally spaced
longitude values $\varphi\sub k$ ranging from $0$ to $2\pi$.  In other words we say
\beqa
f\sub{\odot,i}(\lambda) &\equiv& \frac{1}{\hat t_{{\rm day}}}\int_0^{\hat t_{{\rm day}}}
f\sub{\oplus,i}\Big(\varphi(\hat t),\lambda\Big) d\hat t \\
&\approx& \frac{1}{N\sub k} \sum_{k=1}^{N\sub k}f\sub{\oplus,i}\big(\varphi\sub k,\lambda\big).
\eeqa
In all of our calculations we choose $N_k = 16$.

\begin{figure*}
\begin{center}
\leavevmode
\includegraphics[width=19.5cm]{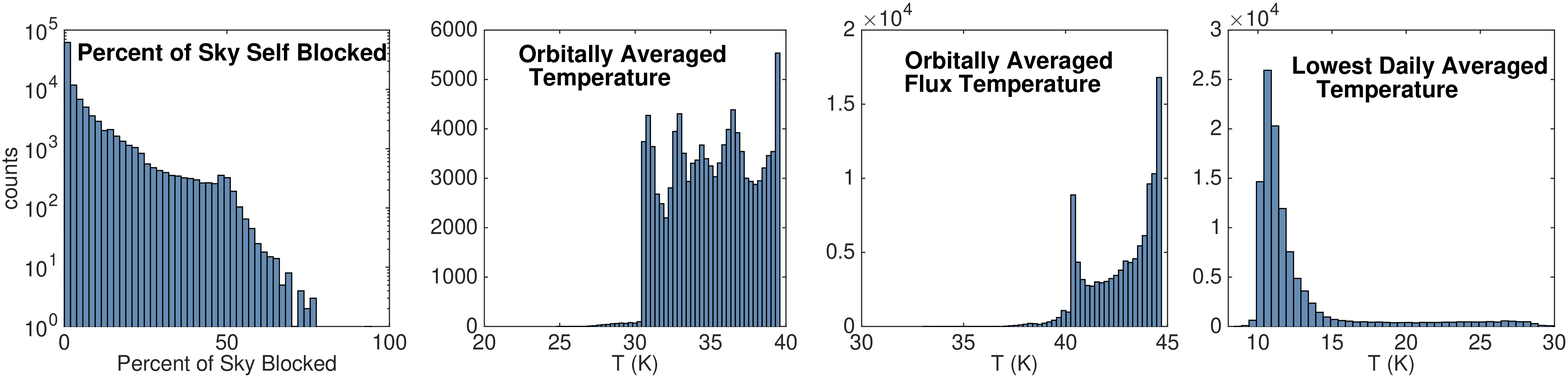}
\par
\end{center}
\caption{Histogram plots for various orbital scale quantities.  Model assumes
${\cal I} = 2.54$ tiu, $A=0.06$, and $\varepsilon = 0.9$.  Vertical axis corresponds to number of facets. The panels are:
(left panel) percent of body self-obscuration in reference to Fig. \ref{sky_coverage}, (middle left) orbitally
averaged surface temperature $T\sub{{\rm int},i}$, (middle right) orbitally averaged flux temperature
$T\sub{{\rm a},i}$, (right) minimum surface temperature over the course of one orbit.}
\label{orbital_timescale_histograms}
\end{figure*}

\begin{figure*}
\begin{center}
\leavevmode
\includegraphics[width=17.5cm]{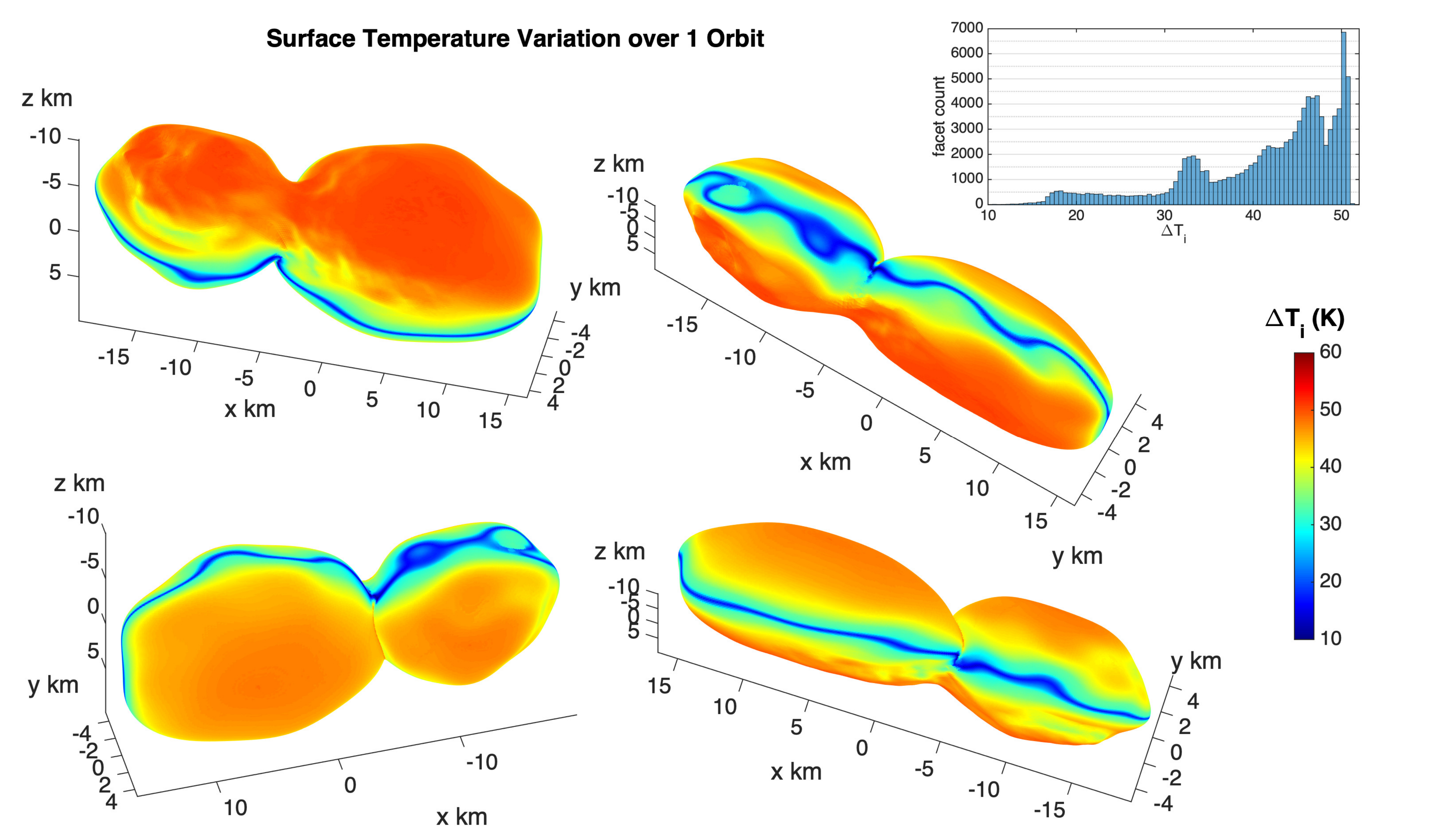}
\par
\end{center}
\caption{Four views of full temperature variation over the course of one orbit ${\Delta T}\sub{i}$.   The four views are the same
orientation as those shown in Fig. \ref{Surface_Temperatures}. Far upper right graph shows a histogram of ${\Delta T}\sub{i}$.}
\label{MinMax_Temperature}
\end{figure*}

\begin{figure*}
\begin{center}
\leavevmode
\includegraphics[width=17.5cm]{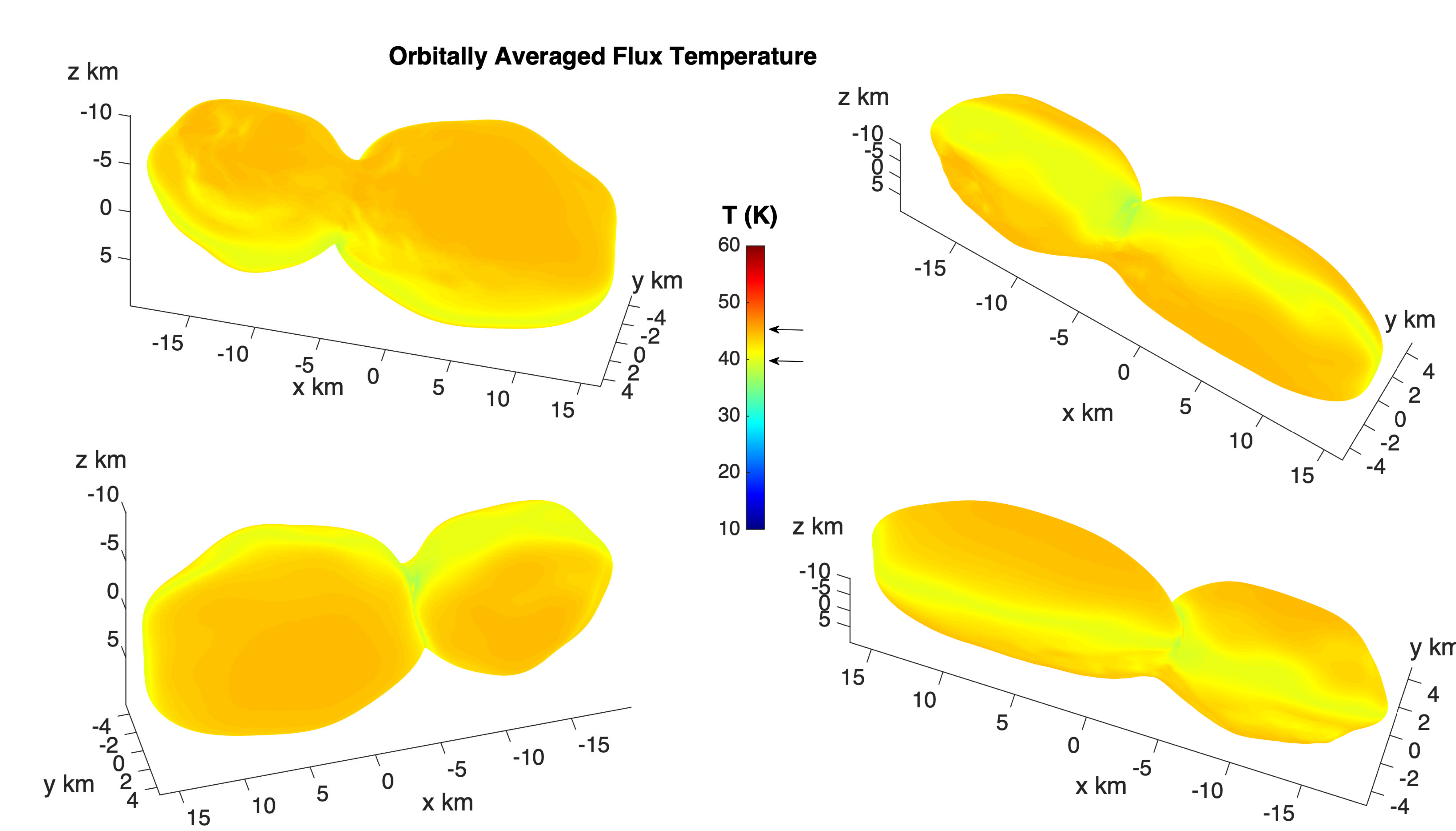}
\par
\end{center}
\caption{Four views of orbitally averaged flux temperature $T\sub{{\rm a},i}$.   The four views are the same
orientation as those shown in Fig. \ref{Surface_Temperatures}.  The range of temperatures are shown with arrows, see also Fig. \ref{orbital_timescale_histograms}.}
\label{Avg_Flux_Temperature}
\end{figure*}

\subsection{Thermal Solution}\label{thermal_solution}
We approach the solution to Eq. (\ref{full_eqn_1}) on the assumption that the thermal response
is purely periodic, which means to say that all transients have died away, therefore making this a so-called time asymptotic solution like used by \citet{Titus_Cushing_2012,Schloerb_etal_2015} and \citet{White_etal_2016}.  We opt for adopting this assumption
and the following Fourier transform based solution method owing to the spectral accuracy in produces,
especially with respect to the importance
of preserving the total conducted energy into and out of the interior over the course of one orbital period. 
As such, the solutions are represented as a truncated Fourier series in
powers of the orbital frequency $\omega$,
\beq
\Theta_i = \sum_{n=0}^N \Theta_{i,n} e^{k_n z} e^{-i\omega n t} + {\rm c.c.},
\label{thermal_solution_1}
\eeq
in which
\beq
k_n = \left(\frac{i \rho C_p \omega n}{K} \right)^{1/2}, \qquad {\rm Re}(k_n) > 0,
\label{kn_definition}
\eeq
where $N$ is an integer that is about the sampling rate of the received flux
$f\sub{\odot,i}$ (see further below).  For each value of $n$, the functional forms
in Eq. (\ref{thermal_solution_1}) are the exact solutions of the Fourier Transform of 
Eq. (\ref{full_eqn_1}) where the amplitudes $\Theta_{i,n}$ are unknown and must be determined from enforcing
boundary conditions.
Eq. (\ref{thermal_solution_1}) automatically solves the lower thermal boundary condition Eq. (\ref{full_eqn_2}).
\REV{We observe that $\Theta_{i,0}$ is half
of the deep interior temperature ($\equiv T_{{\rm int},i}$), and so long as the thermal forcing is not aperiodic we note that
$\equiv T_{{\rm int},i}$ is time independent for a given facet, but may vary from facet to facet.} 
The advantage of this approach is that we do not have to directly calculate
subsurface temperatures numerically as this information is encoded in the vertical structure
function $e^{k_n z}$ for each Fourier mode $n$.  
\REV{For the sake of academic completeness we describe the rudiments of the solution approach in Appendix B, and discuss further in the Appendix C how this approach may be generalizable to configurations in which
the conductivity changes value with depth, as might be the case if one is confronted with
a physical situation in which very low thermal inertia refractories sit atop an otherwise
highly conductive solid ice layer below (see also section \ref{connect_to_REX}).  
This multi-layer model was in fact used in generating the Arrokoth thermal solutions
based on the 1962 facet model found on the Small Bodies Node PDS data repository (\texttt{https:\slash\slash pds-smallbodies.astro.umd.edu\slash data\_sb\slash missions\slash newhorizons\slash index.shtml}), 
and is the reason we have included detailing this kind of layered solution approach in Appendix C.}
\par
With Eq. (\ref{thermal_solution_1}) in hand, it is straightforward to assess the surface thermal flux
($\Lambda\sub i$)
\beq
\Lambda\sub i \equiv K\partial_z T\sub i = 
K\sum_{n=1}^N k_n \Theta_{i,n} e^{-i\omega n t} + {\rm c.c.}
\label{surface_flux_solution}
\eeq
The final stage involves determining the coefficients $\Theta_{i,n}$, which, once inverse Fourier tranformed,
gives the time series of $T_i$ over the course of one cycle $\omega$.  The coefficients are determined
using an iterative procedure sketched in the following:
for notational ease we identify $f\sub i \equiv \epsilon_{{\rm ir}} \sigma T\sub i^4$, then the
boundary condition is
\beq
f\sub i = (1-A)f\sub{\odot,i} + K\partial_z T\sub i + {\rm{K}}_{ij} f\sub j.
\eeq
The zeroth order solution is the solution to the above with the conductive term set to zero.  Thus
\beq
f\sub i^{(0)} = \Big(\mathtt{I}\sub{ij} - {\rm K}\sub{ij}\Big)^{-1} (1-A) f\sub{\odot,i},
\label{fi_zero}
\eeq
which involves a matrix inversion -- with $\mathtt{I}\sub{ij}$ the identity matrix --
followed by determining the first iterate temperature solution $T\sub i^{(1)}$ to
\beq
f\sub i^{(1)} = \epsilon_{{\rm ir}} \sigma \left(T\sub i^{(1)}\right)^4 = f\sub i^{(0)} +  K\partial_z T\sub i^{(1)}.
\label{first_iteration}
\eeq
In this equation $f\sub i^{(0)}$ is known at all points along the orbit, thus it becomes a matter of solving
for $T\sub i^{(1)}$.  
This is achieved by performing a fast Fourier Transform of Eq. (\ref{first_iteration}) into frequency domain followed by using a Newton-Raphson routine
\footnote{
For example, \texttt{https:\slash\slash en.wikipedia.org\slash wiki\slash Newtons\_method}.}
 to solve for the Fourier components $\Theta_{i,n}^{(1)}$.  
The received flux, as well as the solution of the surface temperature field, is well represented with a $N=30$ Fourier mode decomposition.  We have assessed the robustness of these $N=30$ results against solutions generated up to $N=300$ finding characteristic differences in the temperature field (for any given facet) differing by less than 1 part in 10$^5$.
For a given face, we demand for convergence that successive changes in each $\Theta_{i,n}^{(1)}$ be less than $10^{-4}$, which
is usually achieved in 6 or less sub-iterations.  
All higher iterations $m \ge 2$ involve calculating
\beqa
\epsilon_{{\rm ir}} \sigma \left(T\sub i ^{(m)}\right)^4 &=& K\partial_z T\sub i^{(m)} + (1-A) f\sub{\odot,i} + {\rm K}\sub{ij}f\sub j^{(m-1)},  \ \ \ \\ 
f\sub i^{(m)} &=& \epsilon_{{\rm ir}} \sigma \left(T\sub i ^{(m)}\right)^4,
\eeqa
where, as before, the first of the above equations is solved in Fourier frequency space, while the second takes that
solution and expresses it the time domain, calculating the matrix product ${\rm K}\sub{ij}f\sub j^{(m)}$ in the time domain, and then feeding the result forward to the next iteration, $m \rightarrow m+1$.
Note that while this procedure involves one large matrix multiplication per iteration, it is not the main bottle-neck in the calculation, but rather the Newton-Raphson stage, for which values of $m\ge 2$ tends to converge in 1-2 sub-iterations owing
to the closeness of the initial guess to a converged solution.
For this particular system, in which the characteristic Spencer number $\Gamma$ is low, we find that satisfactory convergence is achieved in one ($m=1$) iteration.  \par
This calculation is run in parallel using four 3.1 GHz Intel Cores (i7).
The temperature solution for each facet (including all sub-iterations) takes less than 0.004 cpu seconds/processor for $N=30$ and 0.09 seconds/processor for $N=300$.
We find that it takes approximately 40 minutes to generate a full solution with one iteration for the $10^5$ facet model when $N=300$, whereas it takes a little over 3 minutes when $N=30$. All results displayed in this study are done with $N=300$.
\par
We close this section by defining the orbitally averaged {\emph{flux temperature}} $T\sub{{\rm a},i}$, which is, \REV{for each facet}, the orbital average of the solution to Eq. (\ref{full_eqn_3}) in the infinitely insulating limit (i.e., $K\rightarrow 0$).  Given
the preceding discussion, this amounts to the solution to 
\beq
\epsilon_{{\rm ir}} \sigma T\sub{{\rm a},i}^4 = \frac{1}{P}\int_{0}^P f\sub i^{(0)}\Big(\lambda(t)\big)dt,
\label{flux_temperature_def}
\eeq
\REV{in which $P$ is Arrokoth's orbital period, and where $f\sub i^{(0)}$ -- given in Eq. (\ref{fi_zero}) -- is understood to be a time-dependent function of the subsolar latitude $\lambda(t)$.}  This quantity
is a proxy for the total received insolation for each facet.  Given Arrokoth's high obliquity, we expect $T\sub{{\rm a},i}$
to be relatively higher in the regions experiencing extreme polar winters and summers ($|\lambda| 	\gtrapprox 10^\circ$) compared to Arrokoth's perpetually diurnal equatorial zones ($|\lambda| \lessapprox 10^\circ$).  Henceforth, we designate as {\emph{polar region}} or {\emph{polar zone}} latitudes that satisfy $|\lambda| 	\gtrapprox 10^\circ$, while
ascribing {\emph{tropical zone}} or {\emph{equatorial zone}} to 
latitudes satisfying $|\lambda| \lessapprox 10^\circ$ \citep[e.g., as in the usage of][]{Earle_etal_2017}.

\section{Full Body Thermal Model: Results}
\subsection{Encounter Day Views}
In Fig. \ref{Surface_Temperatures} we present an example temperature map of Arrokoth viewed on encounter day (January 1, 2019 or Julian Date 2458485) on the assumption that ${\cal I} \approx 2.5$ tiu, which, in turn, corresponds to an effective Spencer Number $\Gamma\sub{{\rm eff}} \approx 0.0082$ (with $\overline T\sub s \approx 59$K), where we follow the definition found in Eqs. (\ref{scaled_characteristic_temperatures}-\ref{Spencer_Number}) and adopting $\overline f\sub\odot = f\sub\odot\big(r=44 {\rm AU}\big)$.
On this day, the subsolar latitude ($\lambda\sub{ss}$) was $\lambda\sub{ss} \approx -61.86^\circ$ and Arrokoth
was $\approx 43.24$ AU from the sun (see also Fig. \ref{Orbital_Elements}).
Four views are shown in Fig. \ref{Surface_Temperatures}: 
The upper left panel shows the approach view, where the simulated body specific subspacecraft coordinates are at 
$\varphi = -46.9^\circ, \ \lambda = 76^\circ$, while the lower left panel 
shows the departure view with nominal subspacecraft coordinates  $\varphi = 46.9^\circ, \ \lambda = 284^\circ$. The latter
view position closely approximates the view from the spacecraft when the REX (CA08) scan was made.  
\par
For these parameters
the predicted surface temperatures ranges from 10 K to 60 K.  The distribution of temperatures, in terms of facet counts, is shown in the first panel of Fig. \ref{encounter_day_histograms}. A cursory view of the upper left panel of 
Fig. \ref{Surface_Temperatures} displays the propensity for high temperatures (in the vicinity of 57-60K) over most of the flattened and exposed real estate of the encounter hemisphere ranging from the south pole down to $\lambda = -10^\circ$ or so.   The view from the REX observation point (lower left panel of Fig. \ref{Surface_Temperatures}) shows the surface temperature is dominated by the winter side values (around 12-13 K) with a sliver of high temperatures along the top outer rim.  The temperature transitions from the winter side lows to the summer side highs through the equatorial region, with the transition occurring across a narrow band of tropical estate.  

\par
Fig. \ref{Subsurface_Temperatures} shows a predicted temperature map at approximately 1m below Arrokoth' surface.  The input parameters are as those shown in Fig. \ref{Surface_Temperatures}.  The temperatures at this depth range from about 40K down to 25K.  Notably, the temperatures at this depth are slightly cooler at high summer side latitudes than they are at latitudes approaching the equator (the differences being ~3-4 K).  This is to be expected since tropical zone latitudes never go into polar night over the course of one orbit, which means that the temperature there never get nearly as cold
as they do in the poles (see further below). The winter side temperatures, on the other hand, do not show this slight inversion.  The reason for this is likely due to the winter-side part of the shape model not being nearly as flattened as the imaged encounter hemisphere -- but we caution against drawing strong conclusions here owing to the true uncertainty of the topography of the unseen hemisphere.\par

Fig.\ref{Subsurface_Flux} shows the instantaneous daily averaged conductive flux at the surface at the day of the encounter.  Once again the view and input parameters are the same as those shown for the two prior figures.  The figure confirms one's intuition, in which the thermal flux is interior directed on the lit summer side of Arrokoth ($\sim$ 0.002 W/m$^2$, i.e., ), while the conductive flux is surface directed on Arrokoth's winter side ($\sim$ -0.0015 W/m$^2$).  We have delineated where the transition occurs from outward to inward flux as well.  One notable feature is the appearance of a relatively high intensity ring of surface directed thermal flux ($\sim$ -0.003 W/m$^2$) along the temperature terminator.  This ring coincides with the location of the pre-sunrise edge of Arrokoth's surface, where the temperature gradient along the surface is greatest (e.g., compare against the surface temperatures shown in  Fig. \ref{Surface_Temperatures}).  We find this ring to closely follow the illumination terminator over the course of \REV{Arrokoth's} solar revolution.\par 
Fig. \ref{encounter_day_histograms} shows several facet count histograms for various quantities at encounter day.
These include a distribution of surface temperatures to be compared with  Fig. \ref{Surface_Temperatures}
and surface fluxes of  Fig. \ref{Subsurface_Flux}.  The intense ring of surface fluxes appears as a long tail \REV{below} the
distribution mode at $\Lambda \approx -0.0015 $W/m$^2$.  We have also shown the distribution of instaneous total received
illumation flux at every facet, which includes both direct received sunlight as well as illumination received from  reradiation from other ``seen" facets.  The distribution shows a strong mode at around 0.6 W/m$^2$, while including a large number of facets receiving nearly zero flux (i.e., only CMB).  We have separated out from the total received flux only the flux received from surface reradiation (labelled as flux received from surrounding landscape).\REV{  
{The irradiation} received in the form of surface re-radiation ($\sim 0.02$ W/m$^2$) amounts to less than 5\% of the total irradiation.}  Once again this appears to be due to Arrokoth's relatively muted local relief despite its otherwise flattened global shape.  \REV{These trends are
consistent with those based on low-order 1962 facet model reported in \citet{Grundy_etal_2020}.}

\REV{As a final reflection, the behavior here ought to be compared to temperature predictions done for a body with a neck but much more marked topography (e.g., 67P/Churyumov-Gerasimenko), would be interesting in its own right, in terms of the role of re-radiation, shadowing, etc. 
\citep[][]{Hu_etal_2017}.}

\subsection{Orbital Timescale Features}

Adopting ${\cal I} = 2.54$ tiu, as in the prior section, we display various orbital timescale properties.  In Fig. \ref{Avg_Temperature} we showcase the orbitally averaged surface temperature of each facet.  This quantity also corresponds to the asymptotic interior temperature (i.e., $T\sub{{\rm int},i}$) descriptive of subsurface positions exceeding several thermal skin depths.  For all regions of the body except for a narrow equatorial band ($\sim \pm 10^\circ$), the deep interior temperature is in the low 30K-35K range.  On the other hand, $T\sub{{\rm int},i}$ in the equatorial band 
is closer to 40K.  These trends make sense given that polar winters last long enough for all the received and thermalized summer time solar energy to radiate away.  Since the equatorial ``tropical" zone always sustains a diurnal insolation pattern, the received solar insolation over those regions never sufficiently radiates away to lower the surface temperatures nearly as
much before the Sun rises again.  This effect is most dramatically illustrated in Fig.\ref{Min_Temperature} showing
the absolute minimum temperature any one facet experiences over the course of one orbit.  Almost the entirety of Arrokoth's surface, including substantial parts of the tropical zone, experiences minimum temperatures in the range of 10K-15K.  But, a very thin sliver of real estate following the equator and occasionally outlining cratered morphology regions found within the tropical zone, show
minimum temperatures as high as 30K.  The distribution of these temperatures are shown in Fig. \ref{orbital_timescale_histograms}.  \par
The interior depths of the equatorial zone are thus warmer than the polar region despite the fact that these parts receive overall less insolation over the course
of one orbit.  This is clear when viewing the predicted flux temperature as defined in Eq. (\ref{flux_temperature_def}). $T\sub{{\rm a},i}$ is a proxy for the total \REV{irradiation} received on any given facet and
is shown in Fig. \ref{Avg_Flux_Temperature}. As an orbital average, the tropical zones receive far less solar insolation, with $T\sub{{\rm a},i} \sim 40$K, than the polar regions, where $T\sub{{\rm a},i} \sim 45$K (see also third panel of Fig. \ref{orbital_timescale_histograms}). Yet despite this, the averaged surface temperature in the tropics is much higher than the poles.

\section{Connecting to REX observed brightness temperatures}\label{connect_to_REX}

\begin{figure*}
\begin{center}
\leavevmode
\includegraphics[width=17.5cm]{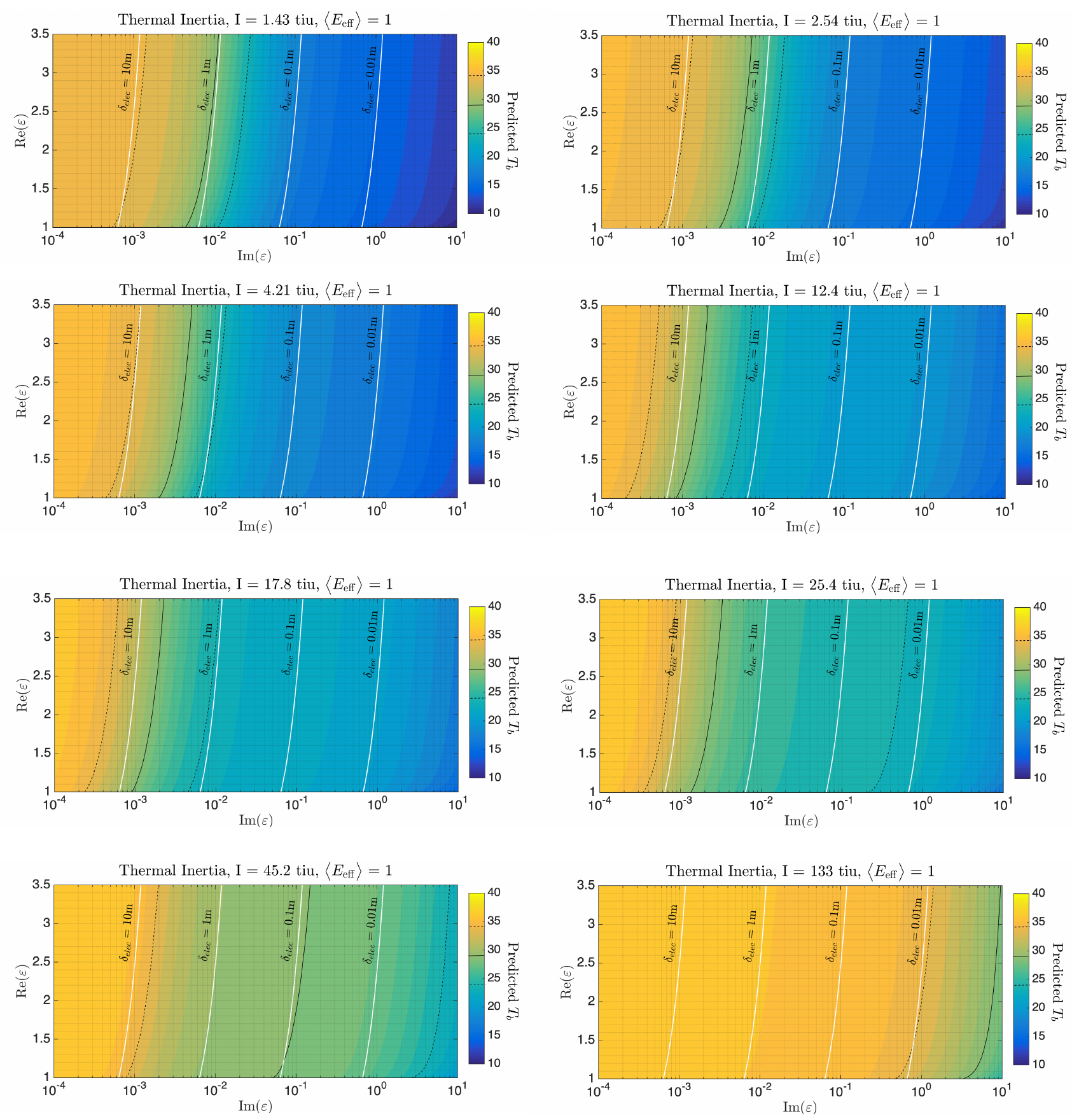}
\par
\end{center}
\caption{Predicted brightness temperatures based on subsurface radiative transfer analysis for 
$\big<E_{\rm eff}\big> = 1$. Several values of ${\cal I}$ shown.  White contours denote several values
of electric skin depth.  Note the weak dependence on the $\varepsilon'$ (donoted on graphs as Re$(\varepsilon)$).}
\label{Predicted_Brightness_Temperatures}
\end{figure*}

\begin{figure*}
\begin{center}
\leavevmode
\includegraphics[width=12.5cm]{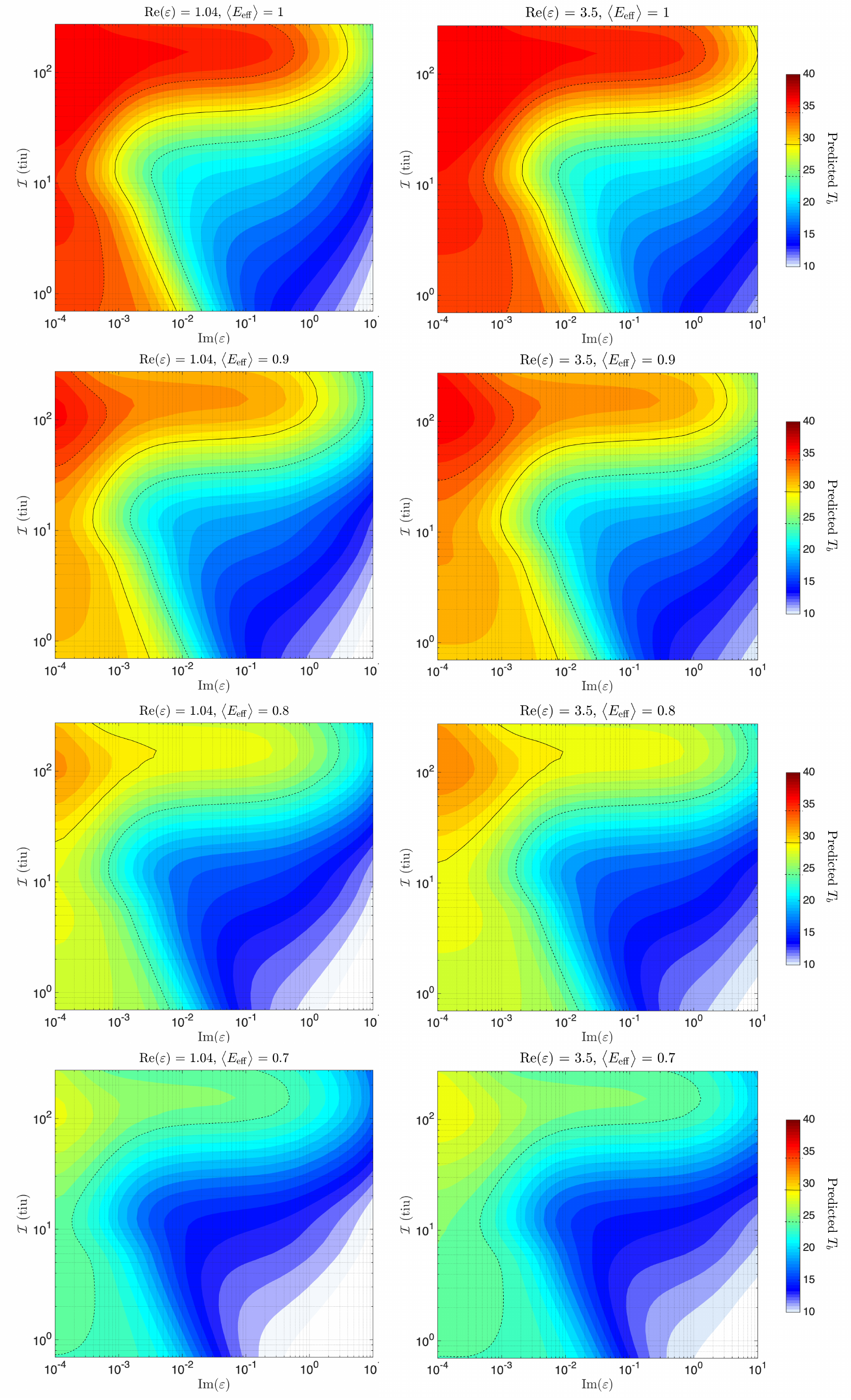}
\par
\end{center}
\caption{Predicted brightness temperatures $T_b$ for the CA08 observation.  For input values of $\varepsilon' \equiv {\rm Re}(\varepsilon)$ and $\Eeff$ predicted $T_b$ shown as a function of $\varepsilon'' \equiv {\rm Im}(\varepsilon)$ and ${\cal I}$:  (left column) $\varepsilon' = 1.04$,  (right column) $\varepsilon' = 3.50$.  Successive rows corresponding to decreasing values of $\Eeff$. Black contours denote $T_b = 29$K while dotted black contours represent $T_b = 24$K and 34K levels.
The right column might be considered representative of pure methanol ice of zero porosity,
while the left column could be the same with porosity of 60\%.}
\label{Tb_Solution_Array}
\end{figure*}

The REX radio flux density measurements made during the CA08 observing sequence, together with the shape model for Arrokoth, leads to an Arrokoth-disk averaged x-band brightness temperature $T_{b,{\rm obs}} = 29 
\pm 5K$ \citep{Grundy_etal_2020}. This result has been reaffirmed in a more comprehensive analysis (Bird et al. 2022). The brightness temperature depends on various properties of Arrokoth’s near surface materials. In order to calculate a model prediction for the brightness temperature that we here simply call $T_b$, it is necessary to perform a radiation transfer analysis that relates Arrokoth’s interior kinetic temperature $T(z)$ -- as detailed in previous sections -- and its X-band thermal and refractive properties to $T_{b,obs}$.
In other words, 
\beq
T_{b,{\rm obs}} = T_{b}\Big({\cal I},\varepsilon',\varepsilon'',\big< E_{{\rm eff}}\big>\Big),
\label{Tb_sol}
\eeq
in which $T_{b}$ depends on knowledge of the material's thermal inertia, the value of the real and imaginary part of its  X-band dielectric constant/permittivity $\varepsilon =  \varepsilon\prime + i  \varepsilon\prime\prime$ 
\footnote{Note, hereafter all values of $\varepsilon$ are scaled in units of vacuum permittivity $\varepsilon_0 
\approx 8.85 \times 10^{-12}$F/m, where F is in units of Farads.},
 and its effective X-band emissivity $\big< E_{{\rm eff}}\big>$.  
These 4 a priori unknown parameters may themselves have some kind of depth-dependence, but for our purposes 
here we take them to be constants.   
In principle, there exists a relationship between 
${\cal I},\varepsilon',\varepsilon''$, and $\big< E_{{\rm eff}}\big>$ that produces values of $T_b$ equal to 
$T_{b,{\rm obs}}$.  Anticipating our discussion in section \ref{transfer_discussion}, at
best we will be able to only circumscribe this relationship between the four unknown parameters.
\par
Utilizing the approach detailed in Section 3, we develop a suite of near surface thermal solutions for several values of the thermal inertia, i.e., in the range
$0.5\ {\rm tiu} < {\cal I} < 275\ {\rm tiu}$ on 35 equally spaced values along a logarithmic scale.  
Together with our adopted values of $\rho$ and $C_p$, 
this range of ${\cal I}$ values corresponds to a characteristic
orbital timescale skin depth values, 
\beq
\ell_{\rm orb} \equiv
{\cal I}\Big/\rho C_p \sqrt{\omega},
\label{lorb_definition}
\eeq
that fall into the range
$
22 \ {\rm cm} \lessapprox \ell_{\rm orb} \lessapprox
120 \ {\rm m},
$
where $\ell_{\rm orb}$'s definition is based on the setting $n=1$ into the absolute value 
of the inverse of the thermal wavenumber defined Eq. (\ref{kn_definition}).
It is sufficient to use $N=30$ Fourier modes to generate this solution array (see section \ref{thermal_solution}).  
We develop solutions in which for depths below 10 meters from the surface the thermal conductivity transitions to a fixed constant value for porous \water ice
\citep[e.g., $K = 1 $W/m/K,][corresponding to a value ${\cal I} = 295$ tiu -- also see sec. \ref{thermal_solution} and Appendix C]{Klinger_1980}.\par
With these temperature solutions in hand, we then employ the radiative
transfer solution method outlined in \cite{deKleer_etal_2021} to produce a set of predicted $\tilde I_\nu$ that finally relate to $T_b$ via Eq. (\ref{Tb_sol}). For the transfer solutions we consider 
the following ranges of the remaining parameters:
$1 \le \varepsilon'\le 3.5$ at 70 equally spaced values on a linear scale, $10^{-4} \le \varepsilon''\le 10$ at 75 equally spaced values on a logarithmic scale,
and four values for, $\Eeff = 0.7,0.8,0.9,1.0$.  These parameter
choices follow those also considered in Bird et al. (2022).  Further details of the solution
method as implemented are found in Appendix \ref{Radiative_Transfer_Solution_Details}.
We note that the electrical skin depth, $\delta_{{\rm elec}}$, approximately relates to $\varepsilon$
via
\beq
\delta_{{\rm elec}} = \frac{\lambda}{2\pi}\frac{\sqrt{\varepsilon'}}{\varepsilon''},
\label{approx_delta_elec_definition}
\eeq
in the $\varepsilon'' \ll 1$ limit of its formal definition found in Eq. (\ref{delta_elec_definition}). 
Finally, we do not account for Fresnel reflection at the interface.
\par
\REVV{In our companion study, Bird et al. (2022, this issue) also develop $T_{b,{\rm obs}}$ predictions based on the thermal solutions.  The differences lie in the approaches taken to solve the radiative transfer problem: As each spacecraft visible facet $i$ has its own unique vertical temperature profile $T_i(z)$, this work develops a facet-by-facet solution to the transfer problem followed by taking a facet area weighted average across the visible disk of Arrokoth to derive $T_{b,{\rm obs}}$.  On the other hand, Bird et al. (2022) develop a simpler estimate for $T_{b,{\rm obs}}$  based on a single effective vertical temperature profile $\overline T(z)$, which is constructed as a facet-area weighted average of all $T_i(z)$, and a surface averaged facet norm with associated cosine of the spacecraft inclination angle ${\cos \theta_t}$.  Formally speaking the two approaches are not necessarily commensurate -- as the average of products is not necessarily equal to the product of averages (also see discussion in Appendix D -- but in this case here we find that the results happen to be in mutual agreement.}
\par
We observe several trends.  Most prominent is that for given values of ${\cal I}$ there is only a weak dependence of on $\varepsilon''$ as $\varepsilon'$ varies from 1 to 3.5 (see Fig. \ref{Predicted_Brightness_Temperatures}).
Fig. \ref{Tb_Solution_Array} displays contour levels of several predicted $T_b$ as a function of $\varepsilon''$ and ${\cal I}$ for several fixed values of $\varepsilon'$ and $\Eeff$. 
Except for $\Eeff = 0.7$, the $T_b = 29$K contour exists for all input parameter values considered.
In section \ref{transfer_discussion}
we further discuss and interpret the implications of these solutions, especially with respect to known candidate materials.  However we can make one clear observation here that solutions in which $\Eeff = 0.7$ appear to be ruled out as the $T_b = 29$K contour fails to even register under these circumstances and, at best, only the low end of the REX observation is predicted ($\sim 24$K) but does so at very low (nearly transparent) values of $\varepsilon'' (< 10^{-3})$ for ${\cal I} \lessapprox 40$tiu, while the predicted $T_b = 24$K contour is plausible only for ${\cal I} > 100$tiu at considerably larger upper bound values for  $\varepsilon''$, i.e., in the range of 0.1 and 1.\par

\begin{figure*}
\begin{center}
\leavevmode
\includegraphics[width=18.cm]{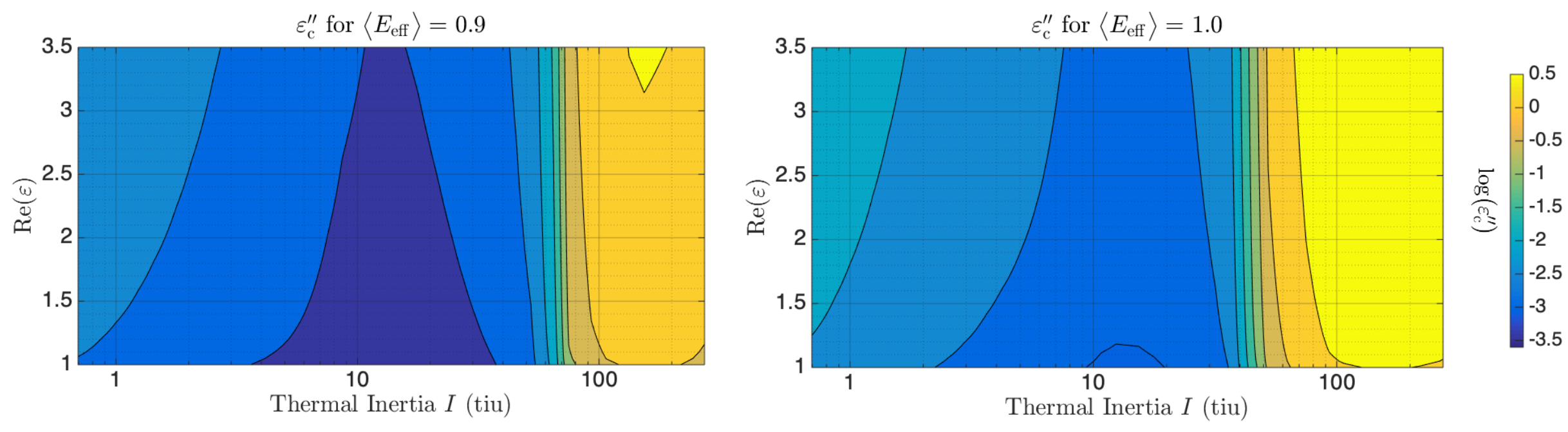}
\par
\end{center}
\caption{Value of $\varepsilon_c$'' that yields $T_b = 29$K during CA08 for $\Eeff = 0.9, 1.0$.  For these values of $\Eeff$ this analysis predicts that the material is very nearly transparent in the X-band ($\varepsilon_c'' \lessapprox 0.002$) for $1\ {\rm tiu}\ <{\cal I}<40$tiu.}
\label{Critical_Im_Epsilon_Array}
\end{figure*}

We pose another question: Given $\varepsilon', {\cal I}$ and $\Eeff$, what value of $\varepsilon''$ yields a predicted $T_b$ during the CA08 observation?  We define this critical value as $\varepsilon_c''$ and we derive these values using interpolation methods based on the solutions determined above.
Fig. \ref{Critical_Im_Epsilon_Array} displays $\log\varepsilon_c''$ as a contour plot for $\Eeff = 0.9,1.0$.  We have left out doing this exercise for the $\Eeff = 0.7$ and $0.8$ cases because $T_b = 29$K registers as a possibility only for a limited range of ${\cal I}$ values for the latter case and none at all for the former case (see Fig. \ref{Tb_Solution_Array}).  It is notable that in both cases $\varepsilon_c'' < 0.05$ for values of ${\cal I} < 70$ tiu.  This analysis also predicts that the material is very nearly transparent in the X-band (i.e., $\varepsilon_c'' \approx 0.001$ or less) for values of $2\ {\rm tiu}\ \lessapprox {\cal I} \lessapprox 40$tiu.  We revisit this in the next section.
\par

\begin{figure}
\begin{center}
\leavevmode
\includegraphics[width=8.5cm]{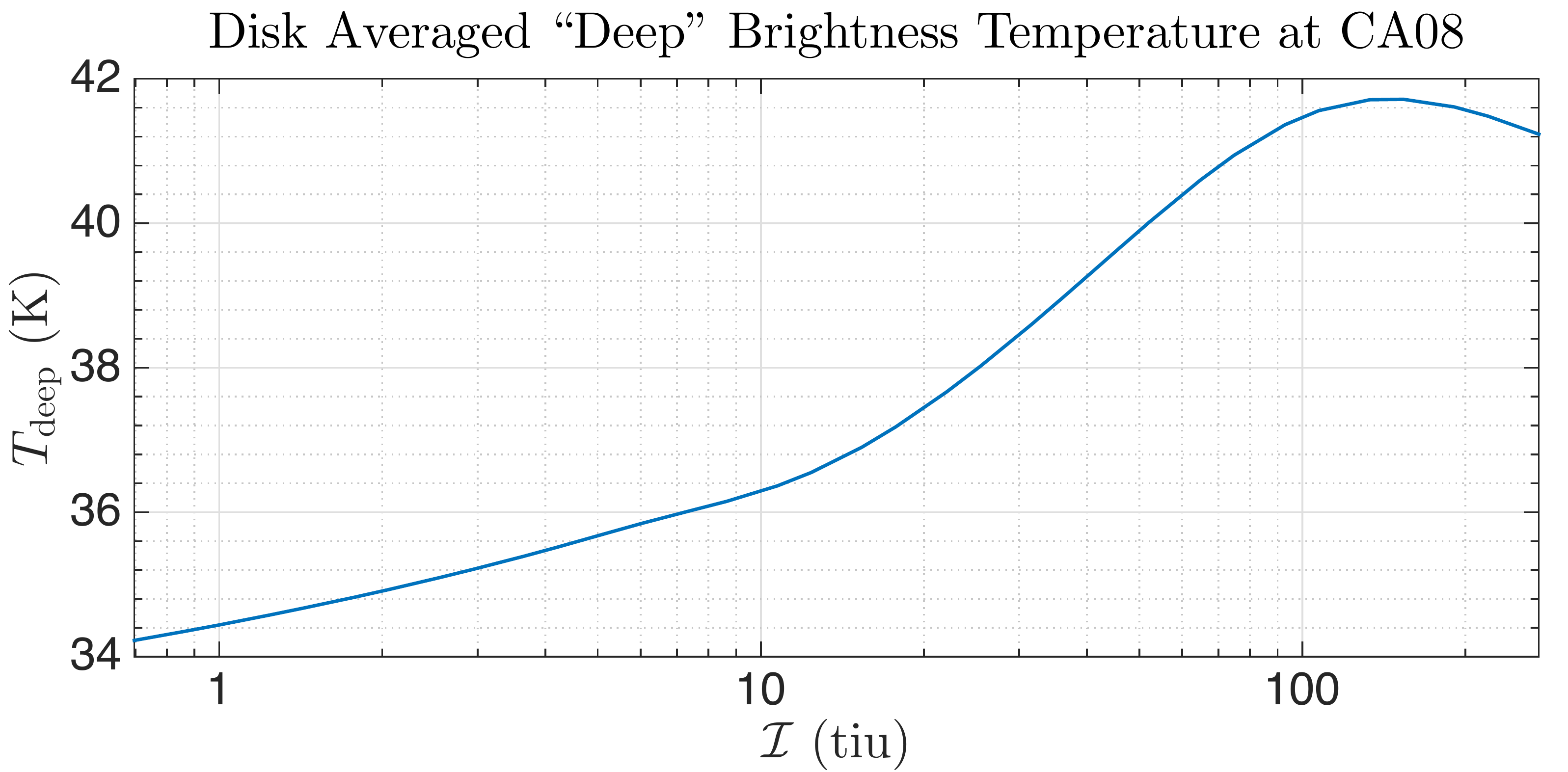}
\par
\end{center}
\caption{The disk averaged ``deep" temperature $T_{{\rm deep}}$ as a function of ${\cal I}$, predicted for CA08. }
\label{Deep_Temperature_CA08}
\end{figure}

A question that will be addressed later is if
$T_{b,{{\rm obs}}}$ might be explained by X-band
radiation emanating from well below the orbital time scale
skin-depth.  To help answer this question, 
we define $T_{{\rm deep}}$ to be the visible-facet weighted observed X-band brightness temperature at $\Eeff = 1$ for values of the dielectric skin depth -- $\delta_{{\rm elec}}$ defined in Eq. (\ref{delta_elec_definition}) -- greatly exceeding $\ell_{{\rm orb}}$. 
In other words, it corresponds to a visible-facet weighted disk averaged {\it deep subsurface temperature} sampled by {\it nearly} transparent X-band radiation sampling the deep interior.
\footnote{We are careful to note that there is a limit here as well.  One might imagine the X-band radiation is so transparent through Arrokoth's interior that it could also sample its  sunlight side.  However this would suggest unrealistically low values of $\varepsilon'' \ll 10^{-6}$.}
From this interpretation it
follows that
\beq
T_b\Big(\Eeff,\delta_{{\rm elec}} \gg \ell_{{\rm orb}}\Big) 
\rightarrow \Eeff\cdot T_{{\rm deep}}.
\eeq
\par \noindent
In Fig. \ref{Deep_Temperature_CA08} we plot $T_{{\rm deep}}$ as a function of ${\cal I}$. 
Inspection of the figure clearly shows
that values of $\Eeff = 0.7$ are clearly precluded from predicting even $T_b = 24$K
on the low ${\cal I}$ end, while possibly permitting $T_b = 29$K on the high ${\cal I}$ end (i.e., $>150$ tiu).  We return
to discussing this in the next section.

\section{Discussion}
\subsection{On the thermophysical properties of Arrokoth's subsurface materials}\label{transfer_discussion}
There are no independent measures of the four input parameters that go into
these thermal solutions constructed here.  
Unfortunately, there is very little one can conclude with certainty about their values or properties based on the New Horizons' single REX scan
(see also Bird et al. 2022). Partly given its laboratory data availability and partly given that planetesimals are thought to be substantially composed of \water ice,  we mainly focus on the property of water ices despite there being no direct evidence for it on the surface.  This \water ice which may also have a tholin covering. We save for future analysis the possibility for other less-well laboratory studied materials to explain the REX observation.
\par
The primary concern here is that there is a relationship between 
$\varepsilon',\varepsilon'', {\cal I}$, and $\Eeff$ that yield a predicted $T_b = 29K$ 
corresponding to the brightness temperature measured during the CA08 REX observation.  One example relationship is the quantity $\varepsilon'' = \varepsilon_c''\left(\varepsilon',{\cal I},\Eeff\right)$ developed in the previous section and, for example,
displayed in Fig. (\ref{Critical_Im_Epsilon_Array}). Aside from Arrokoth harboring surface methanol, there is no other information about the body's surface or subsurface composition. As such, we are here relegated to speculating about the possible values of these quantities in relation to other known bodies of the outer solar system.
 In this section we consider various facets of the problem and we conclude by offering what we consider to be our favored interpretation.
\par

\subsubsection{Permittivity/Dielectric Constant}\label{varepsilon_discussion}
Perhaps most 
confounding for this analysis are the uncertain values of $\varepsilon'$ and $\varepsilon''$, the latter of which strongly controls the material's attenuation of the X-band signal through the subsurface medium especially when $\varepsilon'' \ll 1$, e.g., see Eq. (\ref{delta_elec_definition}).  We consider several possibilities in order of increasing speculation:
\begin{itemize}
\item Methanol, likely a significant
constituent of Arrokoth's near surface, has a value of 
$\varepsilon' \approx 3.52 \pm 0.05$, but with no corresponding measurement of $\varepsilon''$ \citep{LeGall_etal_2016}. This same study also reported a value of $\varepsilon' \approx 3.42 \pm 0.05$ for the similar hydrocarbon ethanol.  
One can estimate an effective $\varepsilon'$ on the assumption a certain ice fraction of vacuum porosity ($p$ hereafter) following the Maxwell Garnet empirical formula for vacuum mixtures discussed in Bird et al. (2022). If, for example, the upper layer is composed of pure methanol with $p=0.6$, then it would have an effective $\varepsilon' \approx 1.04$ corresponding to 
the left column of solutions shown in Fig. \ref{Tb_Solution_Array}, while the $p=0$ solution
would be the same figure's right column.

\item Based on laboratory work of \citet{Paillou_etal_2008}, which examined the dielectric properties of various tholin-like materials,
Bird et al. (2022) compiled a range of possible
 effective $\varepsilon$ values of several types of porous tholins based on the same
 prescription for mixtures.
 For porosity $p$ in the 0.60 to 0.80 range: powdered tholins exhibits
 values of $\varepsilon'$ from $1.05$ to $1.01$, with
   $\varepsilon''$ values ranging from $1\times 10^{-3}$ to 
   $5\times 10^{-4}$; while
   for two types of compact tholins examined $\varepsilon'$ correspondingly
   ranges from  $1.5$ to $1.2$ together with
   $\varepsilon''$ falling somewhere between $4.2\times 10^{-3}$
   and $1.5\times 10^{-3}$.

\item
Arrokoth's subsurface materials might resemble the ``dirty" ice thought to comprise the nucleus of comets like 67P.  \citet{Heggy_etal_2012} have compiled several known measurements of $\varepsilon = \varepsilon_{{\rm mix}}$ of matrix ice with variable dust-to-\water mixtures with mixture fraction $0<\phi_{{\rm mix}}<1$.The measurements were made at wavelengths a factor 10-50 larger than the X-band. 
From these they compile an empirical relationship for $\varepsilon_{{\rm mix}}'$ and $\varepsilon_{{\rm mix}}''$
as a function of $\phi_{{\rm mix}}$, as well as a weak dependence on $T$,  For $T = 30$K
and $p=0.5$ application of their relationship predicts for  $\phi_{{\rm mix}} \ll 1$ that $\varepsilon_{{\rm mix}}' \approx 1.42$ and $\varepsilon_{{\rm mix}}'' \approx 10^{-4}$, while
for $\phi_{{\rm mix}} \rightarrow 1$
$\varepsilon_{{\rm mix}}' \approx 2.90$ and $\varepsilon_{{\rm mix}}'' \approx 4\times 10^{-2}$.  These values will be somewhat smaller in bulk for higher porosities at fixed 
$\phi_{{\rm mix}}$.
Similarly, \citet{Brouet_etal_2015} compiled $\varepsilon'$ values for an ice mixture comprised of \water and the lunar regolith simulant JSC-1A.  For porosities $p\approx 0.6$
they find $1.5< \varepsilon' < 2.25$, depending on mixture fraction \citep[e.g., see Figure 10 of][]{Brouet_etal_2015};  however, no values for $\varepsilon^{\prime\prime}$ were reported.
\end{itemize}

\subsubsection{X-band Emissivity}
Bird et al. (2022) compile a known list of the radio emissivity of the Kronian moons at $\lambda = 2.2$cm based on Cassini RADAR observations.  The color and albedos of these objects vary widely, with corresponding emissivity range of roughly $0.6 < \Eeff < 1$.  However, if one focuses on the darkest of these objects (i.e., those with the lowest albedos comparable to Arrokoth) we find, for example, Iapetus' leading dark side has $\Eeff \approx 0.87$ \citep{LeGall_etal_2014}
while for Phoebe $\Eeff = 0.92$ \citep{Ostro_etal_2006}.  If these bodies's surface materials are analogs to Arrokoth's, then supposing 
$\Eeff = 0.9$ for it would not be unreasonable.
\par
In reference to our preceeding discussion, it seems $\Eeff = 0.7$ is highly unlikely.  One might argue by inspecting the deep temperature solutions shown in Fig. \ref{Deep_Temperature_CA08} that
$T_b = 29$K is achievable for $\Eeff = 0.7$ on the high
${\cal I}$ end (i.e., the product $\Eeff \cdot T_{{\rm deep}}$ yielding 29 K), but cross-referencing that against
the $\Eeff = 0.7$ row of Fig. \ref{Tb_Solution_Array}
would imply that this is feasible only for very large electric skin-depths, corresponding to
$\varepsilon''\ll 10^{-4}$.  Given the discussion in the previous section we know of no plausible materials that would exhibit such highly transparent qualities.  Performing the same cross-comparison a similar line of reasoning applies for rejecting $\Eeff = 0.8$ solutions for values of ${\cal I}< 20$ tiu.  However we see from
the penultimate row of  Fig. \ref{Tb_Solution_Array} that  $T_b = 29$K is achievable for $\Eeff = 0.8$ as it corresponds to physically reasonable values of $10^{-4}<\varepsilon''<10^{-3}$ in the thermal inertia range $20\ {\rm tiu} < {\cal I} \lessapprox 300$ tiu. 
Based on the discussion, of the following section we also lean toward treating this case as unlikely (but not ruled out) on account of identifying plausible materials with such high bulk thermal inertias at such low temperatures and porosities.

\subsubsection{Thermal Inertia}
{\it{TNOs and KBOs as analogs.}}
The 1-10 tiu ${\cal I}$ range of values 
 reported in \cite{Lellouch_etal_2013} for KBOs corresponds to daytime measurements
that probe their surface materials down to
their diurnal skin depths, which are in the vicinity of a few mm.  Such low thermal inertias, which ought roughly be considered body averages, can be interpreted as indicating the presence of a very porous matrix
of weakly loaded grains whose grain-grain radii of contact ($a$) are much smaller than the
grain size (radius $R_g$).  Since the ratio $h\equiv a/R_g$ (sometimes known as the Hertz factor) controls grain-to-grain heat conduction, small values of $h$ -- perhaps due to weak loading -- coupled to high porosity conditions leads to thwarted effective heat conduction
\citep[see further below, and also][]{Howett_etal_2010,Ferrari_Lucas_2016,Ferrari_2018}.  
Therefore, it might not be unreasonable to suppose such fluffy low thermal inertia material extends
much deeper than a few mm on small-sized KBOs like Arrokoth as little material compaction 
is expected under such low gravities ($\sim 1$mm/s$^2$)
over its natural history -- unless its surface experienced sufficient impacting events, but
this appears to be ruled out on the basis of the relatively few observed impact features
\citep{Stern_etal_2019, Spencer_etal_2020,McKinnon_etal_2020}.
\par 
\par
\medskip
{\it{Iapetus and Phoebe as analogs.}}
While daytime observations were used to derive thermal inertias in the previous section, more direct methods have been used for both Iapetus and Phoebe who, at least on their surface, could be considered analogs for Arrokoth:
The three bodies share similar Bond albedos \citep[$\sim 0.05$][]{Ostro_etal_2006} and Phoebe, which likely sources Iapetus' dark side materials, may originate from the same population of KBOs as Arrokoth \citep{Johnson_Lunine_2005,Castillo-Rogez_etal_2012}  
\footnote{Although recent challenges to this view have been raised based on the similarities of
Phoebe's spectral properties
to the C-type class of asteroids \citep{Hartmann_1987,Matson_etal_2009}, 
as well as open questions 
based on internal thermal evolution modeling
of its early history \citep{Castillo-Rogez_etal_2019}.}.
In particular,
Iapetus' leading side's very dark top soil (especially that on Cassini Regio, CR hereafter) has been argued to be sourced from the Phoebe ring, which is itself sourced from Phoebe via micrometeorite impact gardening of its surface \citep{Verbiscer_etal_2009}.   Thus considering Iapetus and Phoebe's surface materials together as an analog for Arrokoth is not wholly unjustified, at the very least. 
\par
Based on observations made by Cassini's Composite Infrared Spectrometer (CIRS), estimates for the thermal inertias of several of the Kronian satellites have been published
\citep[e.g.,][]{Flasar_etal_2005, Howett_etal_2010, Rivera-Valentin_etal_2011}.
Current thermal inertia estimates for Iapetus's darkened side based on CIRS data are  
$6 \ {\rm tiu} \lessapprox {\cal I} \lessapprox 21 \ {\rm tiu}$
\citep{Howett_etal_2010}, with a more recent narrower range
at 
$11.0 \ {\rm tiu} \lessapprox {\cal I} \lessapprox 14.8 \ {\rm tiu}$
\citep{Rivera-Valentin_etal_2011}. For comparison, Iapetus' lighter trailing side surface materials is modeled from CIRS data to have 
$12 \ {\rm tiu} \lessapprox {\cal I} \lessapprox 33 \ {\rm tiu}$
\citep{Howett_etal_2010}, and possibly more tightly constrained at 
$15 \ {\rm tiu} \lessapprox {\cal I} \lessapprox 25 \ {\rm tiu}$ \citep{Rivera-Valentin_etal_2011}.
Analysis of CIRS data provides estimates for Phoebe's
surface thermal inertias around ${\cal I} = 30$tiu 
\citep{Spencer_etal_2004,Flasar_etal_2005}.
\par
It is also important to note that based on thermal modeling leveraged against Cassini RADAR data, which was used to observe CR's subsurface at 2.2 cm, \citet{LeGall_etal_2014}, report
higher thermal inertias of at least around ${\cal I} \approx 50$ tiu and possibly even 
${\cal I} > 200$ tiu.  Given that the electric skin-depths in the radio are vastly larger than in the infrared, this higher ${\cal I}$ estimate is likely characteristic of surface materials far deeper than 20cm.  Given the likelihood that this deeper (presumably) \water ice is both compacted and in cubic crystalline form, we do not consider it as a plausible analog for Arrokoth (see further below).
\par
\medskip
{\it{ The thermal inertia of amorphous \water ice grains.}}
Even though New Horizons did not directly detect \water ice  on Arrokoth's surface, theoretical considerations including global disk evolution modeling indicate that planetesimals formed out beyond the water snowline should be made up of significant amounts either \water ice grains or more complex silicate grains with \water locked in within them
\citep[e.g.,][and many others]{Estrada_etal_2016}.  Perhaps for Arrokoth these \water ice grains are encased by a thin cover of methanol or other hydrocarbons.
Moreover, a recent study of the carbonaceous chondrites NWA5717 and Allende \citep{Simon_etal_2018} reveal that these primitive bodies are comprised of cm-scale aggregates of sub-mm chondrules 
(with average diameters of the distribution mode in the range 0.15-0.70 mm).  These authors find that a variety of subgroups of particles (characterized by different compositions and/or lithologies) manifest similar size distributions and aggregate character, which lead them to posit that the same size sorting process might be operating across the entire early planetesimal formation phase of the solar nebula.
\par
These considerations taken together lead to the not-unreasonable
proposal that small KBO bodies like Arrokoth are composed of 
fluffy aggregates of \water-{\emph{dominated}} sub-mm particles.
Assuming \water to be the dominant component of individual
grains and assuming that the grains were assembled into Arrokoth in situ, then it is also not unreasonable to propose that they are and have always been in an amorphous crystalline form.  This is particularly compelling since the amorphous to cubic crystalline transition in type 1h \water ice becomes important on the timescale of the solar system's existence once temperatures exceed $70$K \citep{Efimov_etal_2011}
-- a condition unlikely to have been possible at 40-45 AU where CCKBOs reside.  For reference we note that amorphous \water ice has a thermal conductivity that is lower than its cubic-phase counterpart 
by an order of magnitude or more when compared under the same thermophysical conditions \citep{Andersson_Suga_1994}.\par
Motivated by a need to explain Mimas' estimated thermal inertia \citet{Ferrari_Lucas_2016} have suggested that the low ${\cal I}$ observed of a large array outer-solar system bodies (Centaurs, TNOs, etc.) might be rationalized as arising from primitive highly porous grains of amorphous \water ice grains.  
We entertain this possibility for Arrokoth starting with a brief theoretical primer on the matter following \citet{Ferrari_Lucas_2016}:  
\par
{\it On an airless body}, the effective thermal inertia of a fluffy aggregate of icy-particles of size $R_g$ and intrinsic ice-grain density $\rho_{{\rm ice}}$ is posited in the form
\beq
{\cal I} = \sqrt{(1-p)\rho_{{\rm ice}} C_p(T)(K_{{\rm r}}
+ K_{{\rm c}})},
\eeq
The effective conductivity, $K_{{\rm eff}} \equiv K_{{\rm r}} + K_{{\rm c}}$, is the sum of the grain-to-grain heat transport via radiation ($K_{{\rm r}}$) and
the thermal conduction from grain to grain across shared contact points ($K_{{\rm c}}$).  For $K_{{\rm r}}$ there are several known approaches \citep[e.g.,][]{Shoshany_etal_2002,Piqueux_2009} but here we adopt the physically tractable model form of \citet{Gundlach_Blum_2012} where
\beq
K_{{\rm r}} = 8\epsilon \sigma T^3 R_g \phi_{r}(p)
, \qquad
\phi_{r}(p) \equiv \frac{e_1 p}{1-p},
\eeq
where $\epsilon$ is the IR emissivity.
Grossly speaking, the expression represents the transport of blackbody radiative energy across
the chasm 
separating grains of similar size
(i.e., $\sim R_g$).  The porosity dependent radiation
exchange factor $\phi_{r}(p)$ accounts for the added effective distance this radiation must
travel due to increasingly rarefied interstitial structure of low porosity grain aggregates. $e_1 = 1.34$ is a structural
constant of the model.  
\par
The contact mitigated conductivity can be described with Johnson–Kendall–Roberts theory of contact mechanics
\citep[JKR hereafter,][]{JKR_1971},
\beq
K_{{\rm C}} = h\phi_{c}(p) K_{{\rm A}},
\eeq
where $\phi_{c}(p)$ is another porosity dependent factor taking into account reduced conduction due to the increase of vacuum space as $p$ approaches 1.  Here we assume the form $\phi_c(p) = 2(1-p)$, which is an averaged approximation based on empirical data across a range of porosities as reported in \citet{Gusarov_etal_2003} \citep[also see][]{Ferrari_Lucas_2016}.  $K_{{\rm A}}$ is the intrinsic conductivity of an amorphous \water ice grain that can be estimated on theoretical grounds to be given by
\beq
K_{{\rm A}} = C_p\rho_{{\rm ice}} v_{{\rm ph}} \lambda_{\rm{mol}}/4,
\label{KS_theory}
\eeq
where $v_{{\rm ph}}$ is its phonon speed, $\lambda_{{\rm mol}}$ is the typical molecule separation in the crystal. 
Adopting this theoretical estimate checks out well against experimental data at temperatures slightly higher than that of interest here \citep[$\sim 70$ K,][]{Andersson_Suga_1994}.  Absent experimental verification in the colder temperature range relevant here, we
assume Eq. (\ref{KS_theory}) is valid for our purposes.
We treat
\water ice's heat capacity -- which is assumed to be the same in its cubic crystalline and amorphous forms -- according to its known temperature dependence, for which \citet{Shulman_2004} has developed an empirical form for based on experimental data in the range of 0 and 273K
\citep[see Eq. 4 of][]{Shulman_2004}. The Shulman approximation is an improvement over
the often used one attributed to \citet{Klinger_1980}, in which
$C_p(T) = 7.49 (T/1{\rm{K}}) + 90$ J/m/kg.
\par
The grain-grain contact radius $a$ reflects the elastic deformation resulting from adhesive ``pinching" forces related to a grain's intrinsic surface tension plus the deformation that occurs under a given load.  For the kinds of low gravities of concern here, it is expected that adhesive forces will dominate overburden load down to at least a few meters on Arrokoth.  Thus
for amorphous water ice grains under negligible load the Hertz factor simplifies to
\beq
h = \frac{a}{R_g} =
\left(\frac{L_{{\rm ec}}}{R_g}\right)^{1/3},
\qquad
L_{{\rm ec}}
\equiv
 \frac{9\pi(1-\nu_p^2)\gamma}{4E},
\eeq
where $E$ is the ice grain's Young's modulus, $\nu_p$ is the Poisson ratio, and $\gamma$ 
is the ice grain's surface tension.  The elastocapillary length scale $L_{{\rm ec}}$
nominally denotes deformation scale at which restorative elastic forces balance adhesive ones arising from surface tension
\citep[e.g.,][]{Jagota_etal_2012,Style_etal_2013}.
\par
\begin{figure}
\begin{center}
\leavevmode
\includegraphics[width=9.cm]{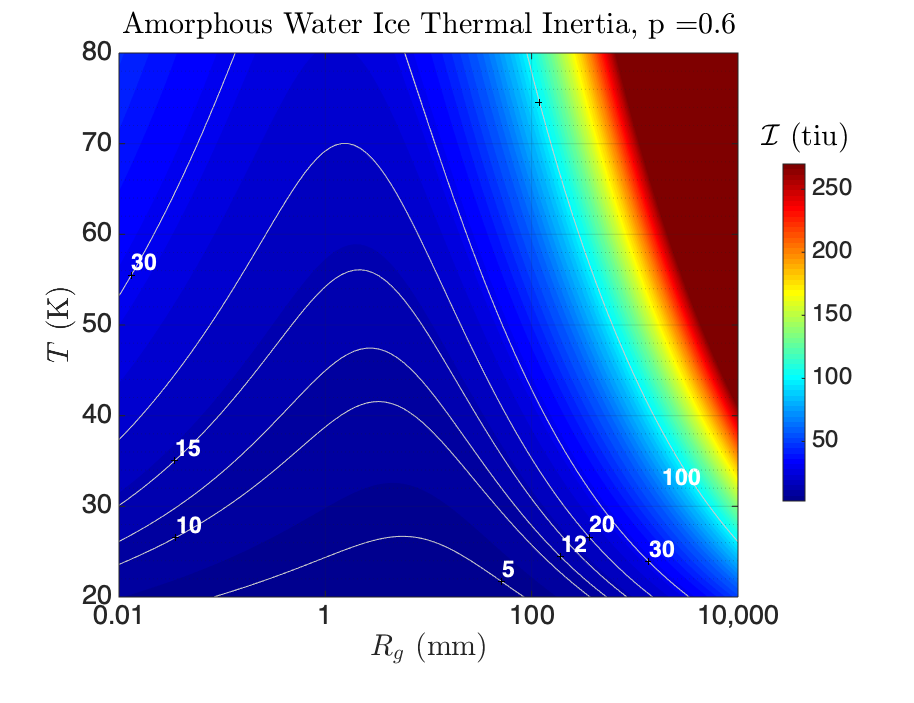}
\includegraphics[width=9.cm]{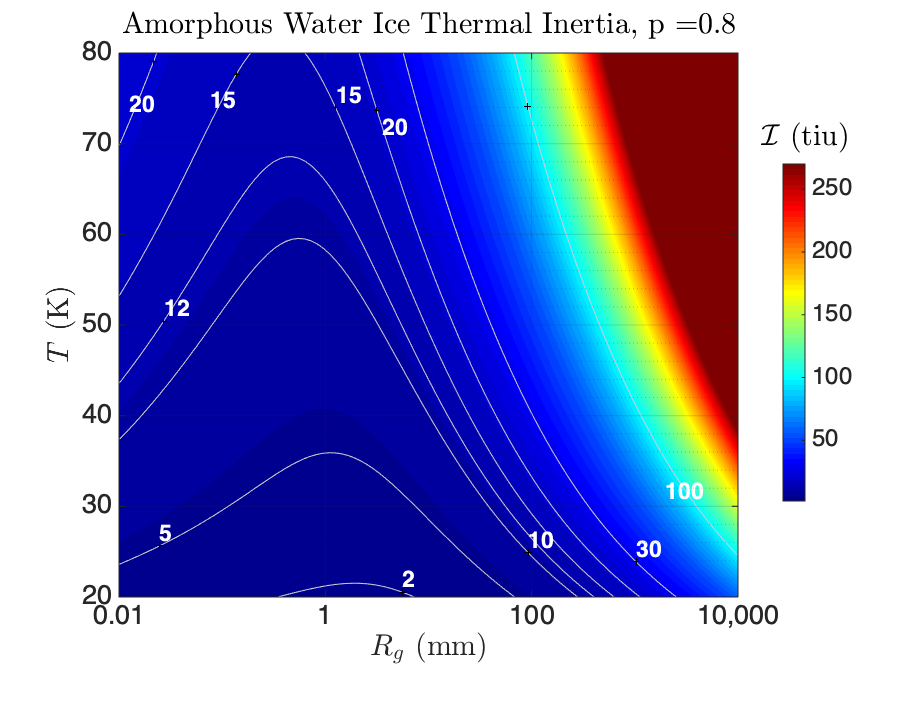}
\par
\end{center}
\caption{Predicted thermal inertia of amorphous ice as a function of grain size and temperature for two porosities: (top) $p=0.6$, (bottom) $p=0.8$ and $\gamma = 0.37$J/m$^2$ and $\epsilon = \epsilon_{{\rm IR}} = 0.9$.  White contours highlight key ${\cal I}$ values for reference }
\label{Thermal_Inertia_Predict}
\end{figure}

Fig. \ref{Thermal_Inertia_Predict} shows theoretical predictions of ${\cal I}$ for several values of porosity in which $\epsilon = \epsilon_{{\rm IR}} = 0.9$ with additional parameter inputs found in Table \ref{thermophysical_inputs_amorphous_ice}.  For putative grain sizes $0.1$mm$<R_g<1$mm and temperature range $20$K $<T<50$K, the thermal inertia of amorphous \water ice with $p=0.6$ broadly falls into the range $5$ tiu $ <{\cal I} <25$ tiu,
while for $p=0.8$ ${\cal I}$ it lies within $2$ tiu $  <{\cal I}<12$ tiu.
\par

\begin{figure}
\begin{center}
\leavevmode
\includegraphics[width=9.cm]{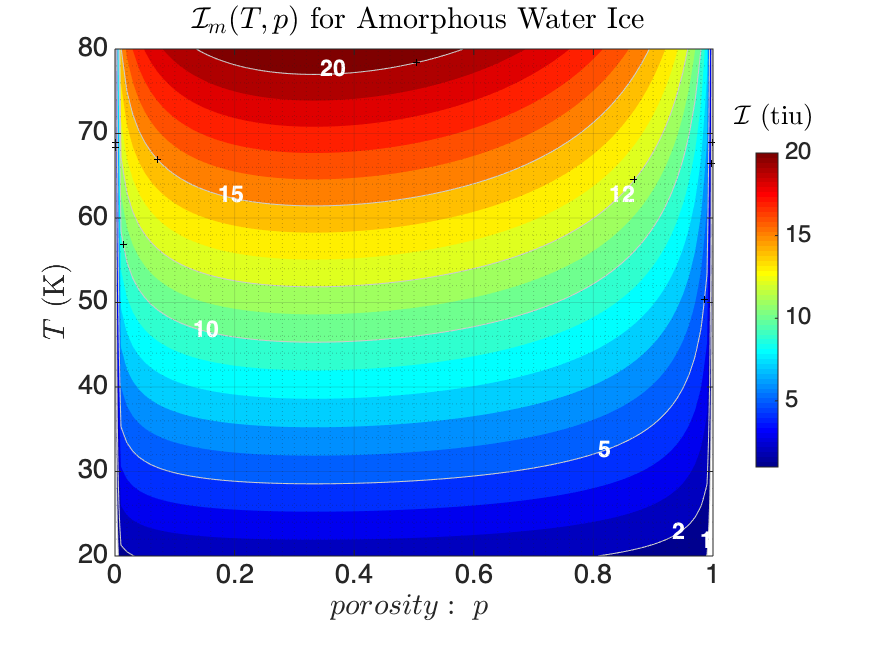}
\par
\end{center}
\caption{Minimum thermal inertia for amorphous \water ice as a function of $T$ and $p$ with $\epsilon = 0.9$.
Under given conditions this quantity implies that ${\cal I}\ge{\cal I}_m(T,p)$.  Note also this quantity is extremized at around $p\approx 0.32$.
}
\label{Extreme_Im}
\end{figure}

These figures also show that for a given $T$ and $p$ there are values of $R_g$ where
${\cal I}$ is minimized.   
This is easy to understand
as the radiative conductivity $K_{\rm r}$ has a linear dependence on $R_g$ while the contact limited ice conductivity $K_{\rm C}$ diminishes as $R_g^{-1/3}$:
the former becomes more efficient as radiation travels across longer distances in the matrix as $R_g$ increases while for fixed $L_{{\rm ec}}$ the latter becomes more effective because smaller grain sizes means that grain-grain contact area approaches the surface area of the grains themselves (i.e., $h\rightarrow 1$).
This means that for given inputs ($T,p,$ etc.) that ${\cal I} \ge {\cal I}_m$, where
\beq
{\cal I}_m = 1.325\left[
(1-p)C_p\rho_{{\rm ice}}
\Big(8\epsilon\sigma T^3K_{{\rm A}}^3L_{{\rm ec}} \phi_{{\rm r}} \phi_{{\rm c}}^3\Big)^{1/4}
\right]^{1/2},
\eeq
at 
 corresponding grain size $R_g=R_m$
 \beq
 R_{{\rm m}} = 
 \left(
 \frac{L_{{\rm ec}}^{1/3} \phi_{{\rm c}} K_{{\rm A}}}
 {24 \epsilon \sigma \phi_{{\rm r}} T^3}
 \right)^{3/4}.
 \label{Rm}
 \eeq
 The numerical factor on ${\cal I}_m$ comes from the minimization procedure and is an approximation to 
 $\sqrt{3^{1/4} + 3^{-3/4}}$. 
 As can be gleaned from Fig. \ref{Thermal_Inertia_Predict} we note that these extreme values occur for grain sizes of interest based on origins considerations (sub-to-few mm size) within the range of plausible ice porosities relevant for Arrkoth ($0.6<p<0.8$).
 Fig. \ref{Extreme_Im} shows ${\cal I}_m(T,p)$, and we observe that the 
 for the relevant temperature and porosity ranges, 
 ${\cal I}\le {\cal I}_m \approx $1-5 tiu are not possible, thus forming a reasonable lower bound
 for ${\cal I}$.  We also observe from the form of $R_{{\rm m}}$ in Eq. (\ref{Rm}), as well as from scanning
 the solutions shown in Fig. \ref{Thermal_Inertia_Predict}, that ${\cal I}_m$
 achieves its minimum in the 20-50 K temperature range for grain sizes between
 $0.7$mm$<R_g<$ 5mm.

\begin{table}
\label{thermophysical_inputs_amorphous_ice}
\caption{Various thermophysical and related quantities for amorphous \water ice.}
\centering
\begin{tabular}{l c l}
\hline
Quantity & Value & Note \\
\hline
$E$ & $\sim 2.4\times 10^{10}$ Pa &  near $T=50$K \\
$\nu_p$ & $0.33$ &  \\
$\gamma$ & $0.07$-$0.37$ J/m$^2$ &  range of measurments \\
$\rho_{{\rm ice}}$ & $940$ kg/m$^{3}$ &  \\
$v_{{\rm ph}}$ & $2500$ m/s &  \\
$\lambda_{{mol}}$ & $5\times 10^{-10}$ m &  \\
\hline
\end{tabular}
\par  
For source references see \citet{Ferrari_Lucas_2016}.
\end{table}

\subsubsection{A speculative synthesis}
Despite the analysis of the previous sections, we can at best offer likely possible candidates for Arrokoth's near surface materials that are consistent
with the $T_{\rm b,obs} = 29 \pm 5$K measurement made by the REX instrument.
This speculation is proffered on the basis of certain uncontroversially sober assumptions that: (1) Arrokoth's surface materials and/or layers are uncompacted
retaining a porosity in the range $0.6<p<0.8$ consistent with Arrokoth's mean density $\sim 250-350$ kg/m$^3$ -- on the assumption that its constituent grains are typical of the mean densities of relaxed KBOs 
\citep[$\sim 1600$ kg/m$^3$, e.g., ][]{Castillo-Rogez_etal_2012,Bierson_Nimmo_2019}; (2)
As being 
a member of the CCKBOs, Arrokoth's surface temperature probably never exceeded 70 K at any time during its natural history owing to its stable location in the Kuiper Belt; (3)
the X-band value of the imaginary part of the dielectric permittivity of Arrokoth's subsurface materials
is no less than that of unpolluted \water ice, i.e., $\varepsilon''>10^{-4}$.\par
By analogy to Phoebe and Iapetus' leading side's dark materials -- mainly owing to their shared low albedos -- we favor the assumption that Arrokoth's X-band emissivity is in the range
$\Eeff = 0.9-1$.  We rule out $\Eeff$ values of 0.7 on the basis that it would correspond to ice with $\varepsilon''$ well below $10^{-4}$.  
We deem $\Eeff = 0.8$ unlikely for similar reasons on the low ${\cal I}$ end ($< 20$ tiu).   $\Eeff = 0.8$ seems unlikely on the high ${\cal I}$ end ($> 80$ tiu) despite its being a possible solution (e.g., see Fig. \ref{Tb_Solution_Array}) as it would correspond to ${\cal I}$ for either very low or very high porosity \water ice (both amorphous or cubic).  According to Fig. \ref{Thermal_Inertia_Predict} the high porosity ice grain solution that predicts ${\cal I} > 50$ tiu would correspond to boulder sized solid ice grains, which is difficult to produce if the grains are formed during the era of planetesimal formation \citep[e.g.,][]{Estrada_etal_2016}.
The low porosity ice grain solution that yields such high thermal inertias
would correspond to sub-micron sized amorphous \water ice grains that, while possible, is hard to reconcile with the meteoritic record and planetesimal formation models. 
\par
If Arrokoth's subsurface materials are composed mainly of mm-scale amorphous \water ice grains, then its thermal inertia is likely below 10-20 tiu since higher thermal inertias are not reasonable to expect for amorphous \water ice at the porosities of interest here.  Of course, thermal inertias in the range 50-100 tiu are possible, but only if the \water ice is in cubic form, which is a possibility only if the near surface materials have experienced sustained periods with temperatures higher than 70-80 K \citep[see extensive discussion in][]{Ferrari_Lucas_2016}.
As mentioned above, high values of ${\cal I}$ are possible
for amorphous ice if the grains are like boulders in size where thermal transport is governed by radiation, but such large constituent surface material is hard to justify from an origins perspective.
The same theoretical concerns suggest that Arrokoth's thermal inertias are no less than $2-5$ tiu, thus forming a reasonable lower bound for ${\cal I}$ as lower values are not predicted for amorphous \water ice with porosities $0.6<p<0.8$.
\par
Taken together we see from inspecting the low $\varepsilon'$ column of Fig. \ref{Tb_Solution_Array} corresponding to its first two rows ($\Eeff = 0.9,1.0$) that for $1\ {\rm tiu} < {\cal I} < 10$ tiu, the $T_{\rm b,obs}
= 29 \pm 5$K prediction occurs in the range 
$10^{-4}<\varepsilon' \lessapprox  10^{-2}$, which is consistent with permittivity values for \water ice in the GHz range as briefly surveyed in section \ref{varepsilon_discussion} and discussed 
further in Bird et al. (2022).  Amorphous \water ice of mm-scale is therefore a strong candidate material to account for the Arrokoth's subsurface materials.
\par
However, the same solution survey cannot rule out tholins -- hydrocarbons produced from plasma discharging in mixtures of
methane and nitrogen -- as candidate materials as they are expected to be ubiquitous reddening agents of the outer solar system \citep{Cruikshank_etal_2005a}. 
Consideration of laboratory data of tholins summarized in section \ref{varepsilon_discussion} as well as in Bird et al. (2022) indicates that high porosity tholins exhibit
$10^{-4}<\varepsilon' \lessapprox  10^{-3}$, which is also consistent with $T_{{\rm b,obs}}$ for both $\Eeff = 0.9,1.0$.  Unfortunately, thermal inertias for tholins are not well constrained and while the $T_{{\rm b,obs}}$
measurement is consistent with the permitted range of thermal inertias for amorphous \water ice (i.e., 1-20 tiu) for $\Eeff = 1.0$,  the full range of thermal inertias are permitted for $T_{{\rm b,obs}}$ in the case where $\Eeff = 0.9$ (second row of Fig. \ref{Tb_Solution_Array}).  
We suspect that the surface ice is likely an admixture of amorphous \water ice and tholins, with the latter forming a small percentage of the composite, much in line with suggestions made for the make-up of the Kronian satellites \citep{Cruikshank_etal_2005b}.  If the dominant constitutent is the \water component, then we suspect the lower range of permissible thermal inertias characterizes the $\Eeff = 0.9$ case too.  
\par We observe that in the recent study
by \citet{Ferrari_etal_2021}, where they examined the thermal response of the Kronian moons -- but not including Iapetus and Phoebe -- that a nominal upper bound conductivity of $K_{e0}\sim 0.001$W/m/K was adopted for their model tholin-covered amorphous \water ice grains.  If such a tholin-\water ice complex were present on Arrokoth, it would correspond to an upper bound value
$\sqrt{(1-p)\rho_{{\rm ice}}K_{e0}} \approx 11$ tiu.

\subsection{On the comparison between simple theory and full results}
\begin{figure}
\begin{center}
\leavevmode
\includegraphics[width=9.5cm]{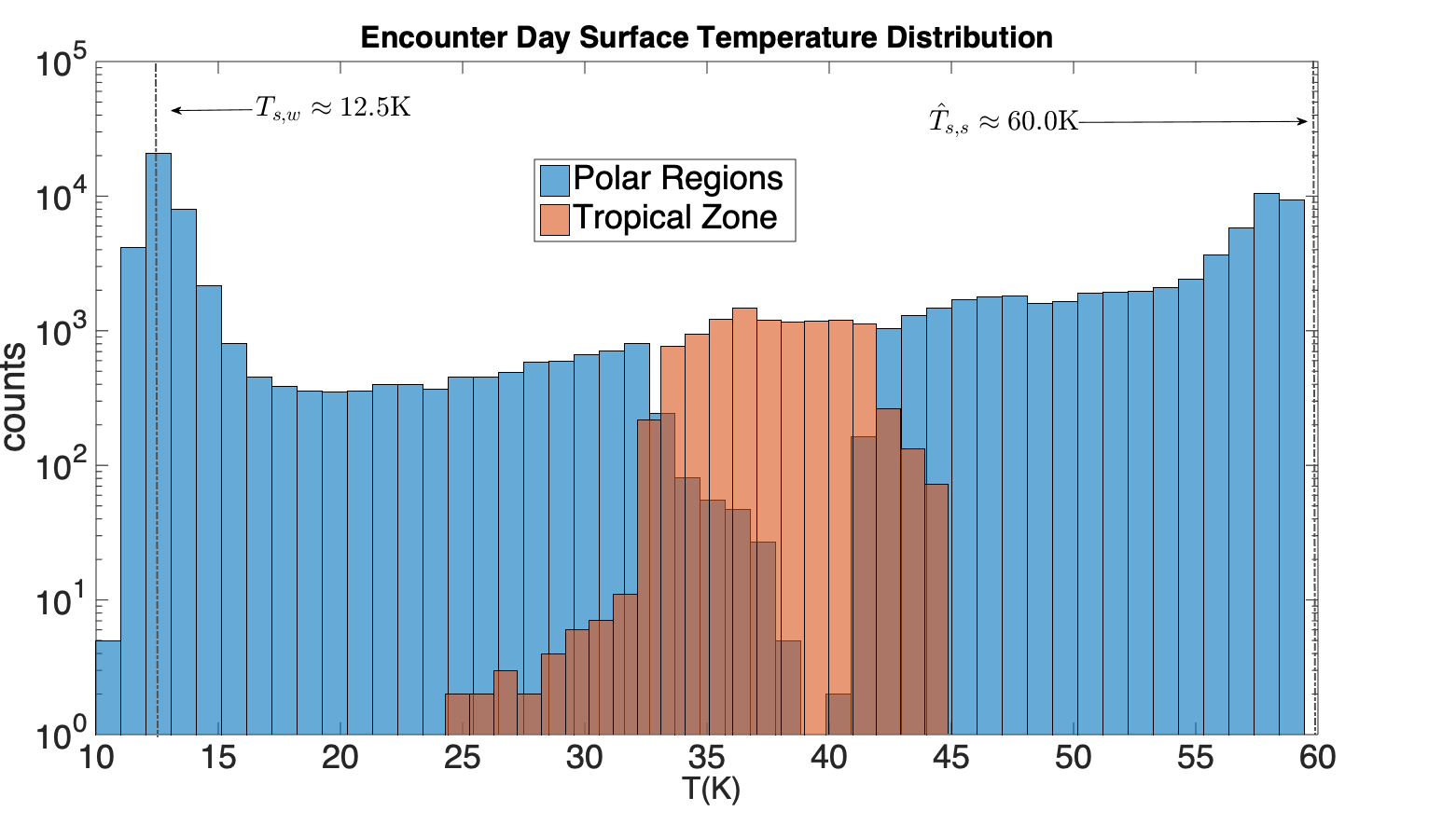}
\par
\end{center}
\caption{A more detailed view of encounter day surface temperatures.  \REV{Comparisons between simple theory predictions
are shown: vertical line at $T\sim 60$K corresponds to the ``summer" side predictions of the simple theory while the vertical line at
$T\sim 12.5$K corresponds to the ``winter" side predictions of the same.}}
\label{detailed_surface_temperature_histogram_analysis}
\end{figure}

In Section 2 we constructed a simplified algebraic theory for the predicted temperatures on the winter and summer sides. It is of interest to see how well this theory works against our predictions based on the more detailed sophisticated model we have developed here.
Examining Fig. \ref{Orbital_Elements} shows that at encounter day Arrokoth was in its late southern hemisphere spring phase of its orbit.  Given that it was near the southern summer solstice, it is useful to compare the distribution of the predicted encounter day surface temperatures against the simplified theory of Sec. 2.
In Fig. \ref{detailed_surface_temperature_histogram_analysis} we re-plot the surface temperature distribution shown
 in the first panel of  Fig. \ref{encounter_day_histograms}, where we have distinguished between surface facets in the polar and tropical zones.  Surface facets are designated as belonging to the polar zone if there exists at least one Arrokoth day in which the sun never rises, while tropical zones are those facets in which the sun rises daily.  The distribution shows two modes.  The low end mode corresponds to approximately 12.5K, which is the predicted value for $T\sub{s,w}\approx 12.2$K based on the
 results shown in Fig. \ref{simple_theory} for ${\cal I} \approx 2.54$ tiu \REV{(the low end of likely ${\cal I}$ values as discussed in the previous section)}.  The high temperature mode shows that the peak temperature sits at around 58.5K.  The simple theory predicts a typical summer time temperature
 of $T\sub{s,s} \approx 60$K, which is only slightly higher than high end mode seen in the distribution, and where this discrepancy is likely due to comparing the simple theory against a solution for a time slightly shy of solstice
 (cf., Fig. \ref{Orbital_Elements}).  
 We conclude that the simple theory 
 of Sec. 2 is overall useful especially with respect to predicting the temperatures on both the summer 
 and winter sides near around solstice. 
 
\subsection{On the value the time-asymptotic solution method}
The traditional approach to solving Eqs. (\ref{full_eqn_1}-\ref{full_eqn_3}), while
similarly involves precomputing the shape model analysis and insolation profile, calls for
developing an initial value solution to Eq. (\ref{full_eqn_1}). Accurately forward time evolving the heat equation can be treacherous and often time-consuming, especially if the parameters of the problem make the system numerically stiff (requiring short time-stepping), although there are many recent methods developed in the astrophysical literature that alleviate this computational overhead while preserving accuracy and stability in various problems involving diffusion operators
\citep[e.g.,][]{Meyer_etal_2012}.  \par
The main issue, however, is developing a solution to a periodically driven problem that is ``time-asymptotic", in the sense that the long-time state of the system -- certainly after tens of millions of orbits since formation \citep{McKinnon_etal_2020} -- is also periodic with the external driver's periodicity (if, however, out of phase, etc.).  Arrokoth is likely in such a time-asymptotic state given that its orbital elements are not thought to have experienced strong chaotic or aperiodic episodes, at least not in recent history \citep{Porter_etal_2018}.  Thus, the problem we have here lies in initiating the initial value problem with the right initial data that sets the solution off into an  exactly periodic transient-free state.  But this is something which, in general, is not known {\emph{a priori}}.  In practice, such initial value problems are setup up with some initial profile (satisfying boundary conditions), and subsequently forward evolved until the time-periodic asymptotic state is eventually reached. But this may take time to achieve (sometimes dozens of forcing periods), which can be expensive.  The Fourier transform method utilized here, as well as
in \citet{Titus_Cushing_2012,White_etal_2016}, while slightly costly due to the fast Fourier transform itself, immediately admits an exactly periodic solution, with no transient structure.  There is no forward time stepping involved and the resulting solutions are accurate to spectral precision, which makes this solution method attractive from the standpoint of generating reliable solutions.

\subsection{On thermally driven mechanical erosion, a speculation}
 \REV{Beyond circumstantial evidence for methanol based on spectral fitting to LEISA data obtained by New Horizons \citep{Stern_etal_2019,Grundy_etal_2020},
what are Arrokoth's near surface materials and what are their thermo-mechanical properties remains unknown to date.}
Based on surveys of other CCKBOs 
\citep{Lellouch_etal_2013}
only bulk thermal inertias probing diurnal skin depths are reasonably known.  For the solutions displayed in Section we have adopted a representative value of ${\cal I}\approx 2.5$ tiu, but we keep in mind that these values may be larger based on the discussion found in section \ref{transfer_discussion}.  \par
Nevertheless, 
these thermal solutions with relatively low values of ${\cal I}$ might suggest something about Arrokoth's erosive character.  Erosion driven by thermal cycling -- wherein cracks nucleate and grow due to a material's non-zero coefficient of thermal expansion -- might be a relevant effect on Arrokoth over the course of its natural history. Such a process has been hypothesized in order to explain the nature and origin of erosion on for cometary bodies
\citep[e.g., for 67P/Churyumov-Gerasimenko,][]{Attree_etal_2018}.  
\par
We focus here on thermomechanical erosion and do not consider thermal erosion driven by sublimation since volatile ices are neither known to be present on Arrokoth's surface nor would be expected to be extant there (see previous section).
Among other things, however, estimating actual rates of erosion requires knowing what is the local temperature temporal profile, the materials coefficient of thermal expansion, its effective porosity, Young's modulus, Poisson's ratio, and if the thermal cycling is on a long enough timescale, its
viscous relaxation properties like the materials activation energy \citep[][see latter reference for a recent comprehensive discussion]{Mellon_1997,Molaro_etal_2015}. Nevertheless, we speculate upon the relative amount of thermally driven mechanical erosion {\emph{over the course of one orbital time}} assuming the near surface of the body is uniformly composed.  Fig. \ref{Min_Temperature} indicates that the highest minimum temperature ($\sim 20-30$K) over the course of one orbit occurs along a narrow line of surface located within the tropical zone and, moreover, appears to trace along the rims of putative craters or depressions found in that region.  
The implied extreme temperature variation is shown in Fig. \ref{MinMax_Temperature}, which shows for each facet the difference of the maximum and minimum temperatures achieved over the course of one orbit, i.e.,
\beq
\Delta T\sub i \equiv {\rm max} (T\sub i) - {\rm min} (T\sub i).
\eeq
The figure indicates that the bulk of the polar regions experiences $\Delta T\sub i$ close to 50 K, while the tropical zones experience $\Delta T\sub i$ closer to 30K.  The aforementioned narrow line within the tropical zone has a temperature variation of about only 10K. With all other variables held fixed, it stands to reason that the thermal erosion is 
greatest across the polar regions where $\Delta T\sub i$ is large, and significantly weakens upon entry into the tropical zone
and even further weakens as one approaches the
narrow line of real estate within in which $\Delta T\sub i \approx 15-20$K.  While this relative prediction is reasonable, the actual amount of erosion that has taken place over Arrokoth's natural history cannot be ascertained until the actual composition of its surface materials are unambiguously determined, together with information about their thermal and viscoelastic properties, with particular attention to its rheology.
\par
As a final remark, we note that the thermal driving is likely most extreme on Arrokoth's equatorial latitudes during its equinoxes since these regions are getting extreme illumination variations on the the short 15.9 hour diurnal period $P\sub{{\rm day}}$ ($\omega\sub{{\rm day}} = 2\pi/P\sub{{\rm day}}$).  
But the thermomechanically driven erosion taking place on this short timescale is also restricted to the upper 1-2mm of surface since the thermal wave lengthscale $\ell\sub{{\rm day}}$ is shortened by comparison to $\ell\sub{{\rm orb}}$ ($\approx 1$m) by a
factor of $\sqrt{\omega\sub{{\rm orb}}/\omega\sub{{\rm day}}} \approx 1/400$.
\par

\subsection{Other Various Caveats}
Despite the intrinsic scientific value of these solutions, several aspects of the thermal model method utilized here could be improvement.  We acknowledge a few here with the aims of developing even more refined solutions in the future.  We have assumed the thermal inertia (and by implication, the thermal conductivity) is a constant independent of temperature and porosity.  Indeed the time-asymptotic method we
have employed relies on this feature, which permits using fast Fourier transform techniques to develop solutions
\citep{Titus_Cushing_2012,Schloerb_etal_2015, White_etal_2016}.  \REV{However, although not done here, a temperature dependence of the thermal inertia may be formally built into the method at the cost of added iterations to the solution process -- a matter we will develop further with future follow-up work.}  We have also assumed constant stratigraphy and/or subsurface porosity, which could also be unrealistic.  But similar to our point above this can be similarly remedied by iteration or by directly building stratigraphy into the solution itself.  We demonstrate how this is done in Appendix C where we develop a subsurface thermal solution for a material composed of two different conductivities where the transition occurs $h$ meters below the surface.  The solution may be developed analytically, as was done for the discussion in the text, and coded directly into the thermal solver.  This procedure naturally generalizes to multiple layers with differing properties. 
\REV{
\subsubsection{On the use of a single albedo value}
New Horizons imaging showed the existence of several bright rings and patches painting its surface \citep[e.g., see the bright patches in Fig. 4, also see][]{Spencer_etal_2020}, with normal reflectances of these bright patches being about twice that of the rest of the terrain.  
Given that these
images were taken at high solar phase angle, the observed relatively enhanced brightness in these regions is likely ``the result of the extreme illumination conditions rather than the intrinsic albedo of the surface," \citep[][]{Stern_etal_2021}.
In light of this and in addition to determining Arrokoth's mean hemispherical albedo, \citet{Hofgartner_etal_2021} have produced an albedo map of the closest approach image (see their Fig. 4) in which the bright patches are assessed to have albedos as high as 0.09.
Provided all other thermophysical properties are the same across bright and dark patches, we find that the temperatures in these bright regions are overestimated by less than 0.5K: if $T_{{\rm dark}}$ is the estimated temperature of the bright terrain assuming the lower albedo $(A_{{\rm dark}} = 0.063$) and $T_{{\rm bright}} \equiv T_{{\rm dark}} - \delta T$ would be its true correct value based on $(A_{{\rm bright}} = 0.09)$, where $\delta T$ is the correction factor, then
\beq
\left(\frac{T_{{\rm dark}} - \delta T}{T_{{\rm dark}}}
\right)^4 = \frac{1-A_{{\rm dark}}}{1-A_{{\rm bright}}},
\eeq
predicting $\delta T\big/T_{{\rm dark}} \approx 0.007$, which for 
$T_{{\rm dark}} \approx 60$K amounts to $\delta T = 0.42$K.}
\REV{
\subsubsection{On the neglect of reflected light}
In our calculations we have included thermal IR in the
re-radiation calculation and neglected reflected light.  This neglect has minor consequences on the final determined temperatures following the same line of thinking as above, where we find the error is even weaker.  If we treat the reflected light as a Lambertian process, then the light received at a given facet will be traced by the IR re-radiated light from all other facets.  We showed earlier that for a given facet $i$ the energy received from all other facets ($f_{i,{\rm rr}}$) amounts to about 10 percent of the locally absorbed solar insolation ($f\sub{i,\odot}$), i.e., $f_{i,{\rm rr}} \approx \chi f\sub{i,\odot}$, where $\chi \le 0.1$. We can estimated the amount of reflected light neglected in this accounting to be about $Af_{i,{\rm rr}}$, such that the correct amount of light received from the surrounding landscape is better estimated by $\tilde f_{i,{\rm rr}} \approx f_{i,{\rm rr}}\big/(1-A)$.
Assuming that all other thermophysical properties remain the same, the correct temperature $T_{i,{\rm cor}} = T_i + \delta T$ compares to the estimated temperature $T_i$ via the same Stefan-Boltzmann law argument applied in the previous section with
\beq
\left(\frac{T_i + \delta T_i}{T_i}
\right)^4 = \frac{f_{i,\odot} + f_{i,{\rm rr}}}{f_{i,\odot} + \tilde f_{i,{\rm rr}}} = \displaystyle \frac{1+\chi}{\displaystyle{1+\frac{\chi}{1-A}}},
\eeq
which, given our estimate for $\chi$, means $\delta T_i/T_i \approx 0.0016$, which for $T_i \approx 60K$, implies that neglecting reflected light on Arrokoth amounts to underestimating the local temperature by $\delta T_i \approx 0.1$K.  In summary it is conceivable that Arrokoth is close to a blackbody, i.e., one that is marginally reflective and radiating most of its heat in IR.}

\subsubsection{On the expiration of surface N$_2$ ice on Arrokoth}
\REV{We can ask about the fate of volatile ices like N$_2$ and CO (with similar properties) on an airless body like Arrokoth following some of the thinking
used to address similar concerns about the nature of Oumuamua \citep{Jackson_Desch_2021}.
With typical solar insolation values based on Arrokoth's semimajor axis, i.e., $\tilde f_\odot \approx $ 0.7 W/m$^2$, a vacuum exposed surface layer of N$_2$ ice on Arrokoth would sublimate at a rate of 3.5m/m$^2$/orbit.  This conservative figure derives from assuming an N$_2$ ice albedo of $A=$0.9 \citep[i.e., based on Pluto observations,][]{Grundy_etal_2016,Schmitt_etal_2017}
and an enthalpy of N$_2$ sublimation ${\cal L} =225$kJ/kg that, if we assume a surface layer of N$_2$ ice converts all of its received insolation into sublimation (and adjusting its temperature accordingly), would result in a mass flux $\dot\Sigma = (1-A)\tilde f_\odot/{\cal L}
=3.4\times10^{-7}$kg/m$^2$/s $\approx 3.24\times 10^3$kg/m$^2$/Arrokoth orbit. Under these vacuum conditions a sublimating block of ice must satisfy the relationship between mass-flux into a vacuum and a material's vapor pressure,$P_{{\rm vap}}(T)$, i.e.,
\beq
\dot \Sigma v_k= P_{{\rm vap}}(T), 
\eeq 
\citep[e.g.,][]{Lebofsky_1975}, where $v_k$ is the typical Boltzmann velocity.  An estimate based on compiled vapor pressure data for N$_2$ \citep[][]{Fray_Schmitt_2009} shows that this balanced state is satisfied for $T\approx 27$K.  Based on the shape model utilized here Arrokoth's total surface $A_{{\rm surf}} \approx 1406$km$^2$. Thus, if Arrokoth were uniformly covered with N$_2$ ice, then it would be losing a total of about 481 kg of N$_2$ per second $\leftrightarrow 1.6\times 10^{28}$ particles per second, well above the mass loss rate upper limits based on New Horizons' observation of the body \citep[$<10^{24}$particles/s,][]{Stern_etal_2019}.  These considerations preclude the plausibility that N$_2$ and CO can remain on the surface of Arrokoth for any appreciable amount of time.  CH$_4$, which is relatively less volatile by a factor of $10^3$ (based on its vapor pressure behavior at similar temperatures) would also result in particle flux rates exceeding those observed by New Horizons.   This analysis says nothing about the plausibility for reservoirs of volatile ices deep within Arrokoth.  Addressing this matter requires further analysis.}

\section{Summary}
We have developed a temperature model of Arrokoth based on the $10^5$ facet model published
in \citet{Spencer_etal_2020}.  For the bulk of the solutions displayed here, especially in Section 4, we have assumed a single value of ${\cal I}\approx 2.5$tiu, together with
a \REV{infrared emissivity} of $0.9$ and an albedo of 0.06.  
\REV{The choice of ${\cal I}\approx 2.5$ is consistent with the low end of thermal inertia values for amorphouse \water ice as detailed at length in section 6.1.}
Although direct temperature measurements were not possible with New Horizons, for the day of the closest encounter we predict that the day side approach hemisphere surface temperatures were in the range of 57-60K. \REV{
According to the simple model of section \ref{sec:simple_model}
(e.g., Fig. \ref{simple_theory})
the predicted daytime (``summer time") surface temperature has a very weak dependence
in the thermal inertia range of 1 tiu $<{\cal I}<$ 10 tiu.
We thus expect this quoted daytime temperature range to hold for the full range of likely  ${\cal I}$ values.
On the other hand, 
for the specific ${\cal I} = 2.5$ tiu value we predict that the obscured winter side of Arrokoth had surface temperatures in the range of 10-15K, but following the same above reasoning, based on the results shown in Fig. \ref{simple_theory}
together with the likely range for ${\cal I}$, we predict
that the night side surface temperature could lie within the slightly wider range of values of
10-20K.}
\par 
Arrokoth's tropical zones receive the lowest yearly averaged insolation but exhibits the highest orbitally averaged temperatures compared to Arrokoth's more extreme polar regions. In fact, the extreme temperature variations experienced by facets in Arrokoth's polar zones ($\Delta T \approx 50$K)
greatly exceed the corresponding variations in its equatorial zones ($\Delta T \approx 10-30$K).
 The property of our thermal model leads us to conjecture that the amount of thermal erosion in the tropical zones should be significantly lower than the erosion suffered in the polar regions.  
 \par
 We also find that the amount of energy received 
on a typical facet that is sourced from surface reradiation typically amounts to about 5\% of the total \REV{irradiation}, although it should be kept in mind -- as a cross comparative inspection of the histograms of Fig. \ref{encounter_day_histograms} shows -- there is a spread in this figure from facet to facet, which ultimately depends upon how much of a given facet is obscured by the rest of the body.  A small adopted value of ${\cal I}$ means that Arrokoth's Spencer Number is also small ($\approx 0.01$). 
\REV{In other words, this highly 
insulating body has very little downward directed thermal conduction from radiation received during total direct illumination (and subsequently returning it at night), amounting to about 0.5\% of the typically received insolation flux budget.}
\par
\REV{Based on a generalized application of our thermal modeling we find that the brightness temperature
$T_{{\rm b,obs}}=29\pm5$K measured by New Horizons' REX instrument
\citep[][also see Bird et al., 2022]{Grundy_etal_2020} is consistent with porous $(0.6 < p < 0.8$) subsurface amorphous \water ice -- possibly coated with tholins -- within the thermal inertia range $1\ {\rm tiu} < {\cal I} < $10-20 tiu
for given X-band emissivity $\Eeff = 1.0$. 
For X-band emissivity $\Eeff = 0.9$, which is similar to the low albedo (and possibly captured) KBO Phoebe,
our thermal modeling predicts a wider range of permissible thermal inertias, with 
$1\ {\rm tiu} < {\cal I} < $ 300 tiu.  However the high-end of the
${\cal I}$ ($>50$ tiu) values are disfavored for two reasons: (i) it would imply that either very large (1-10m) or very small (sub-micron) amorphous \water ice grains are dominant constituents of Arrokoth's near surface -- sizes which are hard to reconcile from an origins planetesimal formation perspective -- and, (ii) it might imply that the near surface \water ice grains are in cubic crystalline form, but this would mean that Arrokoth's surface experienced sustained periods of relatively elevated temperatures exceeding 70K (corresponding to the amorphous-cubic transition occurring on times shorter than Arrokoth's natural age).  Such high thermal inertias might be characteristic of low-porosity tholins, but that is currently unknown and remains to be more completely studied under
relevant laboratory conditions.\par
As such we favor the interpretation that
the REX $T_{{\rm b,obs}}$ measurement indicates that Arrokoth harbours porous hydrocarbon-coated amorphous \water ice grains of rough size $\sim 0.1-1$mm, with $1\ {\rm tiu} < {\cal I} < $10-20 tiu, and characterized by an X-band emissivity
in the range 0.9 and 1.}

\begin{acknowledgements}{{\bf Acknowledgements.}}
O.M.U. and the rest of the New Horizons team acknowledge support from the New Horizons Kuiper Belt Extended Mission grant for support in producing this work.
J.T.K. acknowledges that portion of the research was carried out at the Jet Propulsion Laboratory, California Institute of Technology, under a contract with the National Aeronautics and Space Administration (80NM0018D0004). J.T.K. also acknowledges supported by the California Institute of Technology Joint Center for Planetary Astronomy postdoctoral fellowship.  O.M.U. acknowledges fruitful conversations with A. Jindal and A. Hayes (Cornell University).  We are indebted to insights shared with us by B. Butler (NRAO) and M.A. Gurwell (CFA).
The authors acknowledge and are grateful for the helpful comments made by the two reviewers of the manuscript.
\end{acknowledgements}

\appendix

\section{Details of Shape Analysis of Section 3.2}\label{details_of_shape_analysis}
\begin{figure*}
\begin{center}
\leavevmode
\includegraphics[width=16.cm]{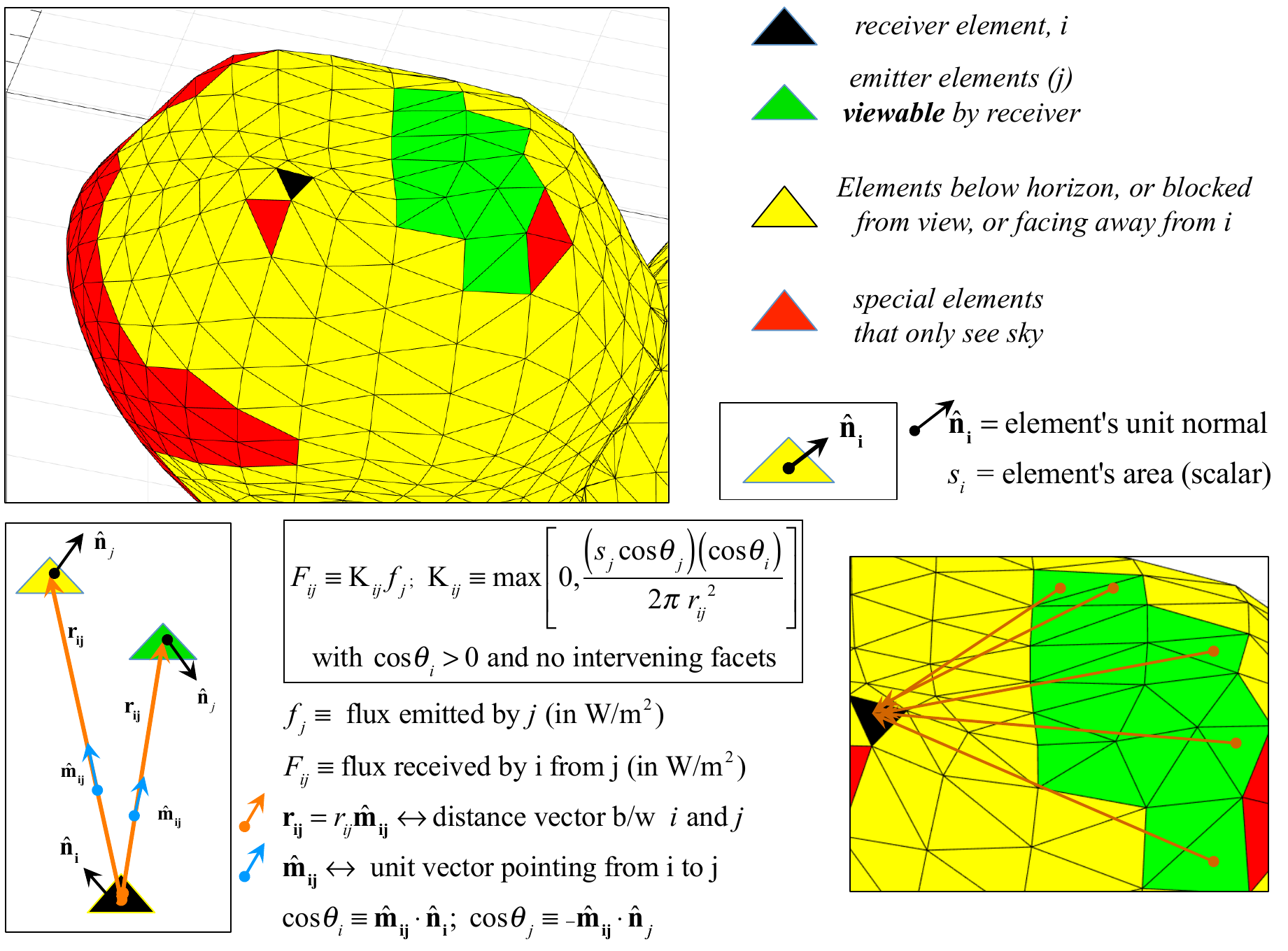}
\par
\end{center}
\caption{This diagram describes the shape analysis discussed in Section \ref{sec:shape_analysis}. The shape model
depicted here is the low order 2000 facet model analyzed in \citet{Grundy_etal_2020}.  However, the methods
described herein are applied to the $10^5$ facet model in \citet{Spencer_etal_2020}.}
\label{Kij_Diagram}
\end{figure*}

This section is concerned with determining \REV{how much a given piece of landscape is visible to the rest of the landscape
(a ``who-sees-who" list)}, and
is meant to address how the integral expression in Eq. (\ref{full_eqn_3}) is calculated.  The
considerations here reference Fig. \ref{Kij_Diagram}.  
Each facet $i$ has an outwardly pointing normal $\hat{\bf n}\sub i$ and associated surface area $s\sub i$
such that $d\hat{\bf s}_i =s\sub i \hat{\bf n}\sub i$.  Our goal is to determine which facets $j$ are visible
to facet $i$.  This involves ruling out all (1) facets $j$ whose unit normals $\hat{\bf n}\sub j$ face away
from facet $i$, (2) rejecting all facets $j$ that are below the facet $i$'s horizon, and finally
(3) discarding all facets $j$ that are blocked by intervening facets.  We treat each of these in order: 
\par We first
determine a separation vector ${\bf r}\sub {ij} = {\bf r}\sub j - {\bf r}\sub i$, and further
parse this in terms of a scalar and unit vector ${\bf r}\sub {ij} = r\sub{ij}\hat{\bf m}\sub {ij}$,
in which $r\sub{ij} = |{\bf r}\sub {ij}|$ and $\hat{\bf m}\sub {ij}$ is a unit vector pointing from the center
of facet $i$ to the center of facet $j$: (1) a facet $j$ is deemed below the horizon if $\cos\theta\sub i = 
\hat{\bf m}\sub {ij} \cdot \hat{\bf n}\sub i < 0$, (2) a facet $j$ is pointed away if 
$\hat{\bf m}\sub {ij} \cdot \hat{\bf n}\sub j > 0$, for which we define $\cos\theta\sub j \equiv -\hat{\bf m}\sub {ij} \cdot \hat{\bf n}\sub j$.  Therefore in order for a facet $j$ to be a candidate for visibility by facet $i$, both  $\cos\theta\sub i >0$
and $\cos\theta\sub j > 0$.  It then follows that all candidate facets $j$ passing the above two criteria are tested for actual visibility using a
ray-tracing routine.  The prototype code we have built is run on Matlab 2014a and we have therefore used a community wide
tested ray-tracing Matlab routine called \texttt{inpolyhedron}, which ascertains whether not elements of a given ray of points are inside a closed surface.\par
  For all the facets $j$ visible to facet $i$, we determine the local altitude of the highest visible facet
by identifying for which facet $j$ is $\cos\theta_i$ maximum, and we assign the value of this
special altitude $\beta\sub i = \pi/2 - \theta_i({\rm max})$.  This information is used in speeding up the solar insolation calculation of the next section.
In some notable instances a given facet $i$ sees no other facets of the body, in which case its $\beta\sub i = 0$.
\par
The luminosity apparent to facet $i$ of a {\emph{visible}} flat Lambertian facet $j$, with area $s\sub j$ and inclined at an angle
$\cos\theta_j$, emitting with a flux $f\sub j$ is $f\sub j s\sub j \cos\theta_j$.  Defining $F\sub{ij}$ as the flux absorbed by facet $i$ from facet $j$, it follows that
\beq
F\sub{ij} = f\sub j {\rm K}\sub{ij}, \qquad {\rm K}\sub{ij} = \frac{s\sub j \cos\theta_j \cos\theta_i}{2\pi r\sub{ij}^2},
\label{Fij_Facet_Formula}
\eeq
where the factor $2\pi$ accounts for the fact that emitted radiation spreads out only across 2$\pi$ steradians.  \REV{We note that the expression in Eq. (\ref{Fij_Facet_Formula}) likely badly quantifies the received irradiation from (say) neighboring facets especially if the angles formed between their planes of intersection are highly acute.  This problem would be especially acute if the quality of the local topography {\emph{on the scales of interest}} appeared
to be fractal.  This is not the case here.}
By identifying $f\sub j = \varepsilon \sigma T\sub j^4$,
then it also follows that the body integral term in Eq. (\ref{full_eqn_3}) becomes
\beq
\int_{\partial S} f\sub j S\sub{ij} d\hat{\bf s}_j
\rightarrow \sum_{\forall j \ {\rm visible}} f\sub j {\rm K}\sub{ij},
\eeq
where, by implication, $S\sub{ij} d\hat{\bf s}_j \longrightarrow {\rm K}\sub{ij}$.
In practice ${\rm K}\sub{ij}$ is constructed as a sparse matrix in which the column elements $j$ are non-zero
for those facets that are visible to $i$.  By its construction a given facet is not visible to itself, therefore
${\rm K}\sub{ii} = 0$.  Finally, we note that although this analysis is expensive ($\sim 5$ hours when run as a parallel computation on a 4-core 3.1 GHz Intel system), it is only done once for a given {\emph {static}} shape model.  All other calculations, including parameter studies and so forth that are done hereafter, utilize ${\rm K}\sub{ij}$.
\par
In Fig. \ref{sky_coverage} we show the results of the shape analysis based on the 10$^5$ faceted shape model
described in the previous section.  The figure depicts a facet-by-facet map of the percent of open sky self-obscured by Arrokoth.  The percent is based on the fraction of $2\pi$ steradians occupied by other surface facets.  We see how the routine identifies craters as places of relatively high obscuration ($\sim \% 30$).  The neck region on the well-imaged
approach trajectory (top panel) also shows similar degrees of self-obscuration.  However, the neck region on the
departure trajectory, which was not directly imaged and whose structure is inferred from model fitting procedures described in \citet{Spencer_etal_2020} and is probably less reliable as a result, is identified as having much greater obscuration in the vicinity of $\sim \% 60$ or so.  It is not a surprise given Arrokoth's flattened structure that the relative amount of self-obscuration is generally very weak (also see the distribution of these percentages in the first panel of Fig. \ref{orbital_timescale_histograms}).

\section{An example calculation using the Fourier method}\label{simple_demo}
We detail the Fourier method solution procedure on a relatively simplified problem.  We suppose we have a 1 dimensional
layer with the same setup as described in the text. We imagine that the surface ($z=0$) is illuminated by a 
 function $f$ with period $2\pi/\omega$.  The surface material's emissivity is 1.  The interior temperature satisfies the heat equation
 \beq
 \rho C_p \partial_t T = K\partial_z^2 T,
 \eeq
The boundary conditions at the top and bottom (respectively) are
\beq
\sigma T^4\Big|_{z=0} + K\partial\sub z T\big|_{z=0} = f, \qquad \partial_z T\big|_{z\rightarrow \infty} = 0.
\eeq
We assume all other coefficients in the above equations are constant.  We represent the 
interior solution by a Fourier series
\beq
T = T\sub 0 + \sum_{n=1}^{\infty}\Theta\sub n e^{k_n z - i\omega n t} + {\rm c.c.}.,
\qquad k\sub n = \sqrt{\rho C_p i \omega n \Big/K}.
\eeq
The above solution satisfies the heat equation for all points $z\le 0$, including the bottom boundary condition.  What remains to be determined are the coefficients $\Theta\sub n$, which can be solved for by satisfying the surface boundary condition.  We approach this in the following way.  The vertical thermal gradient at $z=0$ is given by
\beq
K\partial\sub z T\big|_{z=0} = 
K\sum_{n=1}^{\infty}k\sub n \Theta\sub n e^{- i\omega n t} + + {\rm c.c.}.
\eeq
We can decompose the insolation forcing also as an infinite sum of Fourier harmonics, that is to say
\beq
f = f\sub 0 + \sum_{n=1}^{\infty}f\sub n e^{-i\omega n t} + {\rm c.c.}.
\eeq
The top boundary condition therefore becomes
\beq
\sigma T^4\Big|_{z=0} -f_0 + 
\sum_{n=1}^{\infty}(K k\sub n \Theta\sub n-f\sub n) e^{-i\omega n t} + {\rm c.c.}. = 0,
\label{bc_complex}
\eeq
A complete solution is had once the countably infinite number of complex valued coefficients $\Theta\sub n$ are determined. 
In practice one assumes a maximum number $N$ of these coefficients for a sufficiently converged solution, which is assessed as such once successive changes in the final state meets some requisite set of minimum error criteria. Thus, the solution is determined for the set of $\Theta\sub n$ that solve Eq. (\ref{bc_complex}).  Identifying the function $F(T) \equiv \sigma T^4\Big|_{z=0}$, together with its Fourier decomposition as
\beq
F = F\sub 0 + \sum_{n=1}^{\infty}F\sub n e^{-i\omega n t} + {\rm c.c.}.
\eeq
we find that Eq. (\ref{bc_complex}) becomes after truncating up to $N$ sinusoids 
 \beq
 F\sub 0 -f_0 + 
\sum_{n=1}^{N}(F\sub n + K k\sub n \Theta\sub n-f\sub n) e^{-i\omega n t} + {\rm c.c.}. = 0.
 \eeq
 Taking the (fast) Fourier transform of the above yields a set of $N$ nonlinear algebraic equations that
 must be simultaneously solved, namely
 \beq
 F\sub n(\Theta\sub m) + K k\sub n \Theta\sub n-f\sub n = 0, \qquad k\sub 0 = 0,
 \label{equation_by_NRT}
 \eeq
 for each value of $n = 0,1,\cdots N$.
 The term $F\sub n = F\sub n(\Theta\sub m)$ is a nonlinear function of all the unknown coefficients
  $\Theta\sub m$. As such, the above set of N-coupled nonlinear equations must be solved using a Newton-Raphson (or gradient descent) type of nonlinear equation solver.  In practice, this is done efficiently by assuming an initial guess for 
  $\Theta\sub n$, then backward transforming $T(z=0)$ from frequency space into temporal space, followed by calculating $F(T) = \sigma T^4\big|_{z=0}$, then taking $F$ and forward (fast) Fourier transforming it into frequency space to determine each $F\sub n$, followed by assessing a solution to Eq. (\ref{equation_by_NRT}).  This usually involves an iterative procedure which requires
  re-executing the same set of forward-backward Fourier transform procedures to determine an improved solution, until
  error tolerances are met.  This solution procedure is easily generalizable to nonlinear values of the conductivity coefficient $K$, say, which are dependent on the values of $T$.

\section{Layered thermal solution}
We suppose that the surface is composed of two layers with a transition occurring at $z=-h$ (where $h>0$).  
For the sake of this illustration let us suppose that the only difference between the layers is the thermal conductivity.  In keeping with our convention in the text, we set the conductivity to be $K$ for $-h<z<0$, and $K^{(-)}$ for $z< -h$.  We therefore divide the solution to the heat equation Eq. (\ref{full_eqn_1}) into ones appropriate to these two levels:
\beq
\Theta_i = \left\{
\begin{array}{cc}
\sum_{n=0}^N \left(\Theta_{i,n} e^{k_n z} + \Xi_{i,n} e^{-k_n z}\right) e^{-i\omega n t} + {\rm c.c.}, & -h < z \le 0 \\
\sum_{n=0}^N \Theta\sub{i,n}^{(-)} e^{k_n^{(-)} (z+h)}  e^{-i\omega n t} + {\rm c.c.}, & z \le -h 
\end{array}
\right.
,
\label{thermal_solution_layered}
\eeq
where $\Xi_{i,n}$ and $\Theta\sub{i,n}^{(-)}$ are new frequency amplitudes that must be determined in terms of 
$\Theta_{i,n}, h$ and $K^{(-)}$.  The lower layer wavenumber $k_n^{(-)}$ is analogous to $k_n$, as defined in Eq. (\ref{kn_definition}), where
\beq
k_n^{(-)} = \left(\frac{i\rho C_p \omega n}{K^{(-)}}\right)^{1/2}, \qquad
{\rm with} \ \ {\rm Re}\left( k_n^{(-)}\right) > 0.
\label{kn_neg_definition}
\eeq
The thermal solution form of Eq. (\ref{thermal_solution_layered}) is entirely analogous to Eq. (\ref{thermal_solution_1}).  The solution for the lowest layer is designed to have zero flux as $z\rightarrow -\infty$.  The solutions across the layer separation $z=-h$ are connected to one another for each value of $n$ by imposing the continuity of temperature across the level, i.e.,
\beq
\Theta_{i,n} e^{-k_n h} + \Xi_{i,n} e^{k_n h} = \Theta\sub{i,n}^{(-)},
\eeq
as well as the continuity of thermal flux across the layer, i.e.
\beq
\Big[ K\partial_z T\Big]^{z\rightarrow -h^+}_{z\rightarrow -h^-} = 0,
\eeq
which means that for each index $n$ we require
\beq
k_n\Theta_{i,n} e^{-k_n h} - k_n \Xi_{i,n} e^{k_n h} = k_n^{(-)}\Theta\sub{i,n}^{(-)}.
\eeq
Solving these two simultaneous equations admits
\beq
\Xi_{i,n} = \Theta_{i,n}  \frac{1 - \sqrt{K/K^{(-)}}}{1 + \sqrt{K/K^{(-)}}}e^{-2hk_n},
\qquad
\Theta\sub{i,n}^{(-)} = \displaystyle{\frac{2\Theta_{i,n} e^{-hk_n}}{1 + \sqrt{K/K^{(-)}}}}.
\eeq
Therefore, the temperature at the surface is
\beq
T\sub i = 
\sum_{n=0}^N \Theta_{i,n}\left(1 +  \frac{1 - \sqrt{K/K^{(-)}}}{1 + \sqrt{K/K^{(-)}}}e^{-2hk_n}\right)e^{-i\omega n t} + {\rm c.c.},
\label{surface_temperature_layered}
\eeq
while the conductive flux is
\beq
\Lambda\sub i = K\partial_z T\sub i = 
\sum_{n=1}^N k_n\Theta_{i,n}\left(1 -  \frac{1 - \sqrt{K/K^{(-)}}}{1 + \sqrt{K/K^{(-)}}}e^{-2hk_n}\right)e^{-i\omega n t} + {\rm c.c.}
\label{surface_flux_solution_layered}
\eeq
Consequently, the solution method described in Sec. 3.4 may be executed exactly as it is there described, except for replacing Eq. (\ref{surface_flux_solution}) with Eq. (\ref{surface_flux_solution_layered}), and the surface temperature $T\sub i$
replaced by the expression in Eq. (\ref{surface_temperature_layered}).
In reference to the illustrative case described in Appendix \ref{simple_demo}, Eq. (\ref{equation_by_NRT}) would be rewritten instead
as
\beq
F\sub n(\Theta\sub m) + K k_n\Theta_{n}\left(1 -  \frac{1 - \sqrt{K/K^{(-)}}}{1 + \sqrt{K/K^{(-)}}}e^{-2hk_n}\right)
-f\sub n = 0, \qquad k\sub 0 = 0.
\label{equation_by_NRT_layered_solution}
\eeq 
The solution method described here may be straightforwardly generalized to multiple layers.  The coefficients may be calculated using any symbolic mathematics software or using any prepackaged linear system solver.

\section{Subsurface Radiative Transfer Details}\label{Radiative_Transfer_Solution_Details}
\REV{
In order to connect these interior thermal solutions to the observed brightness temperature, $T_{b,{\rm obs}} = 29 \pm 5$K, measured by New Horizons REX instrument, we have implemented the radiative transfer calculation detailed at length in \citet{deKleer_etal_2021}. A brightness temperature $T_b$ is defined as the temperature of a blackbody that produces the observed radiative intensity at frequency $\nu$.  We designate the Arrokoth disk averaged REX observed intensity by ${I}({\rm REX})$, which is the sum of the disk-averaged Arrokoth emitted intensity at frequency $\nu$ ($\tilde I_\nu$) plus the intensity received from the cosmic microwave background ($I_{\nu,{\rm cmb}}$).  The identification between 
${I}({\rm REX})$ and $T_{b,{\rm obs}}$ comes from the solution of
\beqa
{\cal I}({\rm REX}) = \tilde I_\nu+I_{\nu,{\rm cmb}}
&=&
\frac{\displaystyle 2h\nu^3\big/c^2}{\displaystyle
{\exp\left[{{\frac{h\nu}{kT_{b,{\rm obs}}}}}\right] - 1}}
+
\frac{\displaystyle 2h\nu^3\big/c^2}{\displaystyle
{\exp\left[{{\frac{h\nu}{kT_{{\rm cmb}}}}}\right] - 1}},
\eeqa
where $h$ is Planck's constant, $c$ speed of light and where the frequency $\nu = c/\lambda$.
Given that the X-band REX observation occurs at a wavelength $\lambda = 0.04$m, places this emission in the Rayleigh-Jeans end of the blackbody spectrum as
$T_\nu \equiv h\nu/k \approx 0.36$K $\ll T_{b,{\rm obs}}$ and $T_{{\rm cmb}}$, which simplifies the above into
\beq
{I}({\rm REX}) = \frac{2k\nu^2}{c^2}\Big(
T_{b,{\rm obs}} + T_{{\rm cmb}}\Big).
\eeq
}
\par
$\tilde I_\nu$ is the average of the emitted intensity ($\tilde I_{\nu,i}$) across the set of all spacecraft-viewable facets $i$ {\emph{weighted}} by each facet's spacecraft projected area $\sigma_i$.  We do not include any Gaussian weighting that would take into account of the REX beam.  For each facet $i$ with unit normal vector ${\bf \hat n}_i$ we define $\theta_i$ as the emission angle of the rays emitted from the surface that reach the spacecraft, from which it follows that if each facet has total area $s_i$, then $\sigma_i = s_i\cos\theta_i$. For geometrical reference see both Fig. \ref{Radiative_Transfer_Diagram} as well as Fig. \ref{Kij_Diagram} in Appendix A.   By construction it follows that
\beq
\tilde I_\nu =
 \frac{\displaystyle\sum_{\forall i \ {\rm visible}}\tilde I_{\nu,i} \sigma_i}{\displaystyle \sum_{\forall i \ {\rm visible}} \sigma_i},
 \label{tilde_I_nu_def}
\eeq
where all facets viewable by the spacecraft at the time of observations is derived directly from the network analysis that went into developing Arrokoth's thermal solutions as described in Section 3 and Appendix A.

\begin{figure}
\begin{center}
\leavevmode
\includegraphics[width=8.5cm]{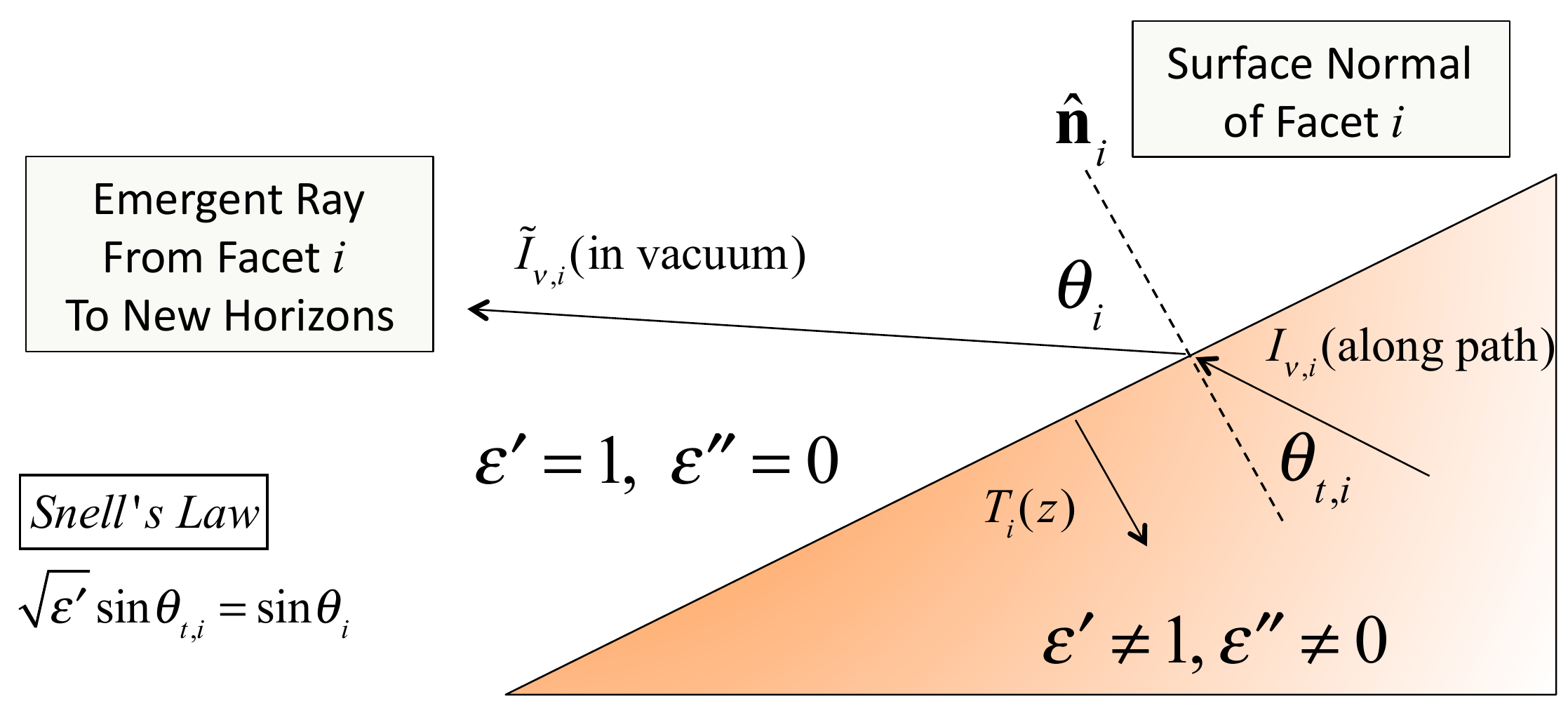}
\par
\end{center}
\caption{Diagram for simple radiative transfer calculation.}
\label{Radiative_Transfer_Diagram}
\end{figure}

\par
\REV{
The intensity $\tilde I_{\nu,i}$ emerging from the surface is the resulting radiation as having propagated from the interior.  This radiation originates from all points beneath the surface as attenuated on a scale set by the electric skin depth $\delta_{{\rm elec}}$, where
\beq
\delta_{{\rm elec}} \equiv \frac{\lambda}{4\pi\kappa},
\label{delta_elec_definition}
\eeq
in which $\kappa = {\rm Im}(n)$ and where $n$ is the complex index of refraction of the medium that is related
to the dielectric constant $\epsilon$ via $\epsilon \equiv n^2$, and is generally a function of a photon's wavelength.
While the electric skin depth is temperature dependent, for simplicity hereafter we will treat it as insensitive to temperature.
A material's $\epsilon$ is usually measured in the laboratory \citep[e.g., for cometary ices
at radar wavelengths see][]{Heggy_etal_2012} and is tabulated in terms of its real and imaginary parts for which we use the convention in \citet{deKleer_etal_2021},  $\varepsilon' \equiv {\rm Re}(\epsilon)$ and 
$\varepsilon'' \equiv {\rm Im}(\epsilon)$.  In all of our calculations below, we assume that $\epsilon$ has no subsurface variation. As such
\beq
\tilde I_{\nu,i}(\cos\theta_i) = 
\frac{\displaystyle \left<E_{{\rm eff}}\right> \int_0^\infty  I_{\nu,i}(z) e^{-z\big/\left(\delta_{{\rm elec}}\cos\theta_{t,i}\right)}dz}
{\displaystyle \int_0^\infty e^{-z\big/\left(\delta_{{\rm elec}}\cos\theta_{t,i}\right)}dz}
\label{tilde_I_nu_i}
\eeq
where $z$ is depth as measured normal to facet $i$'s surface,
$\theta_{t,i}$ is the angle formed between a subsurface propagating photon's path and facet $i$'s unit normal
(for reference see Fig. \ref{Radiative_Transfer_Diagram}). 
We have formally included an X-band emissivity factor
$\left<E_{{\rm eff}}\right>$ as is common practice when dealing
with solar system ices of unknown structural and transmission properties.
Because we have assumed $\epsilon$ is constant in the subsurface
the photon's subsurface slant-path trajectory changes upon exiting into vacuum and, therefore, 
$\theta_{t,i}$ relates to $\theta_{i}$ via Snell's Law, which following
its reformulation in \citet{deKleer_etal_2021}, is given by
\beq
\cos\theta_{t,i} = \sqrt{1- \frac{\sin^2\theta_i}{\varepsilon'}}.
\label{SnellsLaw}
\eeq
The subsurface intensity $I_{\nu,i}$ is 
\beq
I_{\nu,i}(z) = \frac{\displaystyle 2h\nu^3\big/c^2}{\displaystyle
{\exp\left[{{\frac{h\nu}{kT_i(z)}}}\right] - 1}}
\approx
\frac{2k\nu^2}{c^2}T_i(z),
\label{subsurface_intensity}
\eeq
The Rayleigh-Jeans limit is the approximate form given on the RHS of Eq. (\ref{subsurface_intensity}).
The subsurface temperature profile beneath facet $i$, $T_i = T_i(z,{\cal I},\cdots)$, are those as determined in previous sections, and we note that
these temperature solutions depends upon the material's thermal inertia (${\cal I}$), which we shall keep general for our considerations forthwith.  }
\par
\REV{
Using the Rayleigh-Jeans limiting form of Eq. (\ref{subsurface_intensity}) in 
(\ref{tilde_I_nu_i})
permits the re-expression of Eq. (\ref{tilde_I_nu_def}) in terms of a model prediction brightness temperature, hereafter simply as $T_b$,
\beqa
\tilde I_\nu &=&  \frac{2k\nu^2}{c^2} T_b\big({\cal I},\varepsilon',\varepsilon'',\big<E_{\rm eff}\big>\big),
\eeqa
where
\beq
T_b = T_b\Big({\cal I},\varepsilon',\varepsilon'',\big<E_{\rm eff}\big>\Big) =
\overline{T}_{b,i} \equiv
\frac{\displaystyle{\sum_{\forall i \ {\rm visible}} \sigma_i T_{b,i}}}{\displaystyle{\sum_{\forall i \ {\rm visible}} \sigma_i}},
\label{predicted_Tb}
\eeq
with a predicted X-band brightness temperature for each visible facet $i$, $T_{b,i}$ defined by
\beq
 T_{b,i}\Big({\cal I},\varepsilon',\varepsilon'',\big<E_{\rm eff}\big>\Big) \equiv
\frac{\displaystyle{\big<E_{\rm eff}\big>\int_0^\infty  T_{i}(z,{\cal I}) e^{-z\big/\left(\delta_{{\rm elec}}\cos\theta_{t,i}\right)}}dz}{\delta_{{\rm elec}}\cos\theta_{t,i}}.
\label{predicted_facet_Tb}
\eeq
}
\REVV{The overline symbol is intended here as a short-hand representation of the facet-area weighted averaging.}
\REV{
For any given instantaneous temperature solution determined using the methods of section 3, we can produce a model prediction $T_b$
that can be compared directly against the REX observed value $T_{b,{\rm obs}}$.
$T_b$ is a function of four unknown parameters, i.e.,  ${\cal I}$ (through $T_i$), the real and imaginary parts of the dielectric constant $\epsilon$, and the X-band emissivity $\big<E_{\rm eff}\big>$.  The dielectric constant determines the electric skin depth ($\delta_{{\rm elec}}$) and the photon's subsurface propagation angle $\theta_{t,i}$ that leads to a photon emission angle $\theta_i$ that, upon leaving Arrokoth's surface, leads directly to the spacecraft. Each $\theta_{t,i}$ is calculated by Snell's Law, Eq. (\ref{SnellsLaw}), which is a known function of the set of viewing angles $\theta_i$ formed between each facet's unit normal ${\bf{\hat n}}_i$ and the line connecting the spacecraft's position to each facet's center.}

\par
\REVV{ We note that
using an averaging approach Bird et al. (2022) develop $T_b$ along a parallel route that short-circuits the large scale computations involved in developing Eq. (\ref{predicted_Tb}) from Eq. (\ref{predicted_facet_Tb}).  Instead, $T_b$ is estimated straight away from (\ref{predicted_facet_Tb}) where $T_i$ is replaced by a visible facet surface area weighted temperature, $\overline{T(z)}$, and $\theta_{t,i}$ is replaced with a similarly facet area weighted average of the spacecraft-surface-normal inclination angle, $\overline{\theta_t}$.
In other words
\beq
T_b \approx 
\frac{\displaystyle{\big<E_{\rm eff}\big>\int_0^\infty  \overline{T_{i}(z,{\cal I})} e^{-z\big/\left(\delta_{{\rm elec}}\overline{\cos\theta_{t,i}}\right)}}dz}{\delta_{{\rm elec}}\overline{\cos\theta_{t,i}}}.
\eeq
Formally speaking, the two approaches are not-commensurate as the average of products -- namely that of the temperature and the depth dependent exponential factor -- is not necessarily equal to the product of their averages.  However, we do find that in this application the final results are largely consistent with one another. 
}
\par
\REV{
Thus, Following the procedures described in Section 3, we construct a set of model temperature solutions for several values of ${\cal I}_j$
contained in the range of $0.5$tiu and $300$tiu.  We collect the set of predicted temperature solutions for encounter day $T_i(z,{\cal I}_j)$ and generate predicted values of $T_b$ for the range of dielectric values $1<\varepsilon'< 3.5$ and $10^{-4} < \epsilon '' < 10$, which are motivated by those values of $\epsilon$ thought to be relevant to cometery ices.  The numerical solutions determined for each $T_i(z,{\cal I}_j)$ are tabulated only down to a depth $z_b = 10$m because we find that the temperatures below that depth have asymptoted to their orbitally averaged values, i.e., $T_i(z>z_b,{\cal I}_j) \approx T_i(z_b,{\cal I}_j)$.  As such, the integral
in Eq. (\ref{predicted_facet_Tb}) is well-approximated with
\beqa
& &\frac{\displaystyle{\int_0^\infty  T_{i}(z,{\cal I}_j) e^{-z\big/\left(\delta_{{\rm elec}}\cos\theta_{t,i}\right)} dz}}{\delta_{{\rm elec}}\cos\theta_{t,i}} \approx   \frac{\displaystyle{\int_0^{z_b}  T_{i}(z,{\cal I}_j) e^{-z\big/\left(\delta_{{\rm elec}}\cos\theta_{t,i}\right)}dz}}{\delta_{{\rm elec}}\cos\theta_{t,i}} 
+ T_{i}(z_b,{\cal I}_j)e^{-z_b\big/\left(\delta_{{\rm elec}}\cos\theta_{t,i}\right)}.
\eeqa
We observe that the added correction term on the RHS of the above expression is important only when $\delta_{\rm{elec}} \cos\theta_{t,i} \ll z_b$.}
\par
\REV{Fig. \ref{Predicted_Brightness_Temperatures} displays predicted $T_b$ for eight selected values of ${\cal I}$.  Each color contour plot shows $T_b$ as a function of $\varepsilon'$ and 
$\varepsilon''$.  Drawn atop each are black contour lines representing $T_{b,{\rm obs}} = 29$K, with dashed black contours showing the error bounds at 34K and 24K.  Also drawn with white contours a level sets for four corresponding values of $\delta_{{\rm elec}} = 10^{-2},10^{-1},1, 10$m, nominally indicating corresponding depths from which the bulk of the radiation emerges for a given value of $\varepsilon$ of interest. The qualities and implications of these solutions are discussed further in the body of the text.  However we do observe from Fig. \ref{Predicted_Brightness_Temperatures} that the predicted values of $T_b$ are weakly dependent on $\varepsilon '$ in the range 1 and 3.5.

}

\bibliography{Arrokoth_Paper}{}
\bibliographystyle{aasjournal}

\end{document}